\documentclass[journal, 10pt]{IEEEtran}

\makeatletter
\newcommand\approxsim{\mathpalette\@approxsim\relax}
\newcommand\@approxsim[2]{%
  \mathrel{%
    \ooalign{%
      $\m@th#1\sim$\cr
      \hidewidth$\m@th#1:$\hidewidth\cr
    }%
  }%
}
\makeatother

\ifCLASSINFOpdf
  \usepackage[pdftex]{graphicx}
\else
\fi
\usepackage{amsmath}
\usepackage{float}
\usepackage{soul}
\usepackage[safe]{tipa}

\usepackage{array}
\usepackage{makecell}
\usepackage{multirow}
\usepackage{hyperref}
\usepackage{subcaption}
\usepackage{xcolor}
\usepackage[noadjust]{cite}
\begin{document}

%
\title{On Addressing the Impact of ISO Speed upon PRNU and Forgery Detection}


\author{Yijun Quan, Chang-Tsun Li~\IEEEmembership{Senior Member,~IEEE}
\thanks{Y. Quan is with the Department
of Computer Science, University of Warwick, Coventry, CV4 7AL United Kingdom, e-mail: Y.Quan@warwick.ac.uk.}
\thanks{C.-T. Li is with the School of Information Technology, Deakin University, Geelong VIC 3220 Australia, e-mail: changtsun.li@deakin.edu.au.}
\thanks{}}

\maketitle

\begin{abstract}

Photo Response Non-Uniformity (PRNU) has been used as a powerful device fingerprint for image forgery detection because image forgeries can be revealed by finding the absence of the PRNU in the manipulated areas. The correlation between an image's noise residual with the device's reference PRNU is often compared with a decision threshold to check the existence of the PRNU. A PRNU correlation predictor is usually used to determine this decision threshold assuming the correlation is content-dependent. However, we found that not only the correlation is content-dependent, but it also depends on the camera sensitivity setting. \textit{Camera sensitivity}, commonly known by the name of \textit{ISO speed}, is an important attribute in digital photography. In this work, we will show the PRNU correlation's dependency on ISO speed. Due to such dependency, we postulate that a correlation predictor is ISO speed-specific, i.e. \textit{reliable correlation predictions can only be made when a correlation predictor is trained with images of similar ISO speeds to the image in question}. We report the experiments we conducted to validate the postulate. It is realized that in the real-world, information about the ISO speed may not be available in the metadata to facilitate the implementation of our postulate in the correlation prediction process. We hence propose a method called Content-based Inference of ISO Speeds (CINFISOS) to infer the ISO speed from the image content.
%
\end{abstract}


%

\section{Introduction}
\IEEEPARstart{W}{hen} a digital image is used in a forensic investigation or presented as evidence to the court, it is important to authenticate the image to ensure its content is free from manipulation. Thus, image forgery detection draws substantial attentions from researchers. Among different techniques developed for image forgery detection, Photo Response Non-Uniformity (PRNU) based methods have shown their unique strength. PRNU is a sensor pattern noise intrinsically embedded in images. It arises as a result of the manufacturing imperfections of silicon wafers in image sensors. As such, pixels on a sensor would have a non-uniform response to the incident light and introduce a unique pattern noise to the image, which can be treated as the fingerprint of a device. Many different algorithms have been proposed for PRNU-based source camera identification \cite{lukas2006digital,4284577,li2010source,van2012using, cooper2013improved,gisolf2013improving,kang2014context,al2015novel,zeng2016fast, lawgaly2017sensor,6889280} and image forgery detection \cite{lukavs2006detecting,chen2008determining,6004957,6719568,6854802,7776959}. In most of these works, PRNU is utilized by computing the image-wise or block-wise correlations between the source device's reference PRNU and the test image's PRNU. The corresponding image-wise (source camera identification) or pixel-wise decision (forgery detection) can be made by comparing the correlations with a decision threshold.

The PRNU is often estimated in the form of the noise residual of an image. The noise residual can be extracted from an image by simply subtracting the de-noised image from the original image. By nature, PRNU is a weak noise. The existence of camera artifacts and other PRNU-irrelevant noises (e.g. shot noise, thermal noise, etc.) in an image's noise residual can reduce the correlation between the noise residual and the device's reference PRNU. It becomes a non-trivial problem to separate the inter-class (images from different source devices) from the intra-class (images from the same source device) correlations. It becomes particularly problematic when the PRNU quality in the noise residual is poor such that these two types of correlations' distributions can have large overlaps.  

Despite a large number of works that have been done to better extract, estimate and enhance the PRNU \cite{van2012using, cooper2013improved, gisolf2013improving, zeng2016fast, chen2008determining, 7026084, li2010source, 6020790, 7265059, 5934587, 6638213}, the overlap between inter- and intra-class correlations cannot be completely avoided. Thus, many researchers have been working on refining the choice of the decision thresholds to better separate the two classes, especially for image forgery detection \cite{chen2008determining, 6719568, 7776959 }. The decision thresholds are often set with reference to the expected intra-class correlations predicted by a correlation predictor. The correlation between an image's noise residual and the device's reference PRNU reflects the strength of the PRNU in the image. As the strength of the PRNU is multiplicative of the pixel intensity and some highly textured image content or post-processing may damage the PRNU's quality, correlation prediction should be performed in an adaptive manner. A content-dependent correlation predictor is proposed by Chen \textit{et al.} in \cite{chen2008determining}, which formulates the correlation predictor as a regressor model of four image features, namely the \textit{intensity}, \textit{texture}, \textit{signal-flattening} and a \textit{texture-intensity combinative term}. This correlation predictor has been adopted by many PRNU-based forgery detection algorithms (e.g. \cite{chen2008determining,6004957,6719568,6854802,7776959}). Due to the complex nature of the PRNU correlation, despite different attempts to re-engineer the correlation predictor over the past decade, we have not witnessed much success. Thus, the digital forensic community still relies greatly on the correlation predictor from \cite{chen2008determining} for PRNU-based forgery detection.

However, over the last decade, we have also witnessed great advancement in the digital camera industry, especially in sensor design. Such advancement also brings new challenges to PRNU-based digital forensics. Therefore, we have observed a few issues about the correlation predictor proposed in \cite{chen2008determining}. An important feature ignored by the correlation predictor is the camera sensitivity setting, which is commonly known by the name of ISO speed. The ISO speed together with the shutter speed and the aperture size are the three parameters, which control an image's exposure in digital photography. The shutter speed and aperture size control the number of photons arriving at the image sensor during the exposure process while the ISO speed determines the camera's signal gain. In real-life, photographers may face many physical restrictions on the aperture size and shutter speed. Such restrictions require more freedom of choice in ISO speeds to achieve the desired exposure. Thus, many camera manufacturers have been working on improving sensor performance and providing more and higher ISO speeds to digital cameras. While the improvements have been brought to sensor technology, it is also a known fact that high ISO speeds may introduce more noise to an image. As a result, the quality of the PRNU left in the noise residual will be reduced when a high ISO speed is used. A recent work presented in \cite{8451688} empirically shows that different ISO speeds may affect the performance of PRNU-based source camera identification. With camera manufacturers increasingly supporting broader ranges of ISO speed settings on digital cameras and mobile devices, a proper analysis of the ISO speed's influence on PRNU-based image forensics, especially on the correlations, needs to be carried out.

As this work focuses on the correlation between an image's noise residual with its reference PRNU, for simplicity, we will call it the \textit{correlation}. The contribution of this work can be summarized as follows:
\begin{itemize}
\item We first analytically and empirically proved in Section II that the correlation between an image’s noise residual and its reference PRNU is not only content-dependent as previously known, but also dependent on the \textit{camera sensitivity setting} (i.e. the \textit{ISO speed}).
\item We then validate our postulate in Section III that, due to such ISO speed dependency, reliable predictions of the correlation between an image’s noise residual and its reference PRNU can only be accurately made when a correlation predictor is trained on images of similar ISO speeds to the image in question.
\item Base on the postulate, we propose an ISO specific correlation prediction process. Recognizing that in the real-world, information about the ISO speed may not be available to facilitate the implementation of our postulate in the correlation prediction process, we propose a method called Content-based Inference of ISO Speeds (CINFISOS, /\textipa{'sin.f@.s@s}/) in Section \ref{sec:infer}
 to infer the ISO speed from the image content.
\end{itemize}

In order to carry out this in-depth investigation into how the ISO speed can affect PRNU-based image forensics, we use the purposefully built Warwick Image Forensics Dataset \cite{quan2020warwick}. The images in this dataset are taken with diverse exposure parameter settings. The dataset involves 14 cameras and images of various scenes. In particular, for 20 different scenes for each camera, multiple images of the same scene are shot with varying ISO speeds and exposure times. Thus, these images allow us to conduct studies on the ISO speed's influence on the correlation.

\section{ISO Speed Dependent Correlation}
In this section, we demonstrate that an image's ISO speed can affect its correlation. As a general noise model can be complicated, to show the existence of such an \textit{ISO Speed-Correlation} relationship in a concise manner, we use a special case to prove this relationship analytically and then empirically show it with more general cases. The special case considered is a single color channel of a flat-field RAW image, from which we expect the same value for every pixel if they are noise-free. To conduct PRNU-based pixel-wise forgery detection, the correlation between the noise-residual of a block centered at each pixel and the corresponding block of the reference PRNU is calculated. Let $\boldsymbol z$ be a noise residual within a block $N_i$ centered at pixel $i$ and $\boldsymbol \omega$ be the reference fingerprint within the corresponding block. Assume both $\boldsymbol z$ and $\boldsymbol \omega$ are standardized, which means they follow the normal distribution $\mathcal{N}(0,1)$. We can model both signals as the sum of a PRNU component and a PRNU-irrelevant part. At pixel $j\in N_i$:
\begin{equation}
\label{eqn:noise_model}
 	\begin{cases}
 		\omega_j = x_j + \alpha_j \\
 		z_j = y_j + \beta_j
 	\end{cases}
\end{equation}
where $\boldsymbol x$ and $\boldsymbol y$ are the PRNU components of $\boldsymbol \omega$ and $\boldsymbol z$ while $\boldsymbol \alpha$ and $\boldsymbol \beta$ are the PRNU-irrelevant noises. As for a flat-field image, we can approximate its PRNU component, $\boldsymbol x$ in this case, as a  normal distribution $\mathcal{N}(0,{\sigma_x}^2)$ and $\boldsymbol \alpha$ conforms to $\mathcal{N}(0, 1-{\sigma_x}^2)$. For intra-class pairs, $\boldsymbol x$ and $\boldsymbol y$ represent the same PRNU. As they may differ in strength, without losing generality, we can express $\boldsymbol y$ as $\mathcal{N}(0, {\sigma_y}^2)$ with ${\sigma_y} = \sqrt{\lambda} \sigma_x$ and $\boldsymbol y = \sqrt{\lambda} \boldsymbol x$. $\boldsymbol \alpha$ and $\boldsymbol \beta$ are mutually independent. So when we compute the correlation $\rho_i$ of the block $N_i$, the correlation $\rho_i$ becomes:
\begin{equation}
\label{eqn:sim}
\rho_i \sim \mathcal{N}(\mu_i, \Sigma_i)
\end{equation}
with 
\begin{equation}
 \begin{cases}
 \label{eqn:corr}
 \mu_i = \sigma_x \sigma_y =\sqrt{\lambda} {\sigma_x}^2\\
 \Sigma_i = (1+\lambda {\sigma_x}^4)/|N_i|
 \end{cases}
\end{equation}
From the above expression, we can see that the expected correlation value, $\mu_i$, is proportional to the standard deviation $\sigma_y$ of the PRNU component, $\boldsymbol y$, in the image's noise residual, $\boldsymbol z$. Based on the Poissonian-Gaussian noise model \cite{radioCCD,4623175,foi2009clipped}, we can see that the ISO speed would affect this standard deviation $\sigma_y$ and eventually exert influence on the PRNU correlations.

The relationship between the camera gain, $g$, which is directly determined by the camera's ISO speed, and the noisy raw pixel intensity, $I$, is analyzed in \cite{4623175}. The raw pixel intensity is proportional to the number of electrons counted on the sensor. Photo-electron conversion is the main source of the electrons collected from the sensor. \cite{4623175} considers the Poissonian statistics of the incident photon counting process as follows. At pixel $i$, the number of the counted electrons is the sum of the electrons generated from photo-electron conversion ${N_p}_i$  and dark electrons ${N_t}_i$ from the thermal noise. It is assumed that the variance of the thermal noise is uniform across the sensor and all other electronic noises can be modeled as a zero-mean Gaussian noise with variance $s^2$. So the raw pixel intensity, $I_i$, at pixel $i$, can be written as:
\begin{equation}
\label{eqn:intmodel}
I_i \sim g \cdot [p_0 + \mathcal{P}(\eta_i {N_p}_i + {N_t}_i - p_0) + \mathcal{N}(0,s^2)]
\end{equation}
where $\mathcal{P}(\cdot)$ represents the Poisson distribution and $\eta_i$ is the photon-electron conversion rate at pixel $i$. $p_0$ is a base pedestal parameter introduced in the camera design to provide an offset-from-zero of the pixel's output intensity. For each pixel, as a large number of electrons are counted, the normal approximation of Poisson distribution can be exploited. Therefore, $I_i$ can be modeled as:
\begin{equation}
\label{eqn:nosmod}
	I_i \sim \mathcal{N}(\varphi_i,g \varphi_i + t)
\end{equation}
with 
\begin{equation}
\begin{cases}
  t = g^2 s^2 - g^2 p_0\\
  \varphi_i = g \cdot (\eta_i {N_p}_i + {N_t}_i)
\end{cases}	,
\label{eqn:nosmodexp}
\end{equation}
$\varphi$ can be viewed as the expected pixel intensity. Notice that this model from \cite{4623175} has not yet considered the PRNU. To include the PRNU in this model, we write the photo-electron conversion rate $\eta_i$ as the following expression by considering the non-uniform response of each pixel to the photons:
\begin{equation}
\eta_i = \bar{\eta} (1+k_i),
\end{equation} 
where $\bar{\eta}$ is the average photo-electron conversion rate and $k_i$ is the PRNU factor at pixel $i$. $k$ follows normal distribution $\mathcal{N}(0, \sigma_k^2)$. As we are considering the case of a flat-field image here so we can fix the number of photons, $N_{p_i}$, collected at every pixel. By expanding Equation (\ref{eqn:nosmod}), we have:
\begin{equation}
I_i \sim \mathcal{N}((1+k_i)\varphi-gk_i {N_t}_i, g(1+k_i)\varphi+t - g^2k_i {N_t}_i)
\label{eqn:inter}
\end{equation}
As in most cases, both the PRNU and the thermal noise are weak noises. We can ignore the terms involving $k_i {N_t}_i$. When we consider a block $N_i$, often it consists of thousands of pixels (e.g. 4096 pixels for a $64 \times 64$ block). Such a large number of pixels allow us to approximate the overall distribution of the pixel values in this block by another normal distribution. By substituting $t$ of Equation (\ref{eqn:inter}) with the expression for $t$ in Equation (\ref{eqn:nosmodexp}), we approximate the distribution of the pixel values in block $N_i$ as:
\begin{equation}
I_{N_i} \approxsim \mathcal{N}(\varphi, \varphi^2 {\sigma_k}^2 + g \varphi +g^2 s^2 - g^2 p_0 )
\end{equation}
We expect the de-noised version of this block to have pixels of uniform intensity, $\varphi$. Thus, we can approximate the variance of the noise residual of this block as:
\begin{equation}
\label{eqn:distri}
{\sigma_{\text{res}}}^2 \approx \varphi^2 \sigma_k^2 + g \varphi +g^2 s^2 - g^2 p_0
\end{equation}
The PRNU component in the noise residual has a variance of $\varphi^2\sigma_k^2$. By normalizing the noise residual, the standard deviation of the PRNU component in the normalized noise residual becomes:
\begin{equation}
\label{eqn:str}
\sigma_{y} =\sqrt{\frac{\varphi^2\sigma_k^2}{\varphi^2 {\sigma_k}^2 + g \varphi +g^2 s^2 - g^2 p_0}}
\end{equation}
Clearly, $\sigma_{y}$ is dependent on the camera gain $g$. By substituting this expression back to Equation (\ref{eqn:corr}), we can conclude that the correlation $\rho_i$ can be affected by the camera gain $g$ and thus affected by ISO speed.

Notice that when we introduce PRNU by considering different photo-electron conversion rate, $\eta_i$, at each pixel to the raw pixel intensity model from \cite{4623175}, the noise residual variance model described in Equation ($\ref{eqn:distri}$) becomes a quadratic function of the expected pixel intensity $\varphi$, which can be expressed as:
\begin{equation}
{\sigma_\text{res}}^2 = A \varphi^2 + B \varphi + C
\label{eqn:qudratic}
\end{equation}
with
\begin{equation}
\begin{cases}
  A = {\sigma_k}^2\\
  B = g \\
  C = g^2s^2-g^2p_0
\end{cases}	
\end{equation}
 It differs from the linear model in \cite{4623175}. We will empirically validate Equation (\ref{eqn:distri}) to show the physical importance of the PRNU term, $\varphi^2 {\sigma_k}^2$, in the equation despite the approximations made.
 
\begin{figure}
\center
\begin{tabular}{cc}
(a) Nikon D7200& (b) Canon 6D MKII   \\
   \includegraphics[width=.23\textwidth]{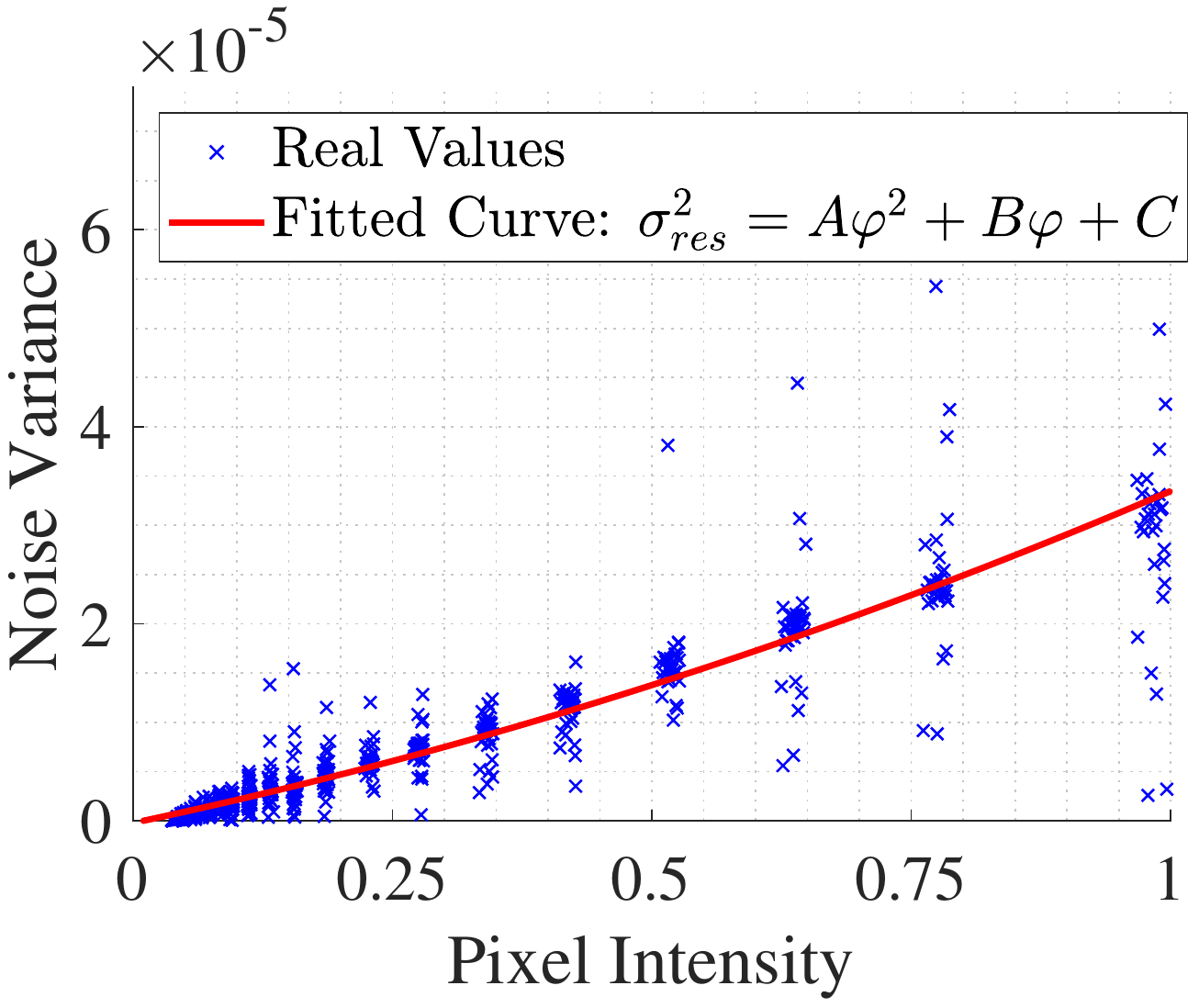} &   \includegraphics[width=.23\textwidth]{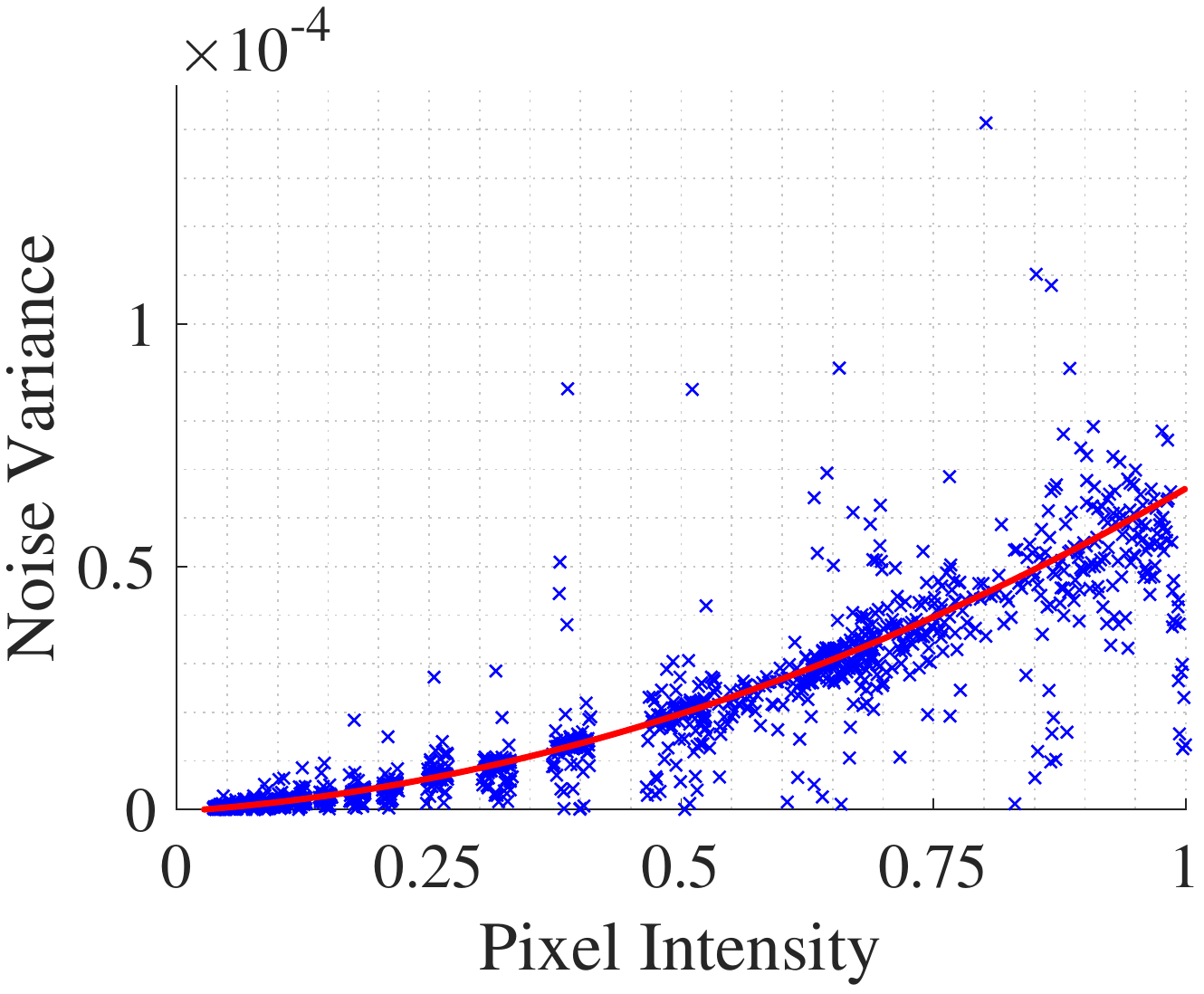} \\
 (c) Canon 80D &  (d) Canon M6  \\
  \includegraphics[width=.23\textwidth]{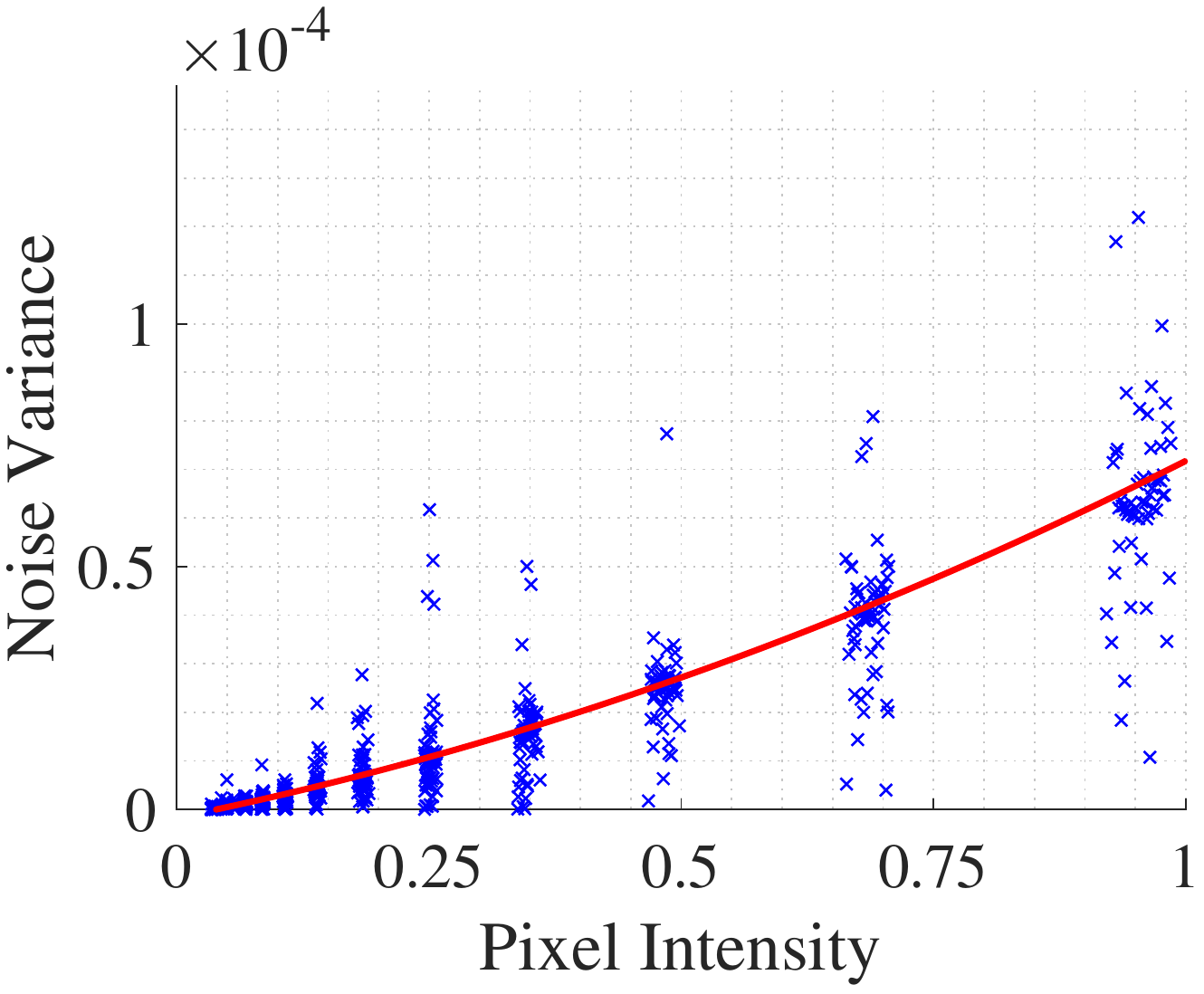} &    \includegraphics[width=.23\textwidth]{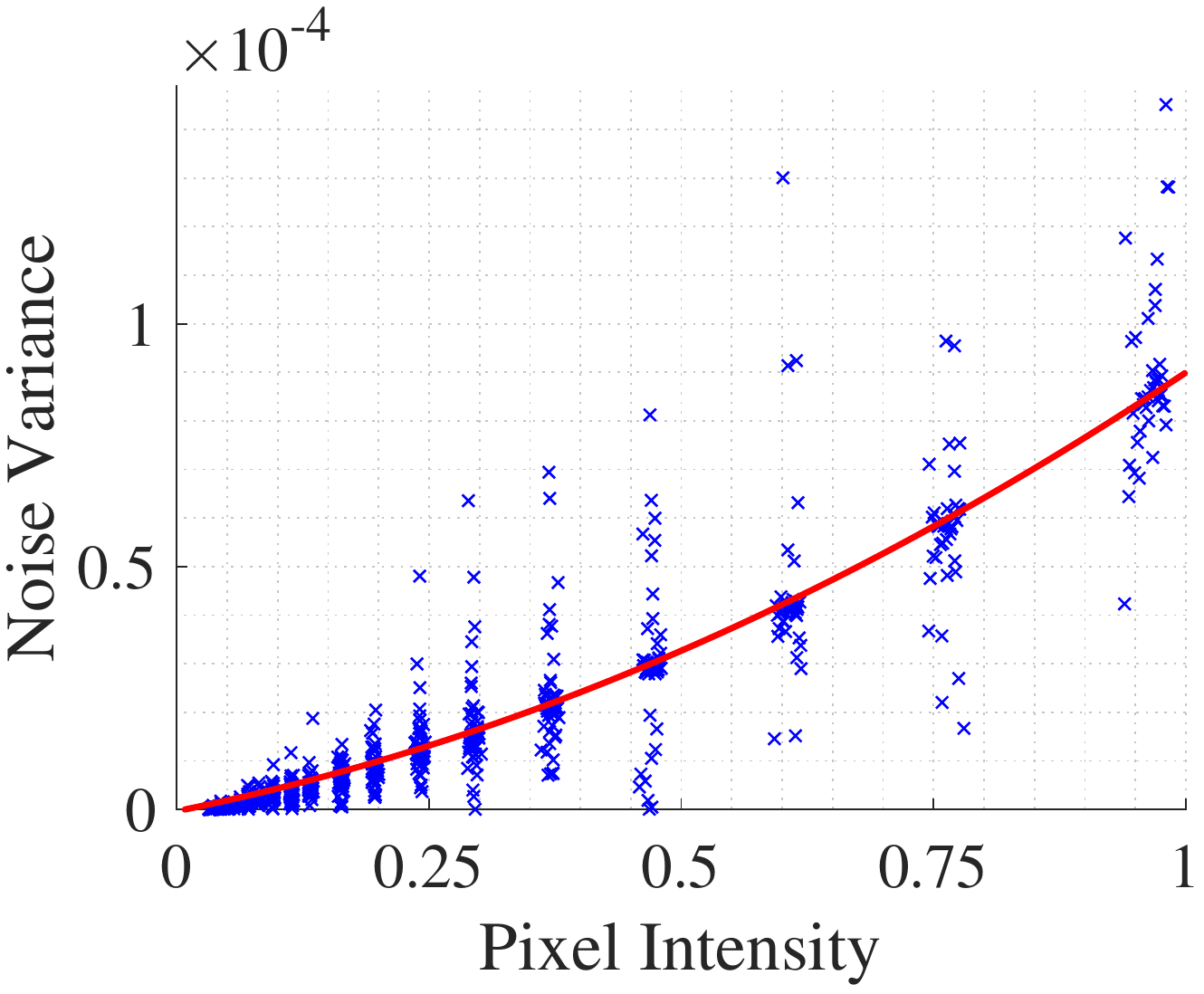} 
\end{tabular}
    \caption{Plots of noise's variance ${\sigma_\text{res}}^2$ against pixel intensity $\varphi$, with a quadratic fitting (red curve) as described by Equation (\ref{eqn:distri}) and (\ref{eqn:qudratic}), of RAW flat-field ISO 100 images from four cameras: (a) Nikon D7200, (b) Canon 6D MKII, (c) Canon 80D and (d) Canon M6. The fitted coefficients for Equation (\ref{eqn:qudratic}) for each image are: (a) $A = 1.14\times 10^{-5}, B = 2.23\times 10^{-5}, C = -2.20\times10^{-7}$, (b) $A=5.24\times 10^{-5}, B=1.41 \times 10^{-5}, C=-4.33\times 10^{-7}$, (c) $A = 3.15 \times 10^{-5}, B = 4.20\times 10^{-5}, C = -1.70\times 10^{-6}$, (d) $A = 4.85 \times 10^{-5}, B = 4.18 \times 10^{-5}, C = -3.51 \times 10^{-7}$}
    \label{fig:munu}
\end{figure}

We use four cameras for the test, namely a Nikon D7200, a Canon 6D MKII, a Canon 80D, and a Canon M6. Each of the four cameras can generate 14-bits RAW images, which means their pixel values can vary between the range of $[0, 16383]$. To better show the physical meaning of the coefficients in Equation (\ref{eqn:distri}), we standardize the pixel values to the range of $[0,1]$. To validate Equation (\ref{eqn:distri}), we plot the variance of the noise in the flat-field images against different pixel values in Fig.\ref{fig:munu}. We use the cameras to take images of a screen of flat color. Each camera's ISO speed is set to 100. The exposure time is varied to change the pixel intensity for different shots. As the cameras use Bayer-filter as their color filtering array (CFA), we subsample the RAW images with a stride of 2 in both vertical and horizontal directions to make sure the pixels we test are from the same color channel. Despite the set-up, the images are not completely flat due to other camera artifacts, e.g. vignetting. Thus, we use the method from \cite{4623175} to estimate the expected pixel value and variance for multiple image blocks from each noisy RAW image. Fig.\ref{fig:munu} shows the fitting of Equation (\ref{eqn:qudratic}) to the experiment data, which is computed using ordinary least squares (OLS) \cite{riley2006mathematical}. A good agreement between the model and the data can be observed.

\begin{figure}
\begin{center}
\begin{tabular}{cc}
(a) ISO200& (b) ISO400   \\ 
  \includegraphics[width=.23\textwidth]{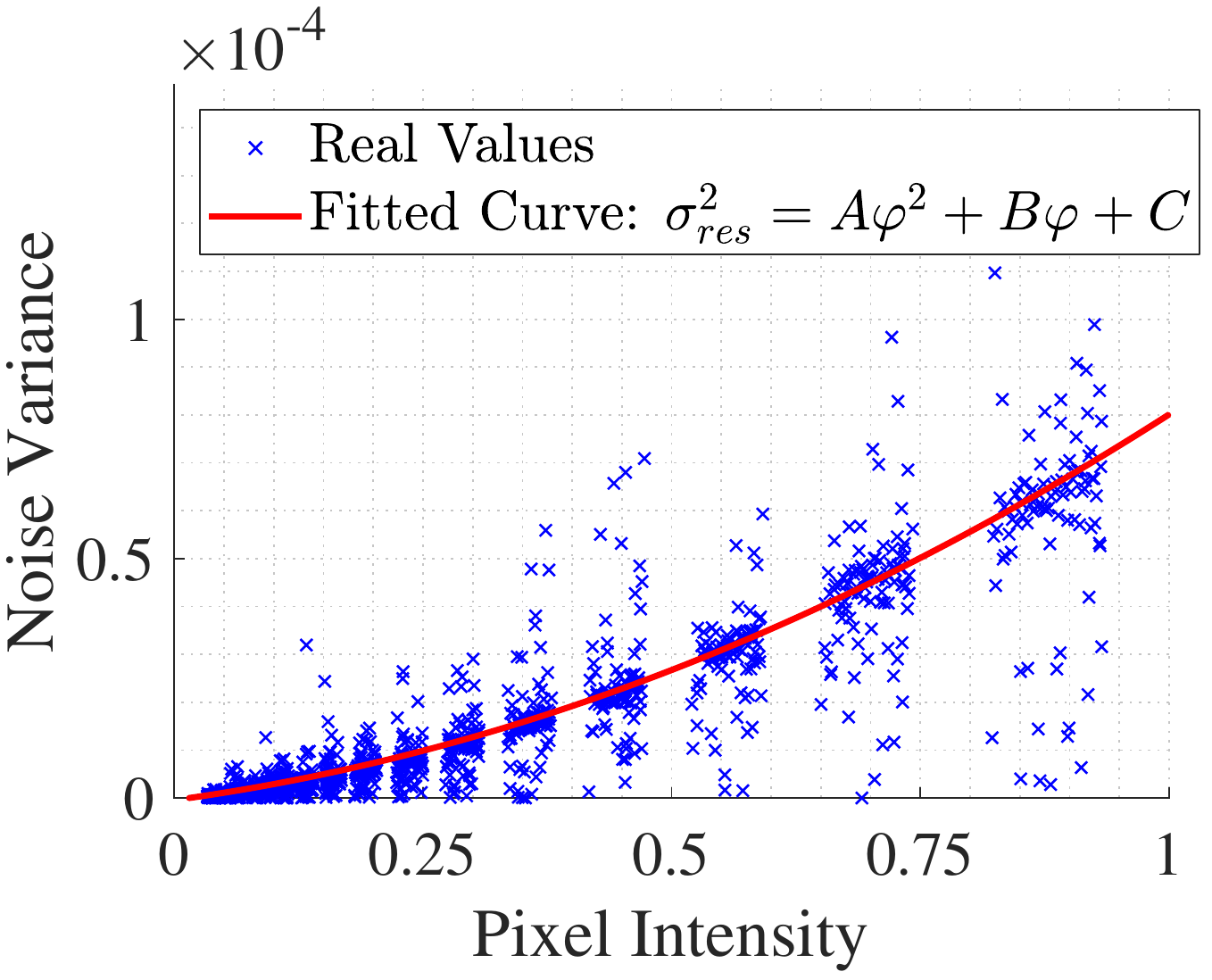} &   \includegraphics[width=.23\textwidth]{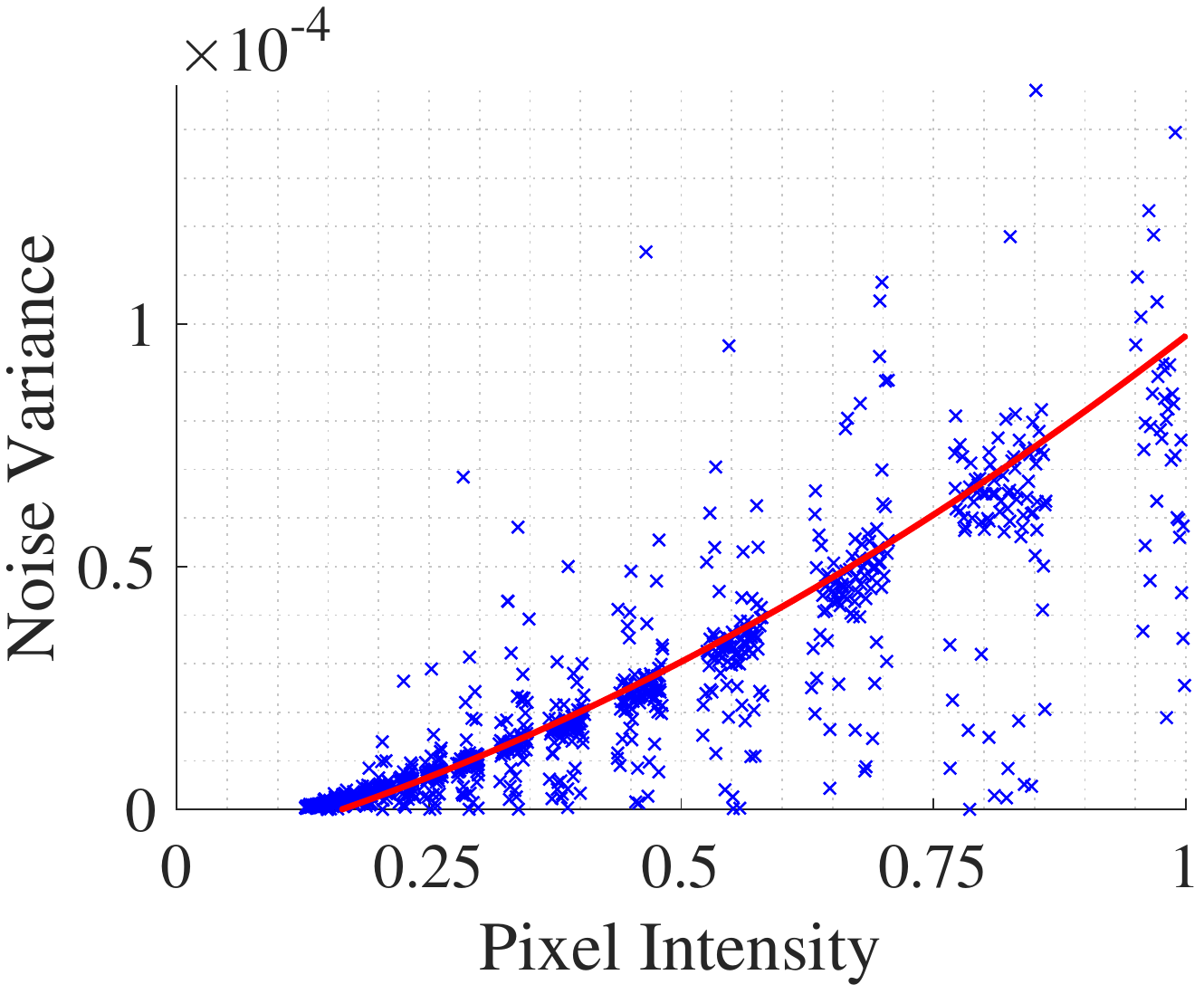} \\
 (c) ISO800 &  (d) IS01600  \\
  \includegraphics[width=.23\textwidth]{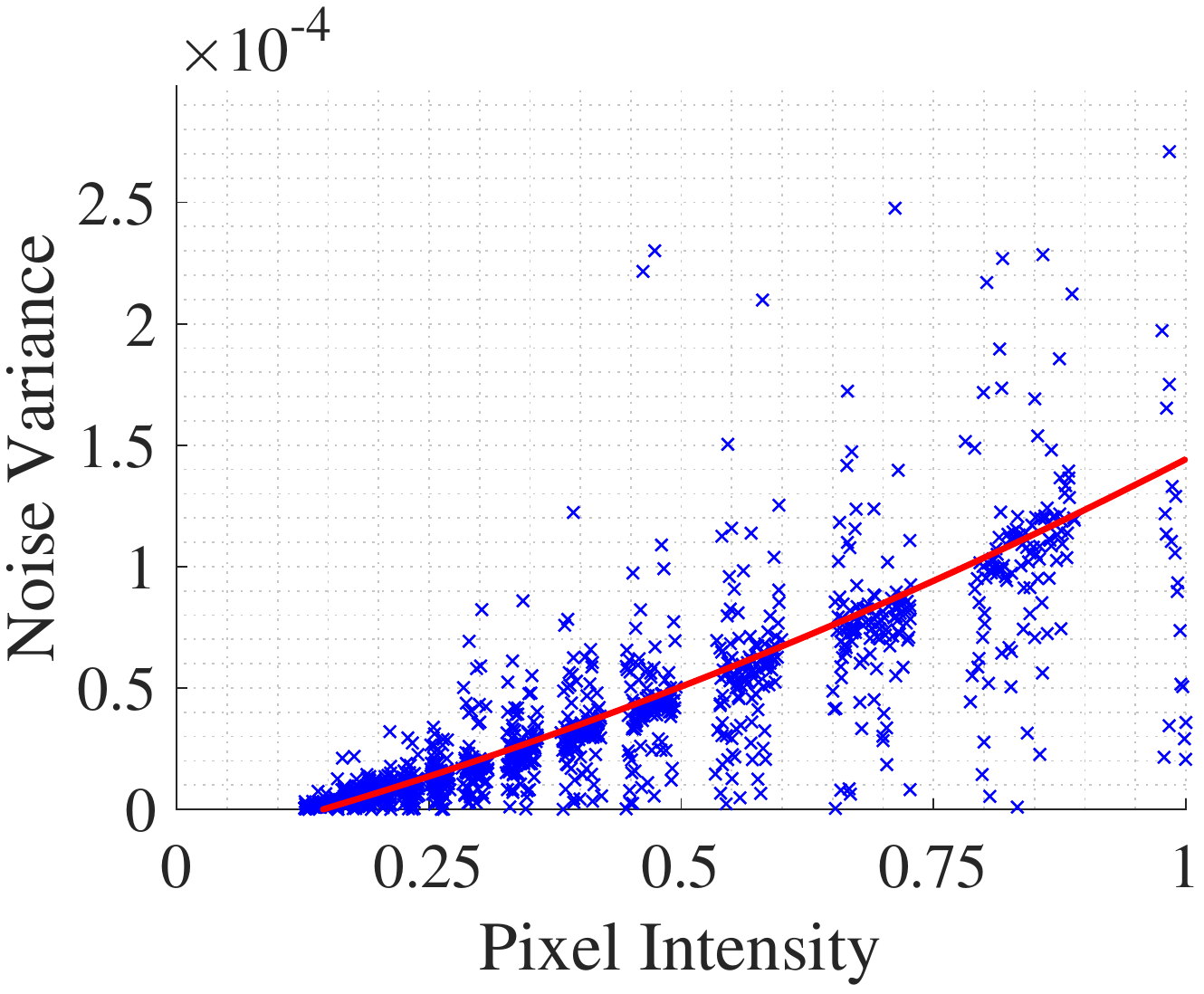} &    \includegraphics[width=.23\textwidth]{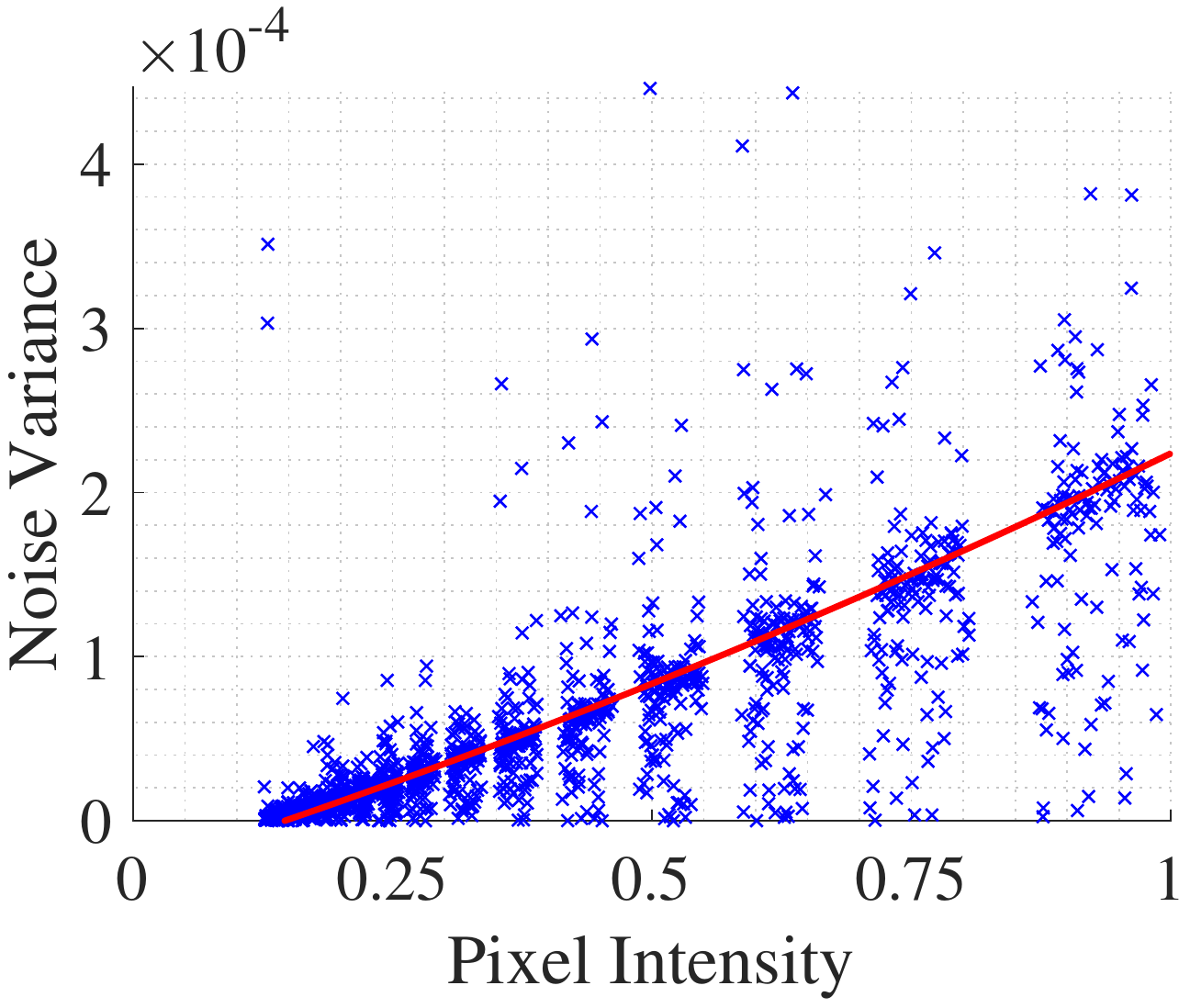} 
\end{tabular}
\caption{Plots of noise's variance ${\sigma_\text{res}}^2$ against pixel intensity $\varphi$ of images with different ISO speed from a Canon 6D MKII. We fit Equation (\ref{eqn:distri}) to the plots with a fixed second order coefficient, $A = \sigma_k^2 = 5.24 \times 10^{-5}$, estimated from Fig.\ref{fig:munu}(b). The first order coefficient $B$, for the four fittings are: (a) $B = 2.81\times 10^{-5}$, (b) $B= 5.56\times 10^{-5}$, (c) $B = 1.09\times 10^{-4}$ and (d) $B = 2.02 \times 10^{-4}$}
\label{fig:6dmunu}
\end{center}
\end{figure}

\begin{figure}
\centering
\includegraphics[width=.35\textwidth]{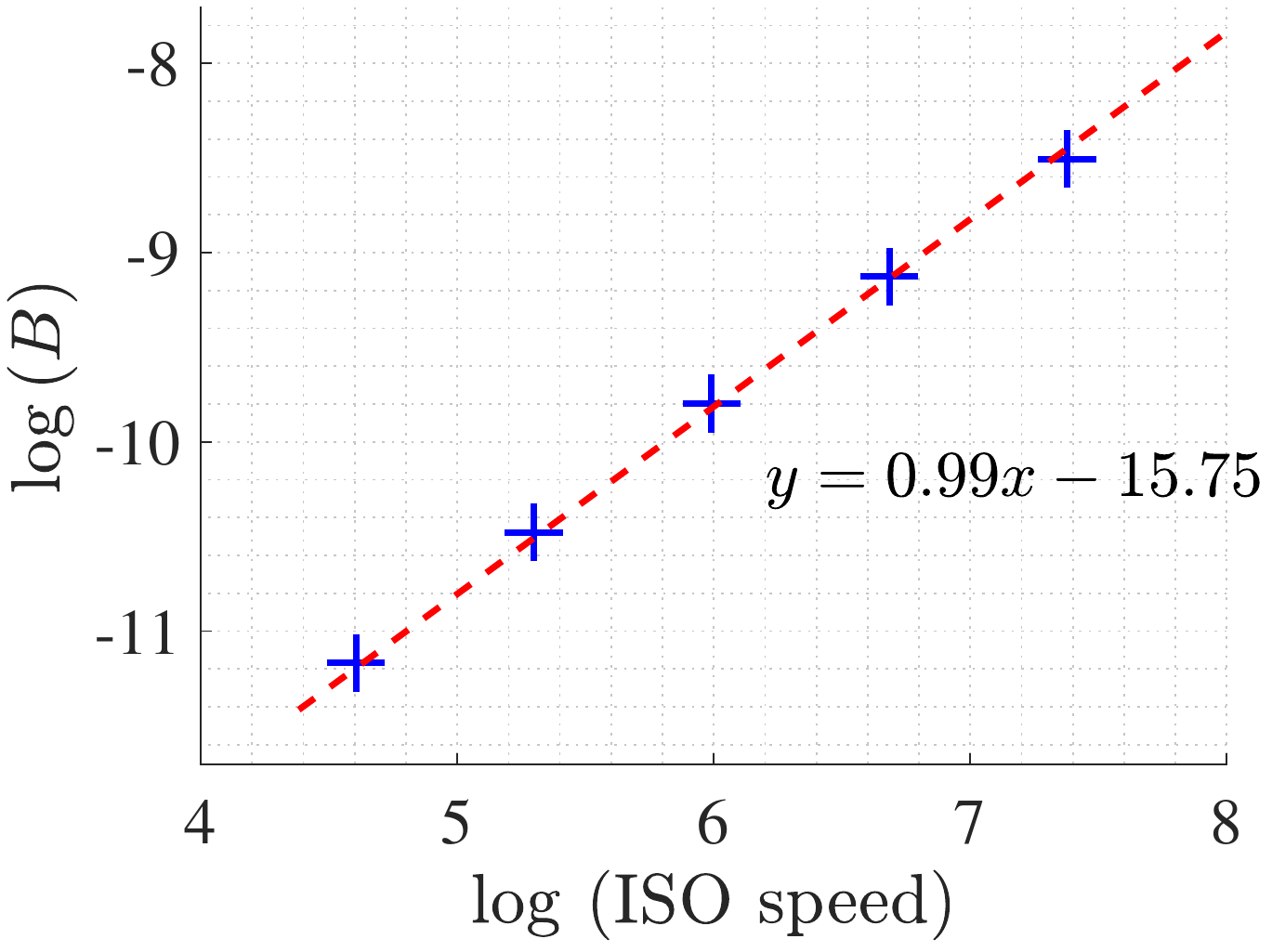}
\caption{$\log$-$\log$ plot of the estimated first order coefficient $B$ against the ISO speeds of the images used to estimate $B$. A straight line is fitted with a slope of $0.99$}
\label{fig:ISO_gain}
\end{figure}

In addition to showing the good agreement of the derived model and the real data, we would like to show the physical meaning of the first order coefficient, $B = g$ in the model as well. We use the RAW images from the same Canon 6D MKII from the previous test for this test. We repeat the previous experiment four times but set the cameras' ISO speed to ISO 200, 400, 800, and 1600, respectively. Again, we fit Equation (\ref{eqn:qudratic}) to the data. As for the same camera, despite the change of ISO speed, we can assume that the PRNU factor on the sensor should remain the same and so does the variance of the PRNU factor, $\sigma_k^2$. Thus, it is reasonable for us to fix the second order coefficient $A = \sigma_k^2$ to $5.24\times 10^{-5}$, the value estimated from Fig.\ref{fig:munu}, in Equation (\ref{eqn:qudratic}) for these fittings and the corresponding fittings generated using OLS are shown in Fig.\ref{fig:6dmunu}. Once again, good agreement between the fitted curve and the data can be observed. In addition, we show a $\log - \log$ plot of the estimated first order coefficients $B$ from Fig.\ref{fig:munu}(b) and \ref{fig:6dmunu} against the ISO speed of their corresponding images in Fig.\ref{fig:ISO_gain}. We fitted a straight line to the plot given slope close to $1$. As a camera's ISO speed is proportional to its camera gain, $g$, Fig.\ref{fig:ISO_gain} validates our noise model from Equation (\ref{eqn:distri}) with $B = g$. Therefore, it confirms that the correlation model is dependent on ISO speed.

\begin{figure*}
\begin{center}
\begin{tabular}{ >{\centering\arraybackslash}m{1.2cm} >{\centering\arraybackslash}m{3.5cm}>{\centering\arraybackslash} m{3.5cm} >{\centering\arraybackslash}m{3.5cm} >{\centering\arraybackslash}m{3.5cm} }
& ISO 100 & ISO 400 & ISO 1600 & ISO 6400\\
\makecell{Scene 1} &\includegraphics[width=.16\textwidth]{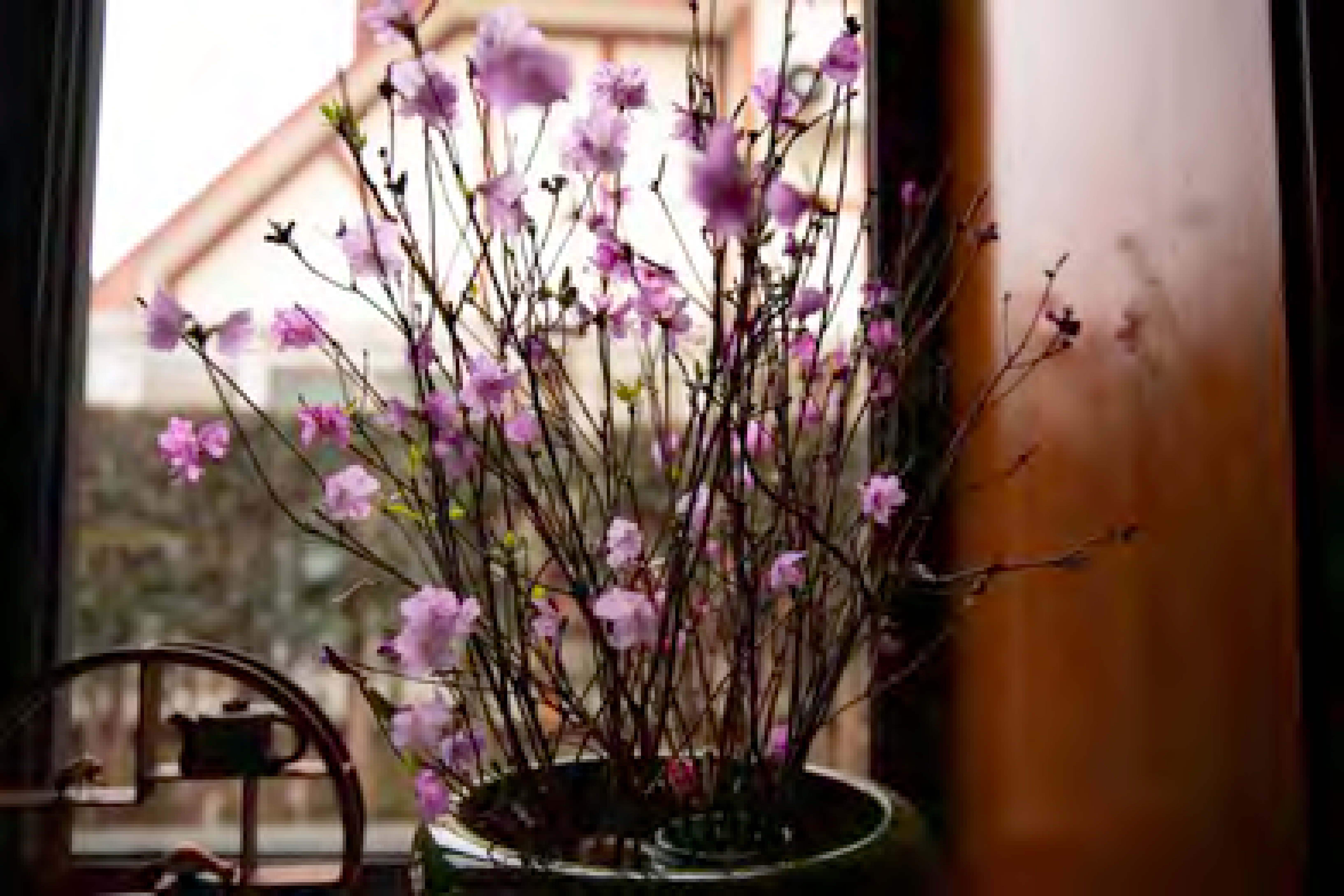} & \includegraphics[width=.16\textwidth]{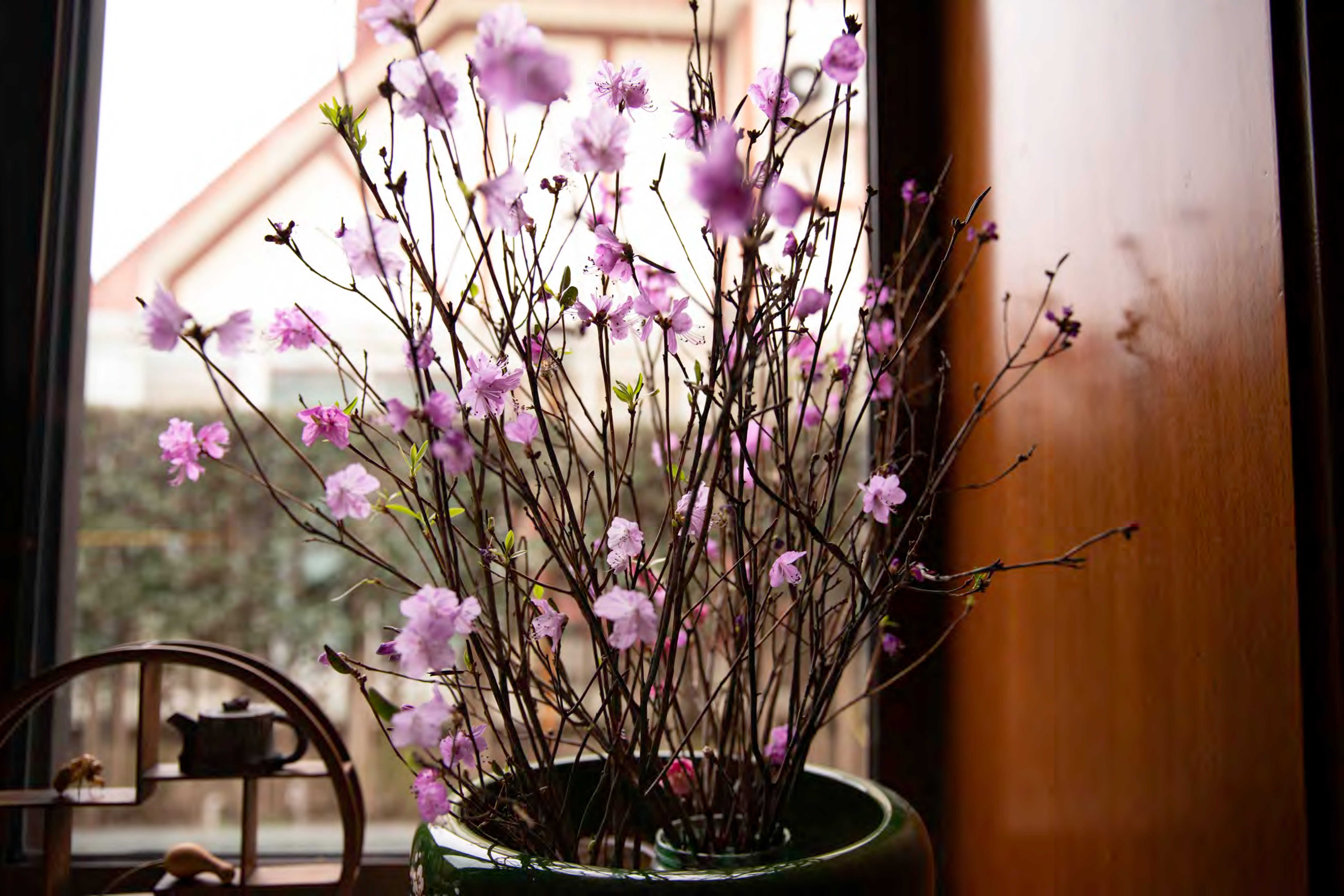} & \includegraphics[width=.16\textwidth]{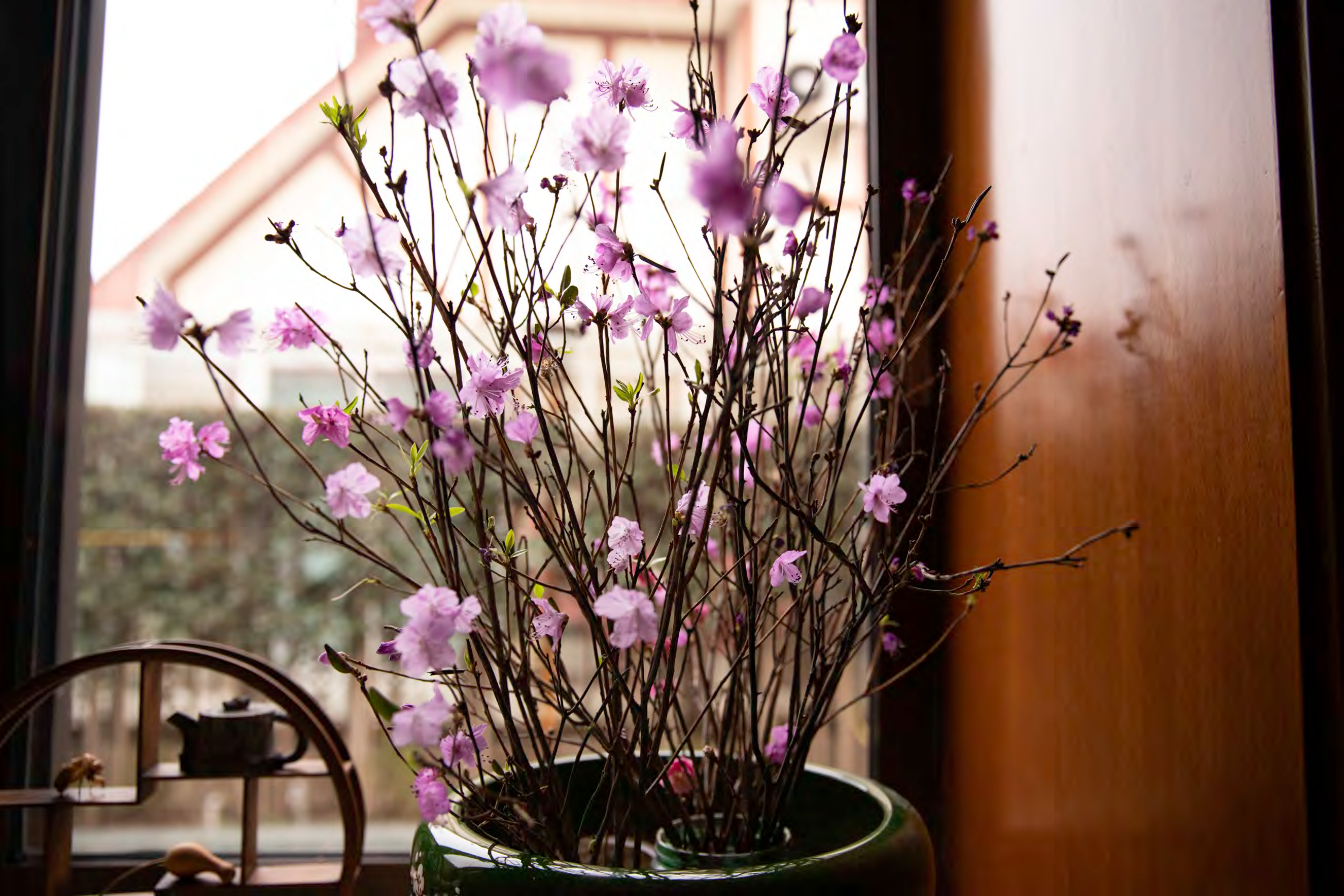} & \includegraphics[width=.16\textwidth]{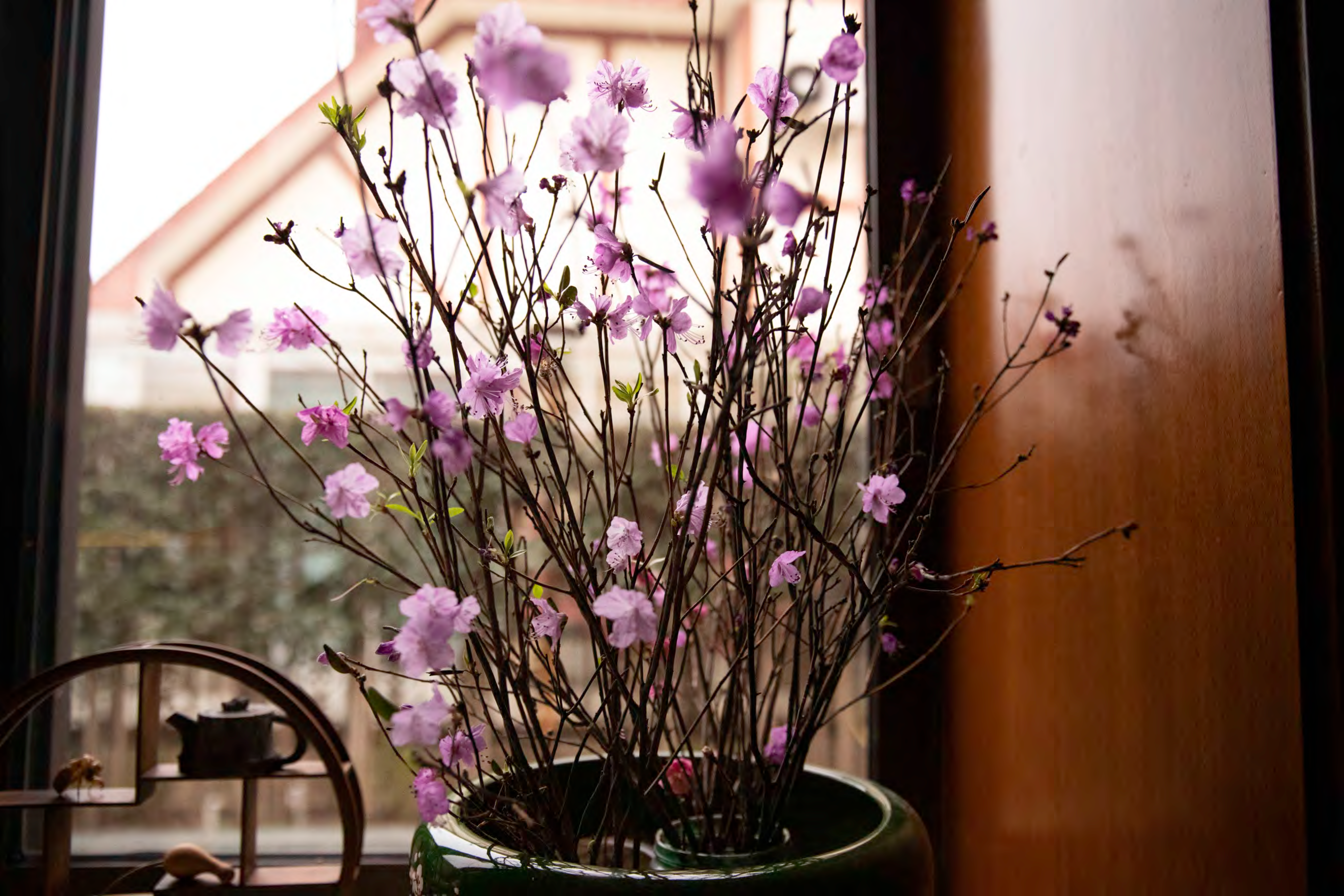} \\  
\makecell{Correlation\\ Map} &\includegraphics[width=.16\textwidth]{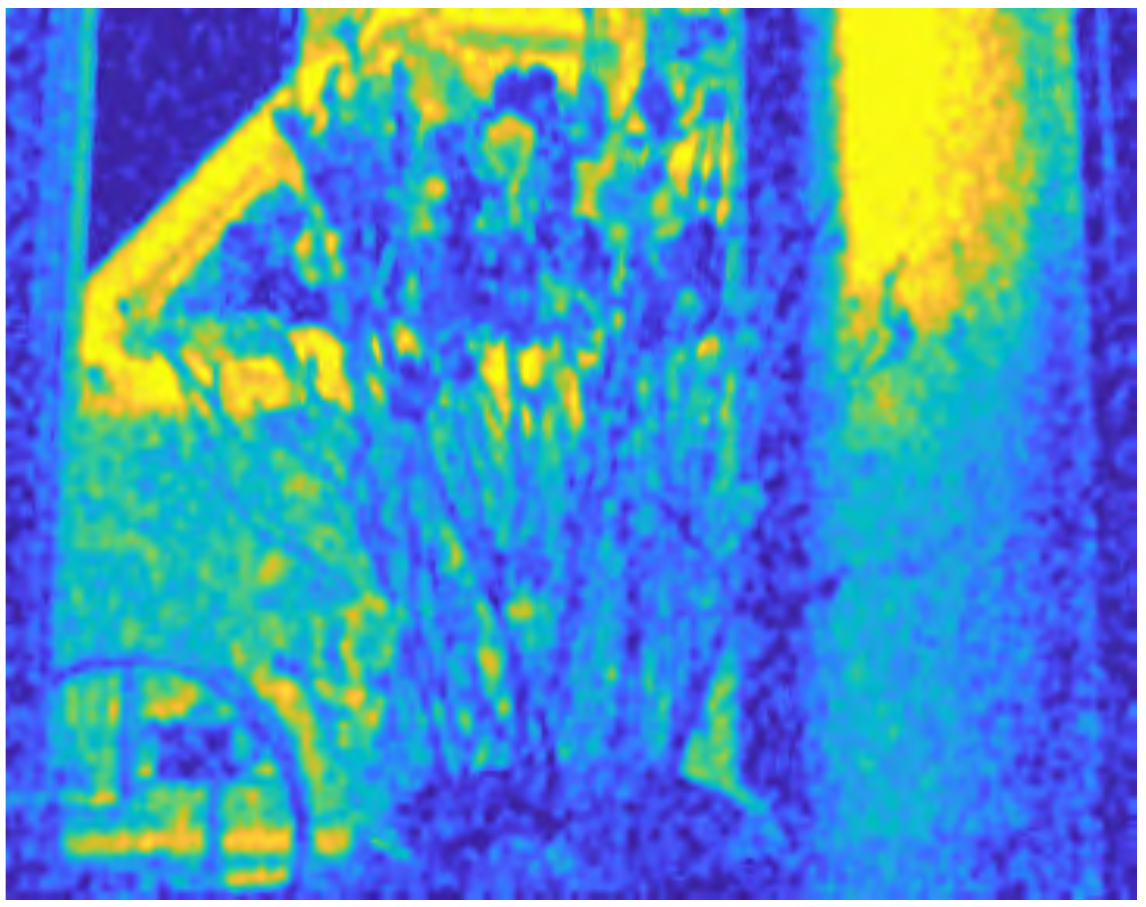} & \includegraphics[width=.16\textwidth]{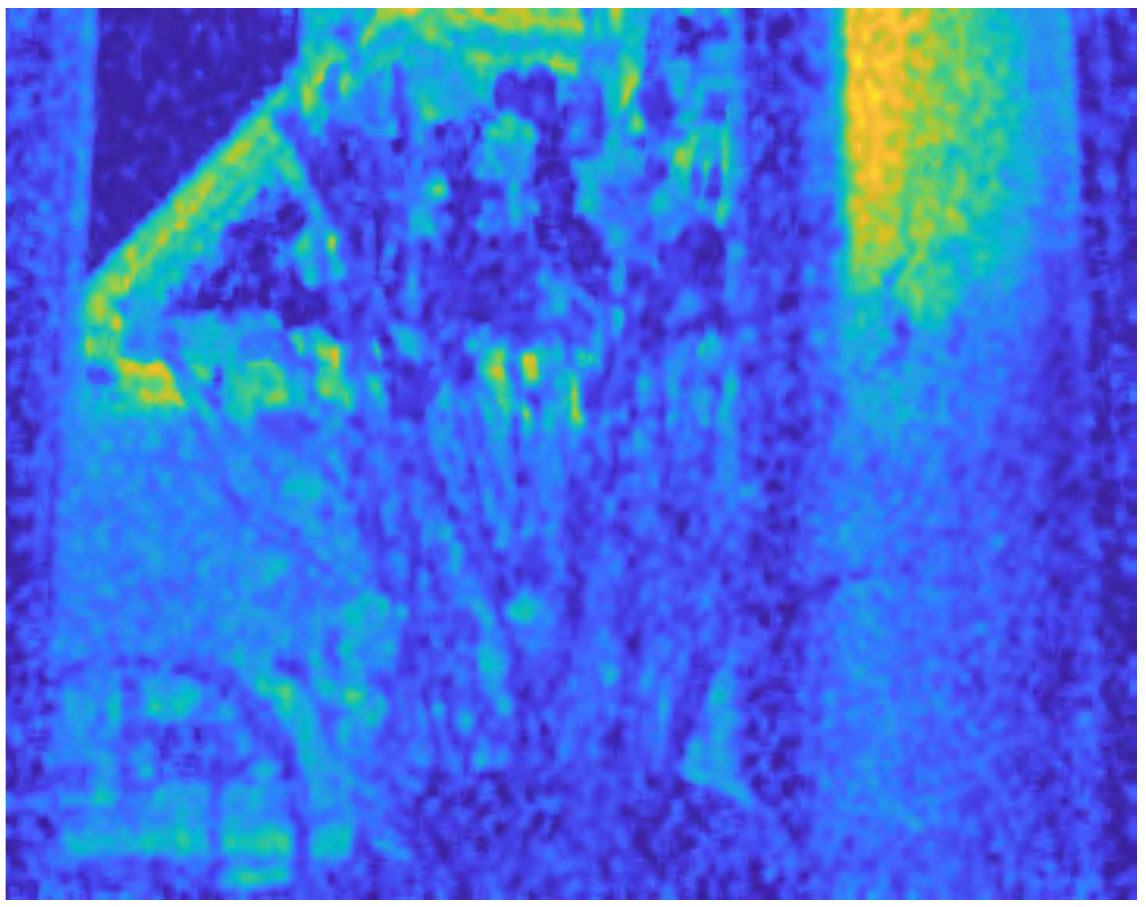} & \includegraphics[width=.16\textwidth]{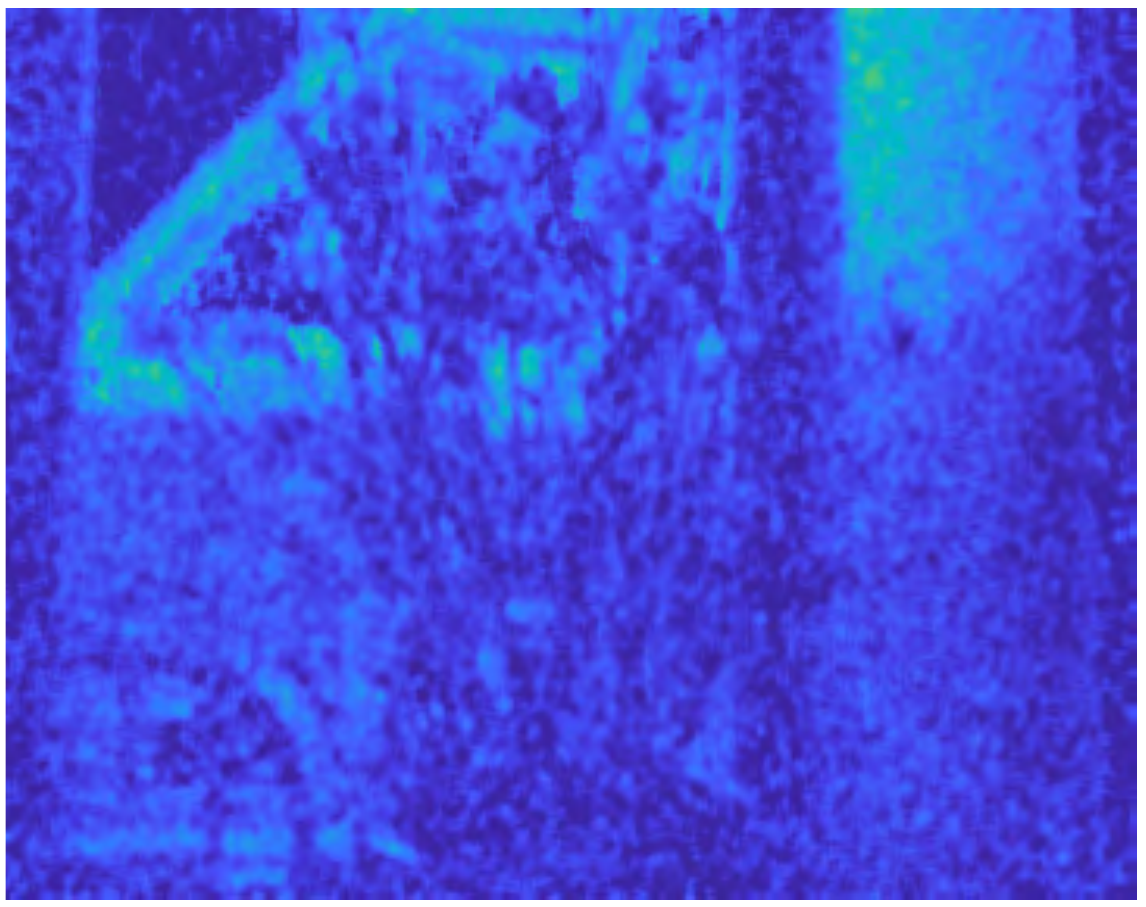} & \includegraphics[width=.18\textwidth]{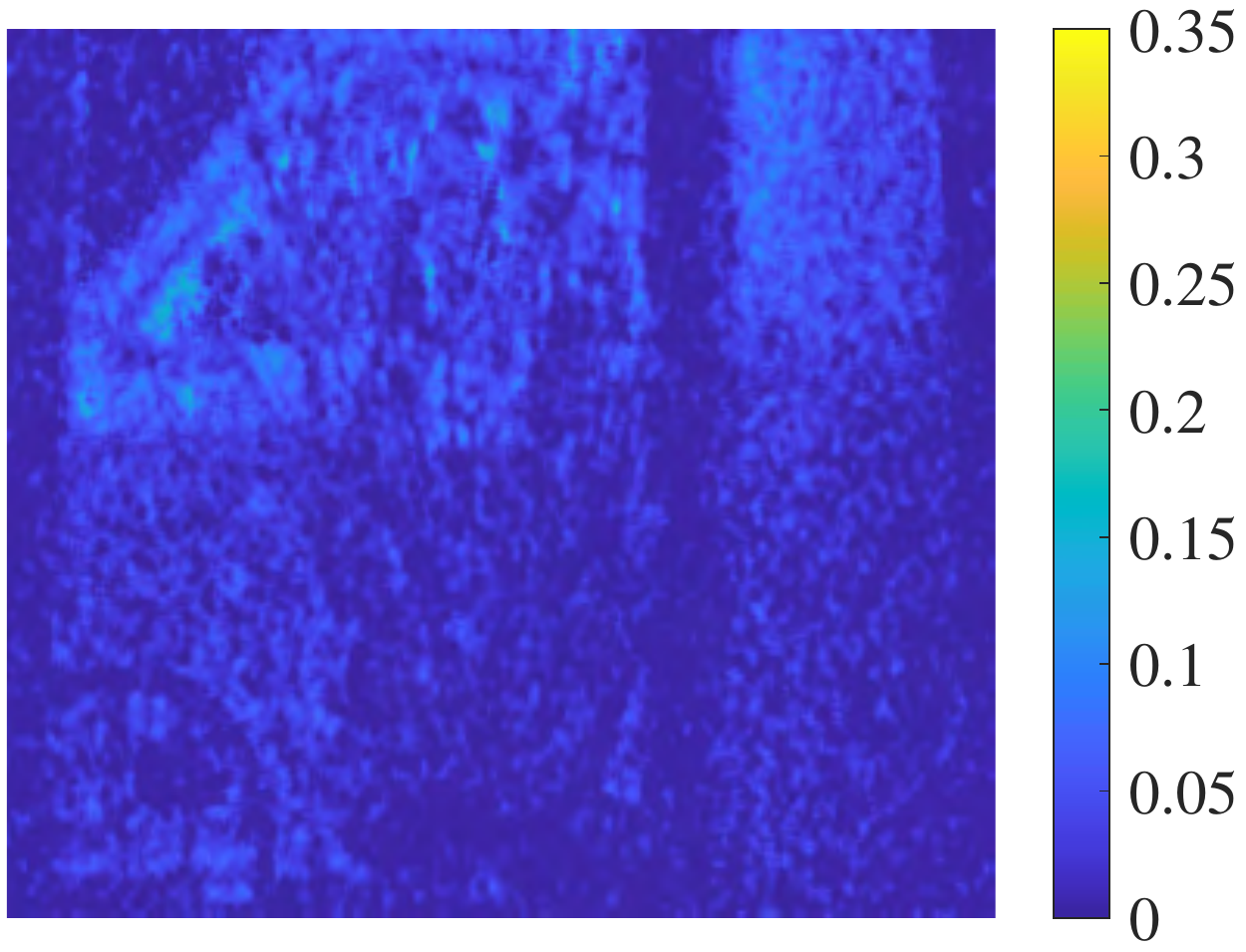}\\
\makecell{Scene 2} &\includegraphics[width=.16\textwidth]{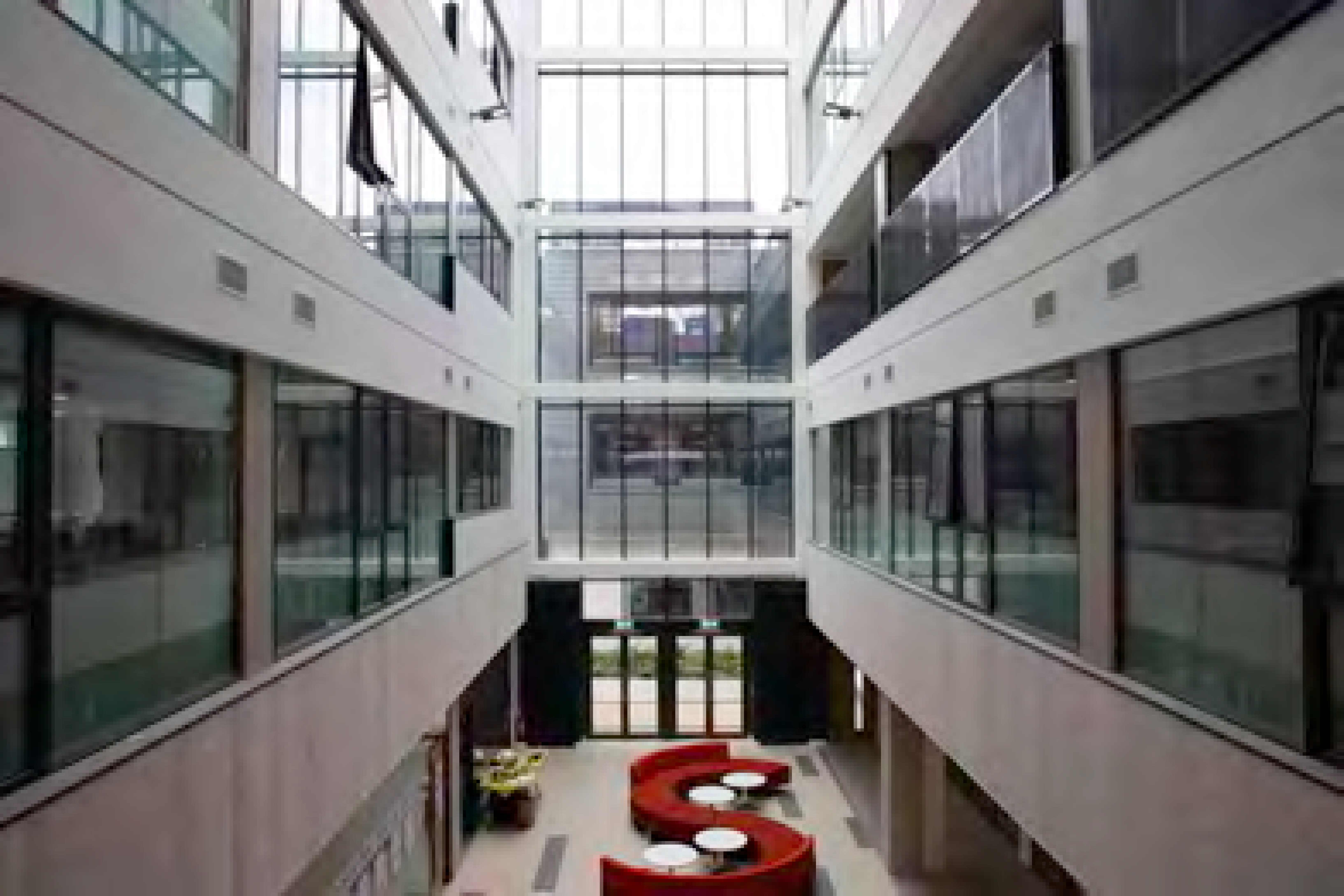} & \includegraphics[width=.16\textwidth]{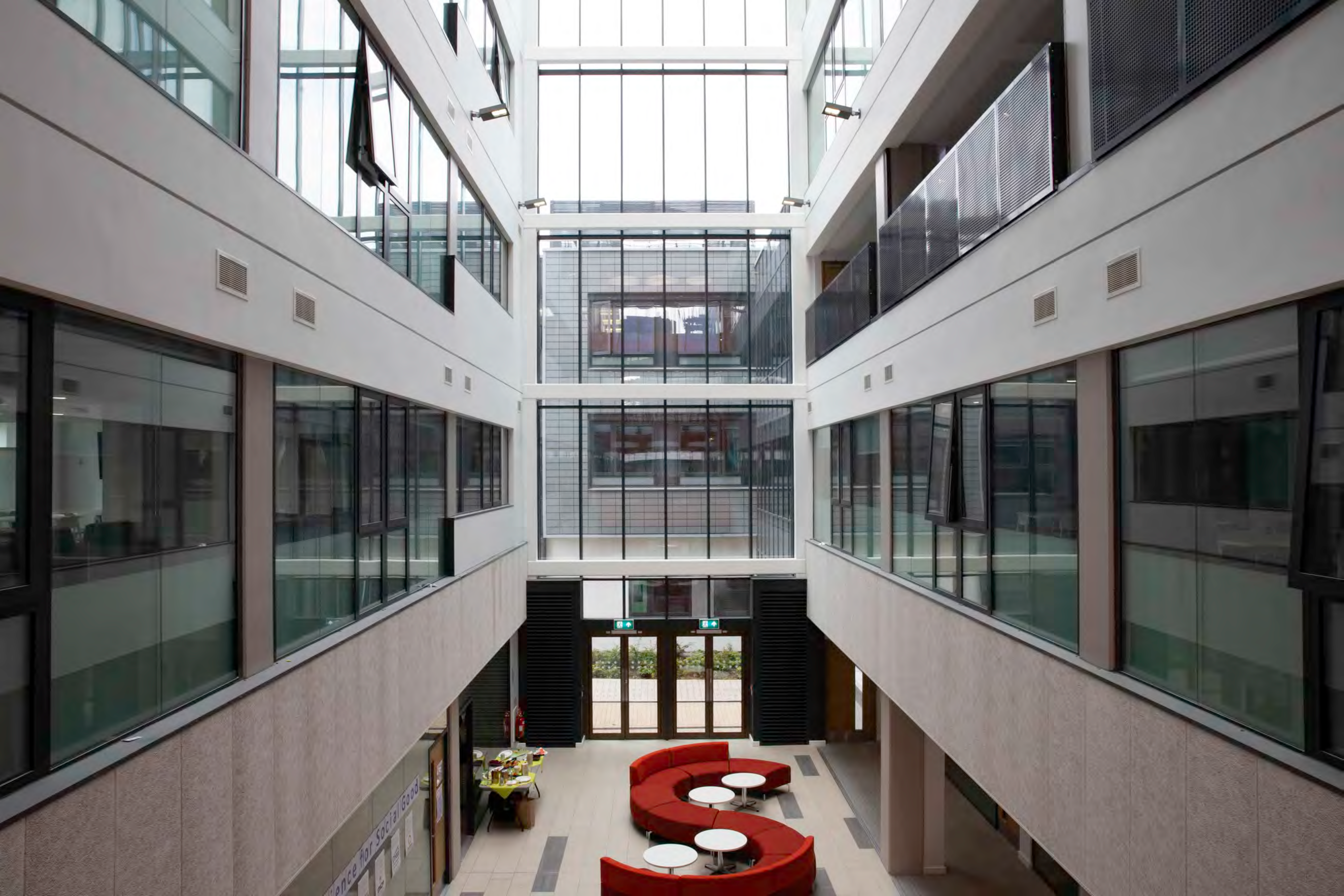} & \includegraphics[width=.16\textwidth]{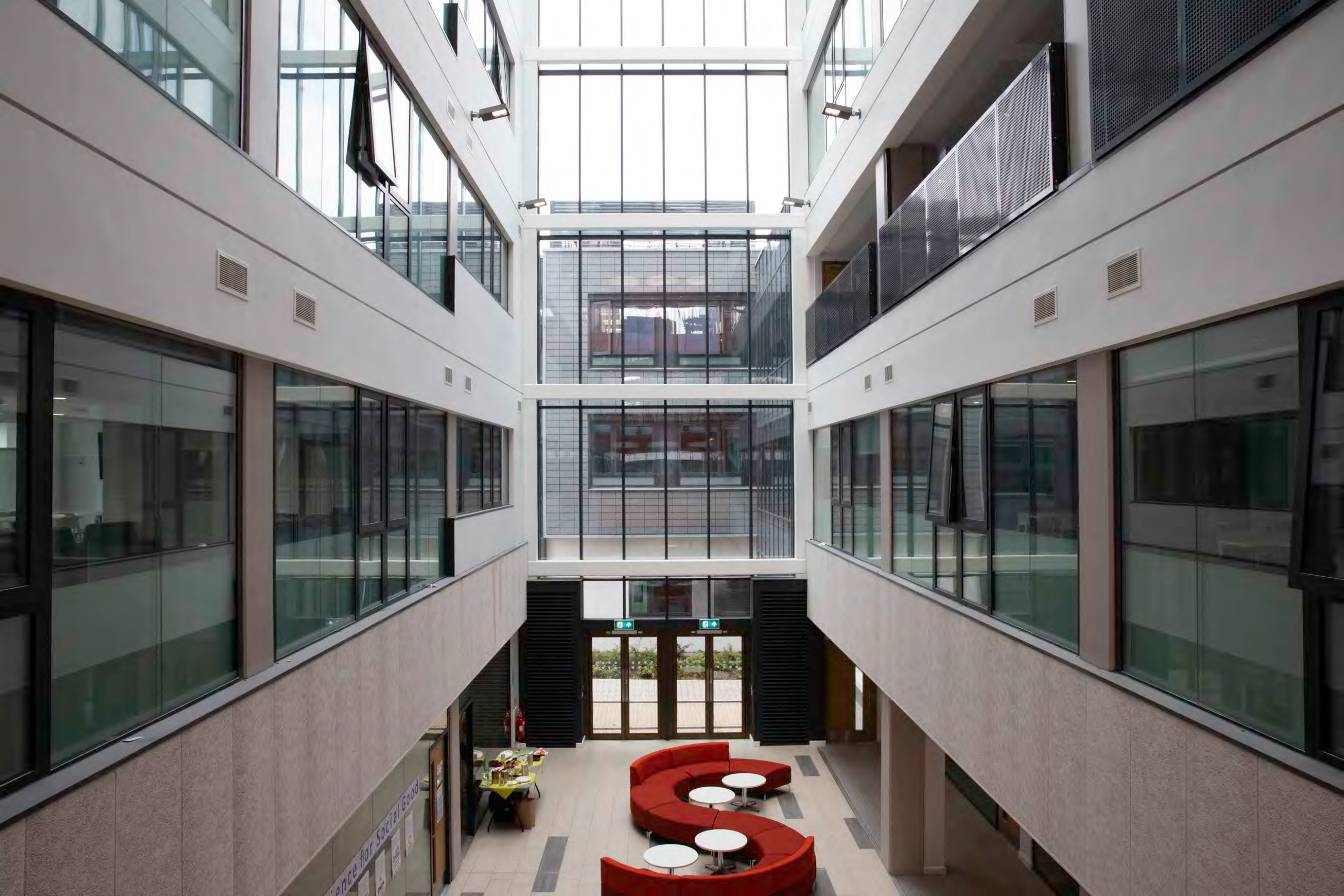} & \includegraphics[width=.16\textwidth]{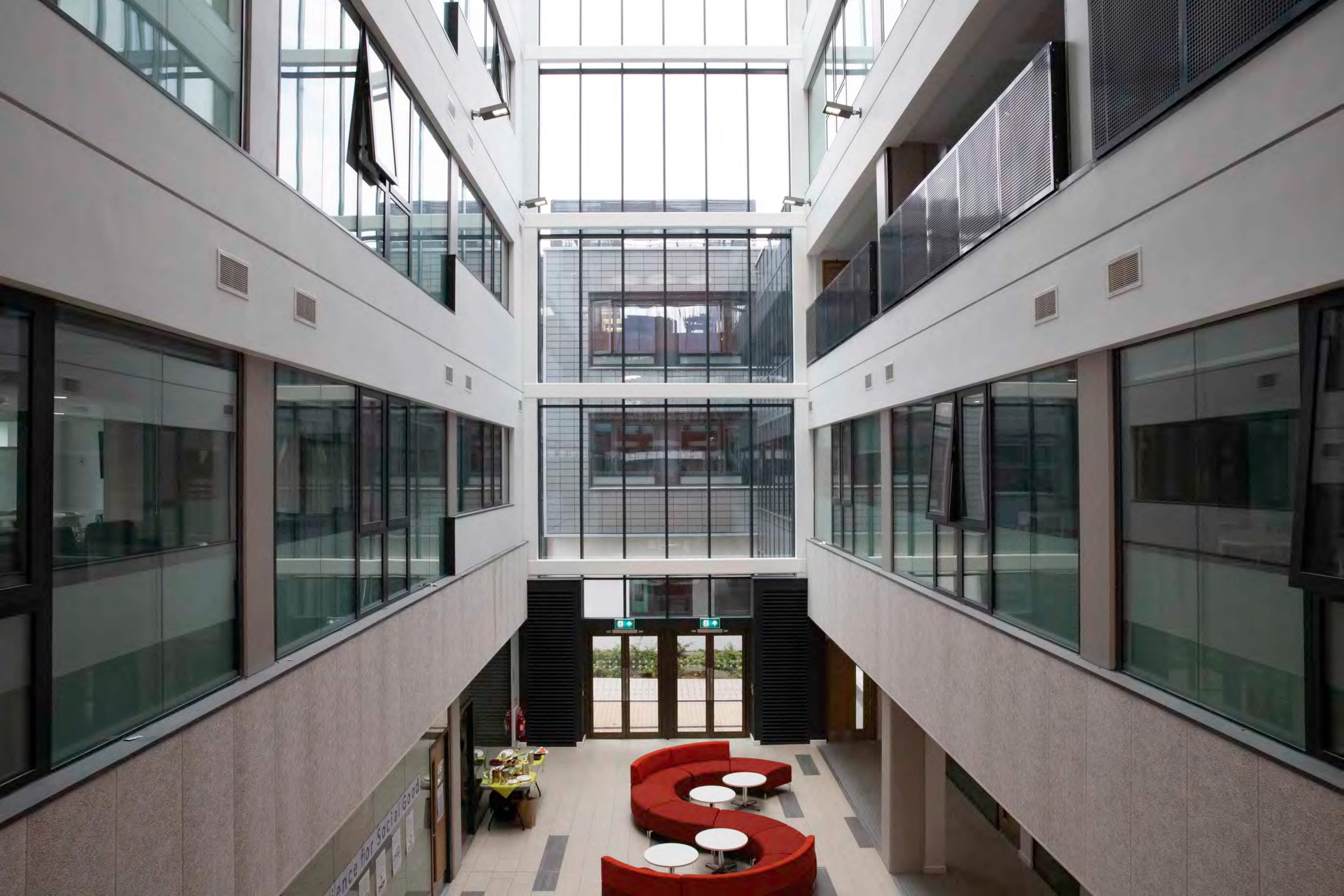} \\  
\makecell{Correlation\\ Map} &\includegraphics[width=.16\textwidth]{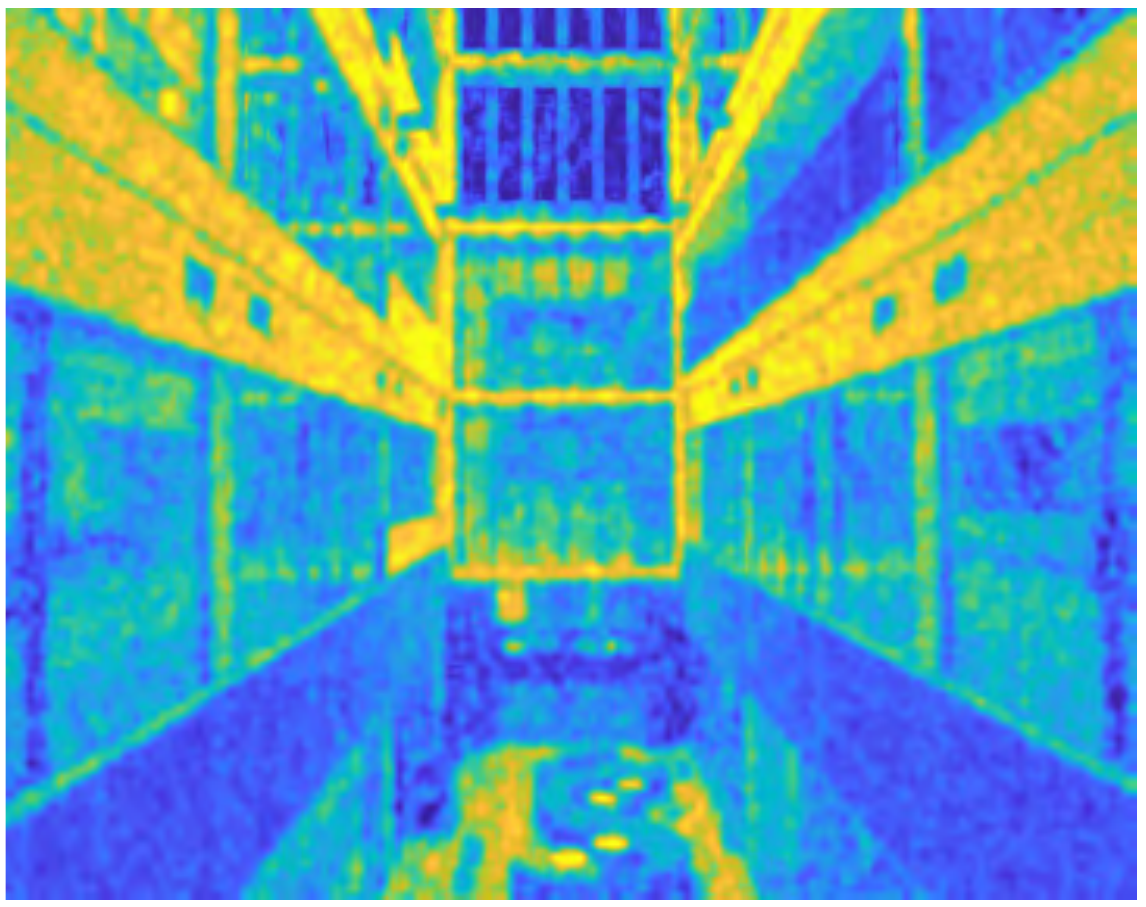} & \includegraphics[width=.16\textwidth]{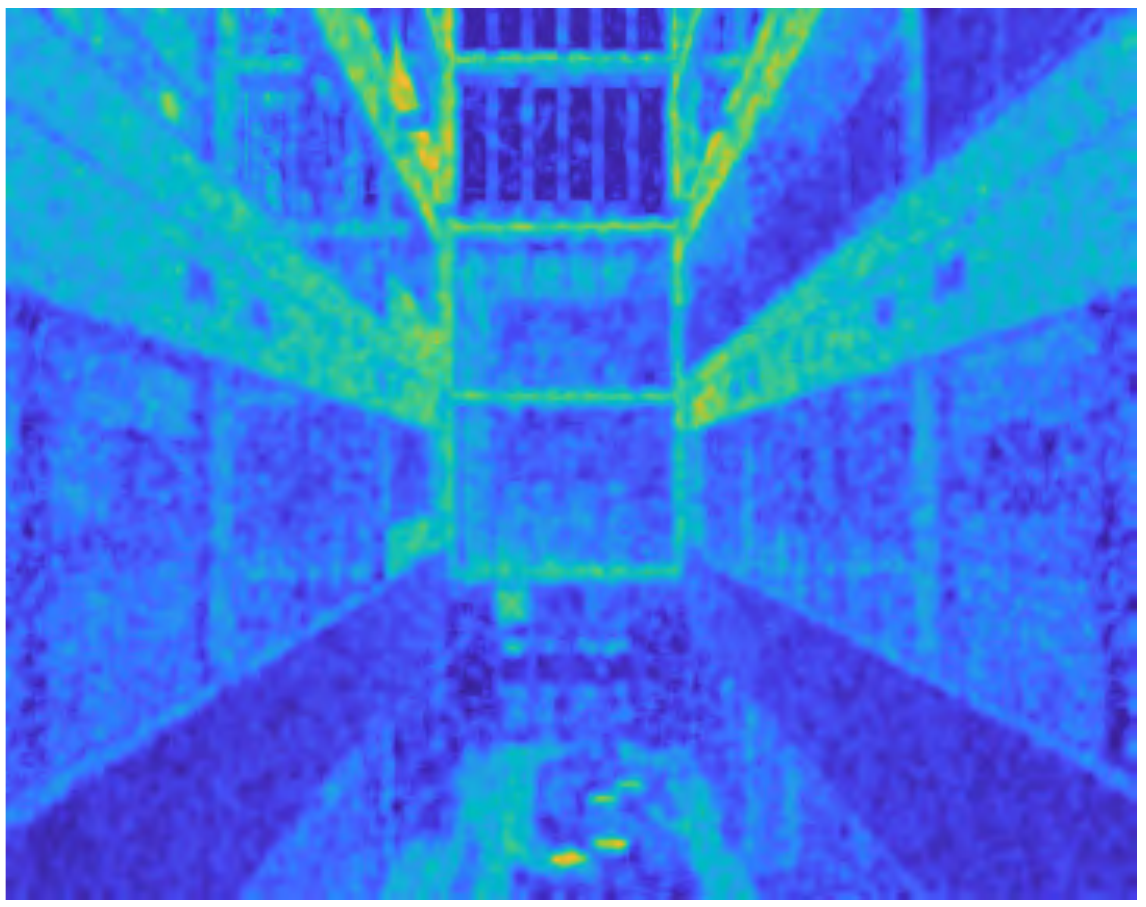} & \includegraphics[width=.16\textwidth]{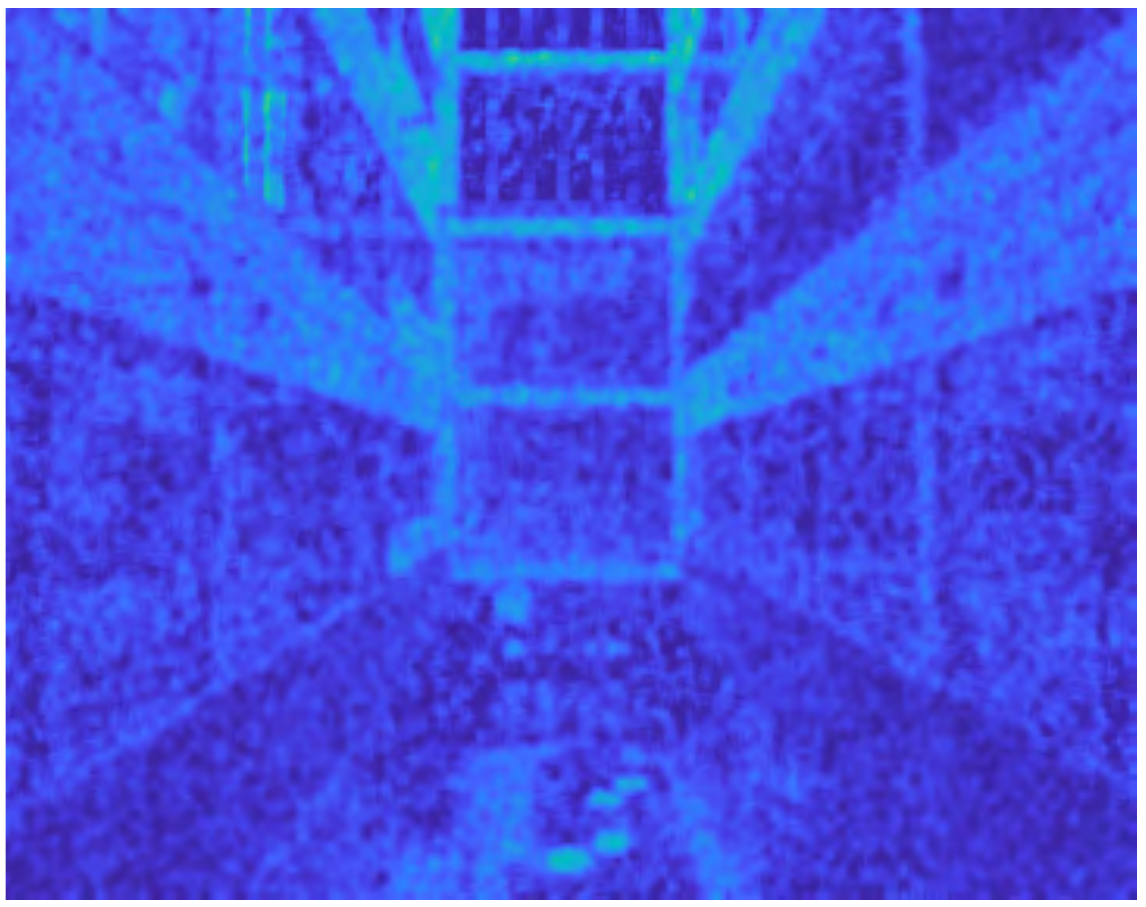} & \includegraphics[width=.18\textwidth]{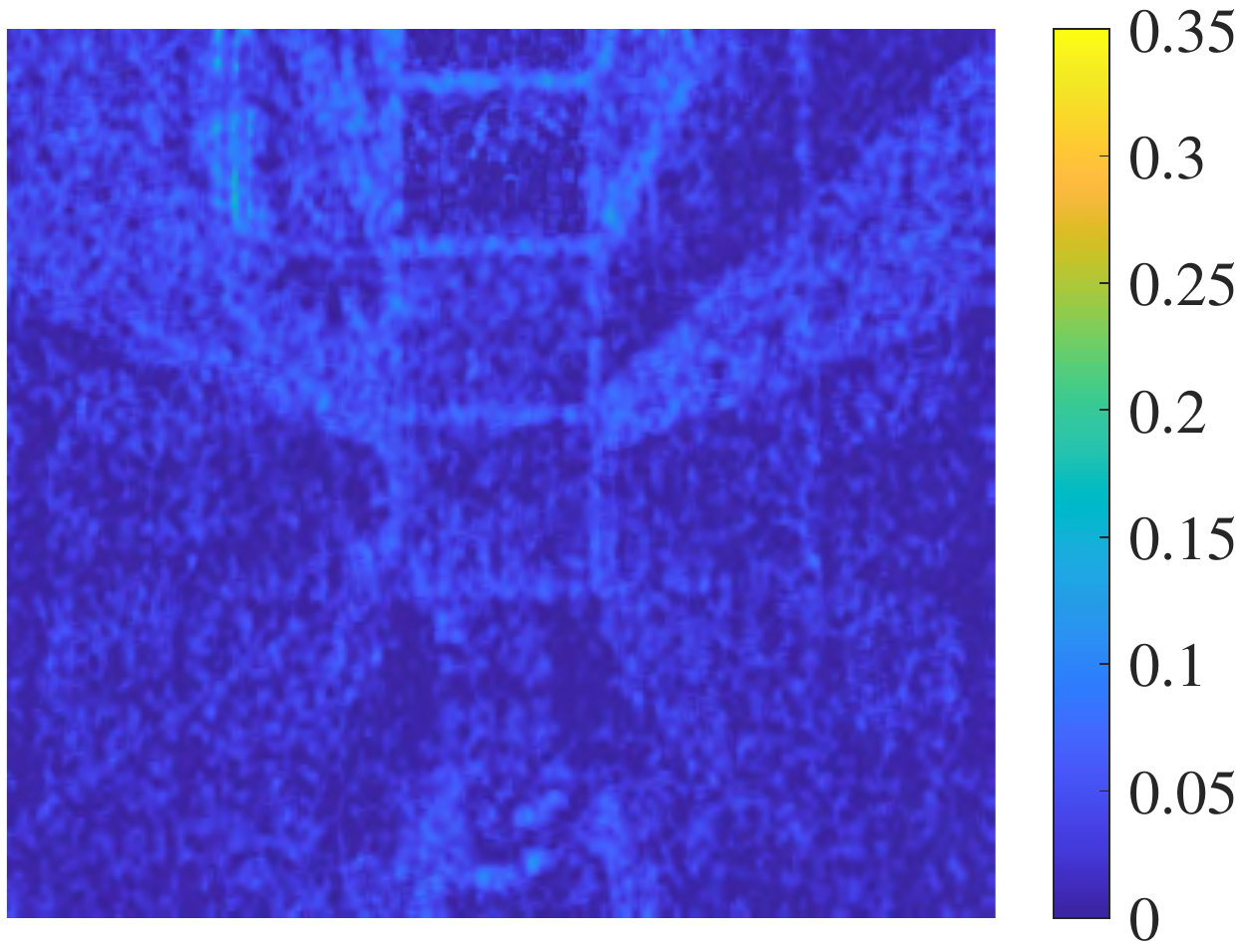}
\end{tabular}
\end{center}
\caption{Image of two different scene from a Canon 6D MKII from the Warwick Image Forensics Dataset. The images are taken with different ISO speeds. The exposure time for each image is set accordingly to let the images of the same scene have similar exposure level. The block-wise correlation maps are computed with a block size of $128 \times 128$ pixels. The color bars used for the correlation maps are at the right hand side, next to the ISO 6400 correlation maps.} 
\label{fig:testSet}
\vspace{-.2cm}
\end{figure*}

The above conclusions are made for the special condition when we consider the images to be RAW flat-field image. When we take post-processings (e.g. color interpolation and JPEG compression) and the influence due to the image content into consideration, the noise model could become rather complicated. This is both because the PRNU is multiplicative of image content and image content may propagate into the noise residual due to imperfect denoising. And actually, higher ISO images are more likely to suffer from strong JPEG compression and imperfect denoising (see supplementary material). Thus, though Equation (\ref{eqn:distri}) cannot be translated directly to the general conditions, all the factors suggest a higher ISO speed can introduce more PRNU-irrelevant noise. As a result, this will reduce the proportion of signals corresponding to the PRNU in the noise residual and eventually reduce the correlation. We use Fig.\ref{fig:testSet} to empirically show that the correlation is dependent on the image's ISO speed when post-processings like de-mosaicing, gamma correction, JPEG compression, etc., are applied to a non-flat RAW image.

The images shown in Fig.\ref{fig:testSet} are from a Canon 6D MKII camera in the Warwick Image Forensics Dataset. All the images shown here are saved in the JPEG format by the camera's default setting. Images of two scenes are taken under different ISO speeds using different exposure times to ensure that every image can reach the same exposure level. Thus, there is nearly no difference in pixel intensity between the images of the same scene. As the PRNU is a multiplicative signal, having images of the same pixel intensity of the same image content allows us to make a fair comparison with ISO speed's impact on the correlation. The correlation heat maps in Fig.\ref{fig:testSet} are computed by correlating the noise residuals from the images' green channel with the device's reference green channel PRNU. The reference PRNU is extracted from 50 flat-field images. The block size for the computation of the correlation at each pixel is $128 \times 128 $ pixels. We use yellow to show high correlation regions and blue to show the opposite. Apparently, as the ISO speed increases, the correlation map shows more regions with low correlation. It can be concluded that despite these images with complex image content have undergone post-processing, their correlation with the reference PRNU is still dependent on the image's ISO speed.\\
\section{ISO Speed's Impact upon correlation prediction}
\label{sec:set}

\begin{table*}
\caption{$r^2$ and RMSE from correlation predictions made from matching ISO and mixed ISO correlation predictors for 13 cameras in Warwick Image Forensics Dataset}
\label{tab:warwickr2}
\begin{center}
\begin{tabular}{c|cc|cc|cc|cc|cc|cc}
\multirow{3}{*}&\multicolumn{6}{c|}{\makecell{Matching ISO\\ Correlation Predictor}} & \multicolumn{6}{c}{\makecell{Mixed ISO\\ Correlation Predictor}}\\ \cline{2-13} 
 & \multicolumn{2}{c|}{ISO 100} & \multicolumn{2}{c|}{ISO 800} & \multicolumn{2}{c|}{ISO 6400\footnote{}} & \multicolumn{2}{c|}{ISO 100} & \multicolumn{2}{c|}{ISO 800} & \multicolumn{2}{c}{ISO $6400^1$}  \\
 & $r^2$ & RMSE  & $r^2$ & RMSE & $r^2$ & RMSE & $r^2$ & RMSE & $r^2$ & RMSE & $r^2$ & RMSE \\ \hline
Canon 6D & \textbf{0.7974} & \textbf{0.0194}& \textbf{0.7196}& \textbf{0.0169} & \textbf{0.5574} & \textbf{0.0116} &0.7839 & 0.0200 & 0.3983& 0.0247 & 0 & 0.0292 \\
Canon 6D MKII & \textbf{0.9518} & \textbf{0.0270} & \textbf{0.6870} & \textbf{0.0251} &\textbf{0.6912} & \textbf{0.0144}&0.9373 & 0.0307&0.2599 & 0.0386 &0 & 0.0420 \\
Canon 80D & \textbf{0.8593} & \textbf{0.0738}& \textbf{0.6920} & \textbf{0.0244}& \textbf{0.4108} & \textbf{0.0124}& 0& 0.1406 &0 & 0.1574 & 0 & 0.0836 \\
Canon M6 & \textbf{0.8584}& \textbf{0.0182}&\textbf{0.9076} &\textbf{0.0125} &\textbf{0.7246} & \textbf{0.0083}& 0.5439& 0.0327&0.8042 & 0.0183&0.2912 & 0.0134 \\
Fujifilm XA-10\_1  & \textbf{0.5562}& \textbf{0.0426} &\textbf{0.0582} & \textbf{0.0203} &\textbf{0.1143} & \textbf{0.0155}& 0& 0.0780& 0.0123 & 0.0208& 0.0809 & 0.0158\\
Fujifilm XA-10\_2 & \textbf{0.4648} & \textbf{0.0394} & 0 & 0.0409 & \textbf{0.1324} & \textbf{0.0151} & 0.1649 & 0.0492 &0 & \textbf{0.0384} &0 & 0.0251\\
Nikon D7200 &  \textbf{0.7344}& \textbf{0.0145} & \textbf{0.6339} & \textbf{0.0116}&\textbf{0.4868} & \textbf{0.0101} &0.3753 & 0.0223& 0.1737 & 0.0174&0 & 0.0170\\
Panasonic Lumix TZ90\_1 & \textbf{0.6878} & \textbf{0.0149} &0 & \textbf{0.0213} &0 & \textbf{0.0125} & 0.2032 & 0.0239& 0&0.0243 &0 & 0.0208 \\
Panasonic Lumix TZ90\_2 & \textbf{0.7766}& \textbf{0.0135} &\textbf{0.1448}  &  \textbf{0.0131}& \textbf{0.0458} & \textbf{0.0122}&0 & 0.0187 &0 & 0.0140 &0 & 0.0130\\
Sigma SdQuttro & \textbf{0.6758} & \textbf{0.0261}& \textbf{0.6404} & \textbf{0.0274}& \textbf{0.6361} & \textbf{0.0102}& 0& 0.0520& 0& 0.0871 & 0& 0.0693 \\
Sony Alpha68 & \textbf{0.8614}& \textbf{0.0171} &\textbf{0.8202} & \textbf{0.0131} & \textbf{0.4684} & \textbf{0.0072}& 0.7578 & 0.0226 & 0.7494 & 0.0155&0 & 0.252\\
Sony RX100\_1 & \textbf{0.4560} & \textbf{0.0446}& \textbf{0.7128}& \textbf{0.0185}& \textbf{0.7393} & \textbf{0.0151}& 0 & 0.1021& 0.5233 & 0.0239 &0.5299 & 0.0203\\
Sony RX100\_2 & \textbf{0.7075}& \textbf{0.0197}& \textbf{0.6528} & \textbf{0.0168}& \textbf{0.4752}&  \textbf{0.0142}& 0.5713& 0.0238& 0.3614 & 0.0227& 0& 0.0208 \\        
\end{tabular}
\end{center} 
\end{table*}
\footnotetext{ISO 3200 for Panasonic Lumix TZ90\_1 and TZ90\_2}
\begin{center}
\begin{table}
\caption{$r^2$ and RMSE for the correlation predictors generated from the matching and non-matching ISO correlation predictors for 9 cameras from Dresden Image Dataset}
\label{tab:r2rmse}
\begin{tabular}{c|cc|cc}
& \multicolumn{2}{c|}{\makecell{Matching ISO\\ Correlation Predictor}} & \multicolumn{2}{c}{\makecell{Non-matching ISO \\Correlation Predictor}}\\ 
 & $r^2$ & RMSE & $r^2$ & RMSE \\ \hline
 Canon\_Ixus55\_0 &\textbf{0.7012} & \textbf{0.0234} & 0.6558 & 0.0251\\
 Canon\_Ixus70\_0 & \textbf{0.7111} & \textbf{0.0297} & 0 & 0.0567 \\
 Canon\_Ixus70\_1 & \textbf{0.7161} & \textbf{0.0267} & 0.2251 & 0.0441 \\ 
 Canon\_Ixus70\_2 & \textbf{0.6631} & \textbf{0.0306} &  0 & 0.0664 \\
 FujiFilm\_FinePixJ50\_0 & \textbf{0.8940} & \textbf{0.0195} & 0.5130 & 0.0417 \\
 FujiFilm\_FinePixJ50\_1 & \textbf{0.8928} & \textbf{0.0190} & 0.8726 & 0.0207 \\
 FujiFilm\_FinePixJ50\_2 & \textbf{0.9013} & \textbf{0.0199} & 0.8326 & 0.0260 \\
 Nikon\_CoolPixS710\_0 & \textbf{0.5400} & \textbf{0.0168} & 0.3005 & 0.0207 \\
 Pentax\_OptioA40\_0 & \textbf{0.3811} & \textbf{0.0315} & 0 & 0.0596
\end{tabular}
\end{table}
\end{center}

A correlation predictor is an important component of many PRNU-based tampering localization methods. Many PRNU-based tampering localization methods are applied by comparing the block-wise correlations with a decision threshold set according to the predicted correlation. As a result, the choice of the decision threshold and the performance of these methods can be greatly affected by the accuracy of the correlation prediction. As the correlation is content dependent, without considering the ISO speed, \cite{chen2008determining} models the correlation as a function of four image features, namely the intensity, texture, signal flattening and a texture-intensity combinative term. However, due to the correlation's dependency on the ISO speed, we postulate that: \textit{a correlation predictor can only produce accurate predictions for images with the same ISO speed as the training images}. And we call such a correlation predictor as a matching ISO correlation predictor.

To show the ISO speed's influence on correlation predictor and validate our postulate, we first compare the performance of the correlation predictors trained with (a) images with mixed ISO speeds and (b) images with the same ISO speed as the test images. We did the test on 13 cameras from the Warwick Image Forensics Dataset (An Olympus EM10 MKII camera from the dataset doesn't show the existence of PRNU. Thus it is not included in this test). 50 flat-field images from each camera are used to extract the cameras' reference fingerprints. For each camera, we select images from three ISO speeds to form three test sets, namely ISO 100, 800, and 6400, apart from the two Panasonic LumixTZ90, which do not have ISO 6400. For these two cameras, we test on ISO 3200 images instead. Accordingly, we trained three matching ISO correlation predictors, each with 20 images of the corresponding ISO speed following the method from \cite{chen2008determining}. The correlations are computed between image blocks of $128 \times 128$ pixels. To make the comparison, for each camera, we trained another correlation predictor with 20 images randomly selected from the 60 images used for the training of the camera's three matching ISO correlation predictors. We call this correlation predictor as a mixed ISO correlation predictor. Block-wise correlation predictions are made for the test sets. For each set, we computed the coefficient of determination ($r^2$) and the root mean square error (RMSE) for the matching ISO and mixed ISO correlation predictors as shown in Table \ref{tab:warwickr2}. We highlighted the better performance for each test set in terms of larger $r^2$ and smaller RMSE with bold font.

The matching ISO correlation predictors show superior performance over the mixed correlation predictors for all test sets except for the two Fujifilm XA-10 and the two Panasonic Lumix TZ90 at high ISO speeds. These two models of cameras are more prone to strong noise at high ISO speeds. As a result, the correlations with their reference PRNU become close to zero despite different image features. Due to the relatively large variance of the correlations introduced by the PRNU-irrelevant signal in the noise residuals, neither of the correlation predictors managed to produce large $r^2$ for the correlation predictions. However, by using the Matching ISO correlation predictor for these cameras, we notice small RMSE still can be observed. This is particularly important as the correlation predictors would not generate predictions that deviate too much from the actual correlation. False positives can be significantly reduced when we apply these correlation predictors for forgery detection.

In addition to the test on the Warwick Image Forensics Dataset, the experiments are extended to 9 cameras from the Dresden Image Dataset \cite{Gloe:2010aa} as well. In the Dresden Image Dataset, about 150 images of natural scenes are produced by each camera. However, as the dataset was created without considering the ISO speed as an influential factor, the images' ISO speeds span over many different values. For most ISO speeds, the number of images available is not enough for us to train a matching ISO correlation predictor using the method mentioned above and to test it with the matching ISO images. So we test the matching ISO correlation predictor on the most popular ISO speed from each camera only, each with 20 test images. For each camera, we trained a matching ISO correlation predictor with 20 images of the same ISO speed as the test images and another 20 images are selected randomly from all the images available for the training of the mixed ISO correlation predictor. $r^2$ and RMSE of the predictions are shown in Table \ref{tab:r2rmse}. Again, the superior performance of the matching ISO correlation predictors can be observed in every case. Both the tests on images from Warwick Image Forensics Dataset and Dresden Image Dataset show that the performance of a correlation predictor may degenerate by completely ignoring the impact of ISO speed and trained images of mixed ISO speed.
%
\begin{figure*}
\begin{tabular}{ccccccc}
 & \makecell{Original \\Image} & \makecell{Forgery\\ Image} & \makecell{Forgery\\ Map} & \makecell{ISO 100\\ correlation\\ predictor} & \makecell{ISO 800\\ correlation\\ predictor} & \makecell{ISO 6400\\correlation\\ predictor} \\
\\
\makecell{  ISO 100 \\ forgery image}
& \raisebox{-.5\height}{\includegraphics[width=0.11\textwidth]{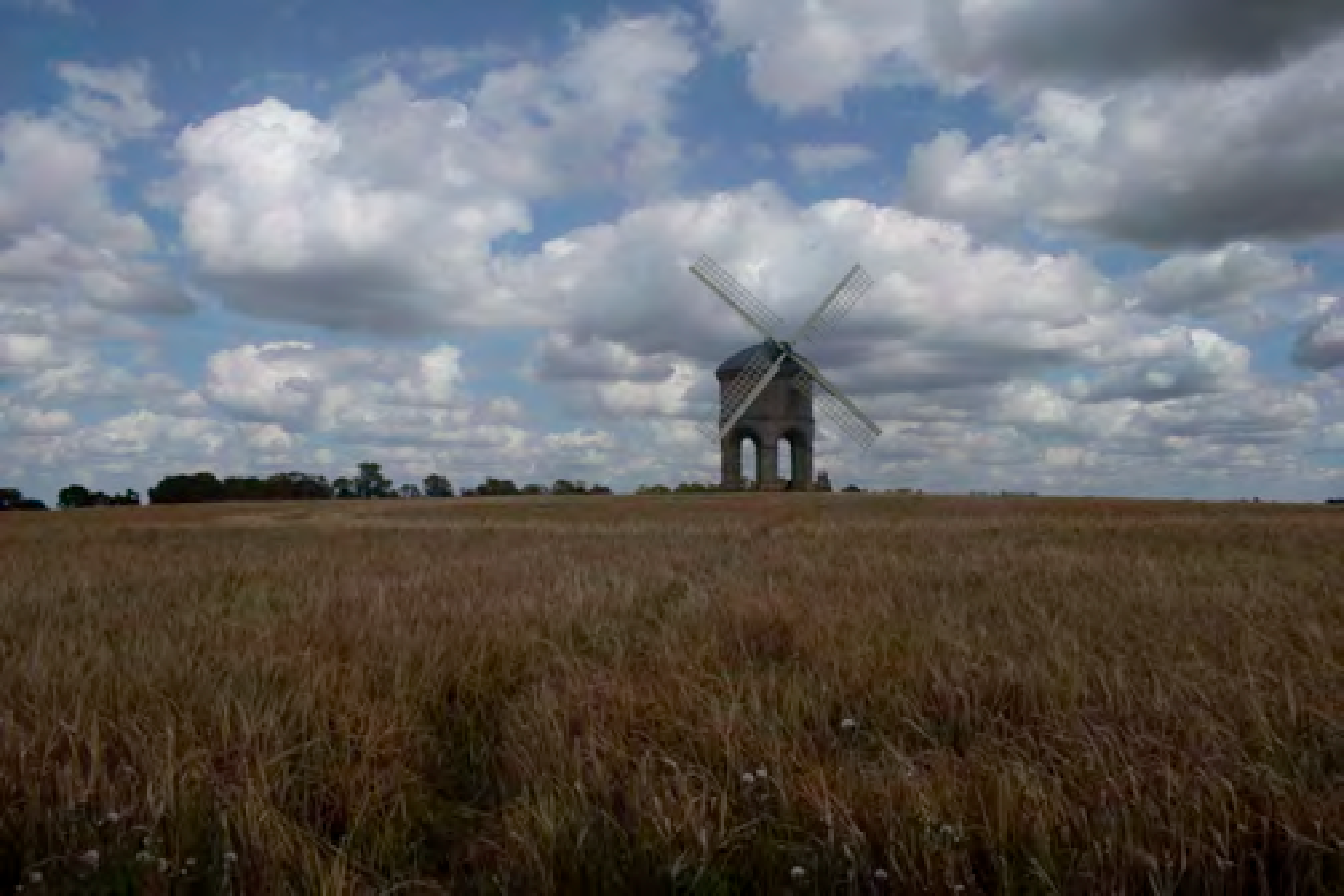}} & \raisebox{-.5\height}{\includegraphics[width=0.11\textwidth]{detection_map/Canonm6/1/ISO100/forgery.pdf}} & \raisebox{-.5\height}{\includegraphics[width=0.11\textwidth]{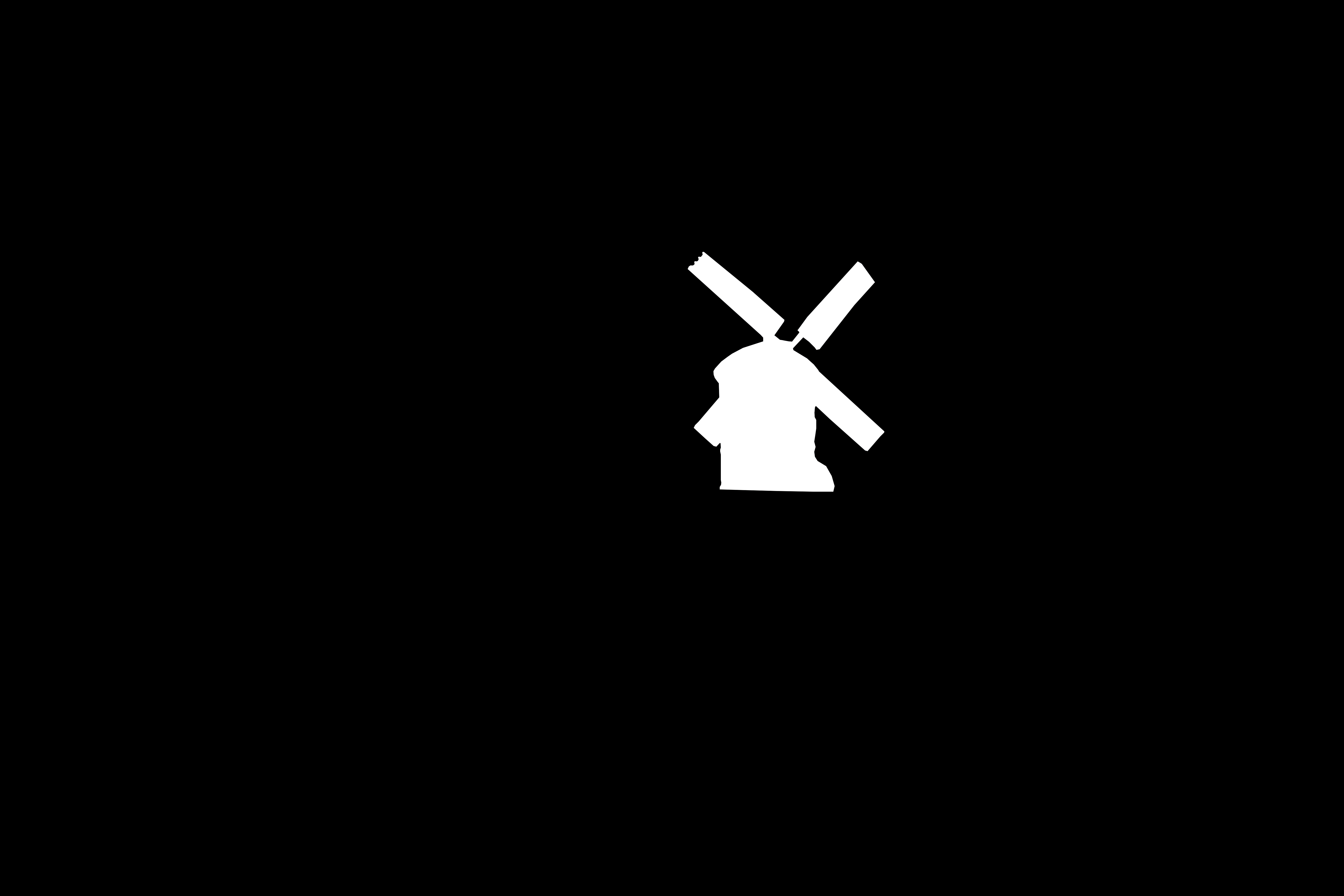}} & \raisebox{-.5\height}{\includegraphics[width=0.11\textwidth]{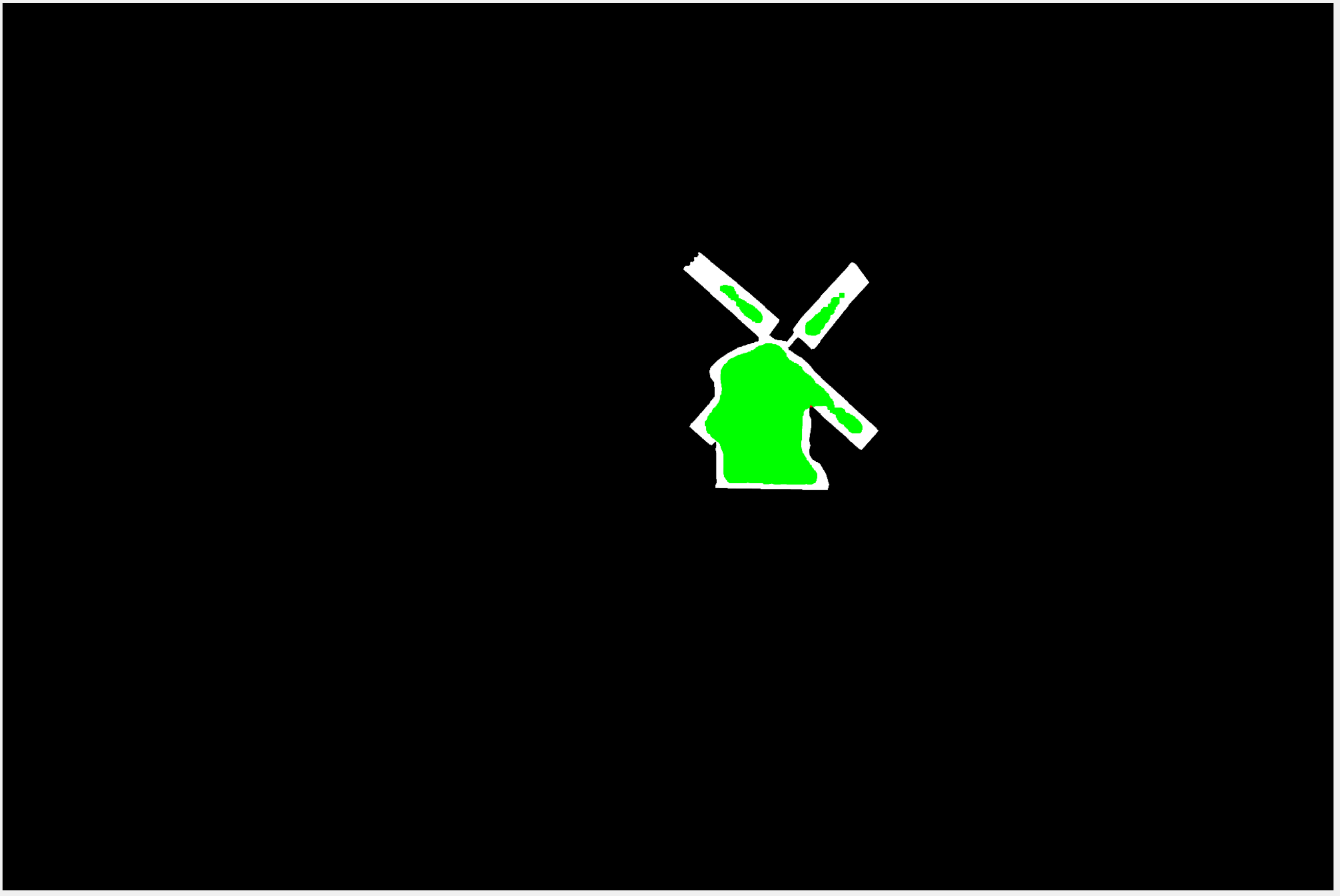}} &\raisebox{-.5\height}{ \includegraphics[width=0.11\textwidth]{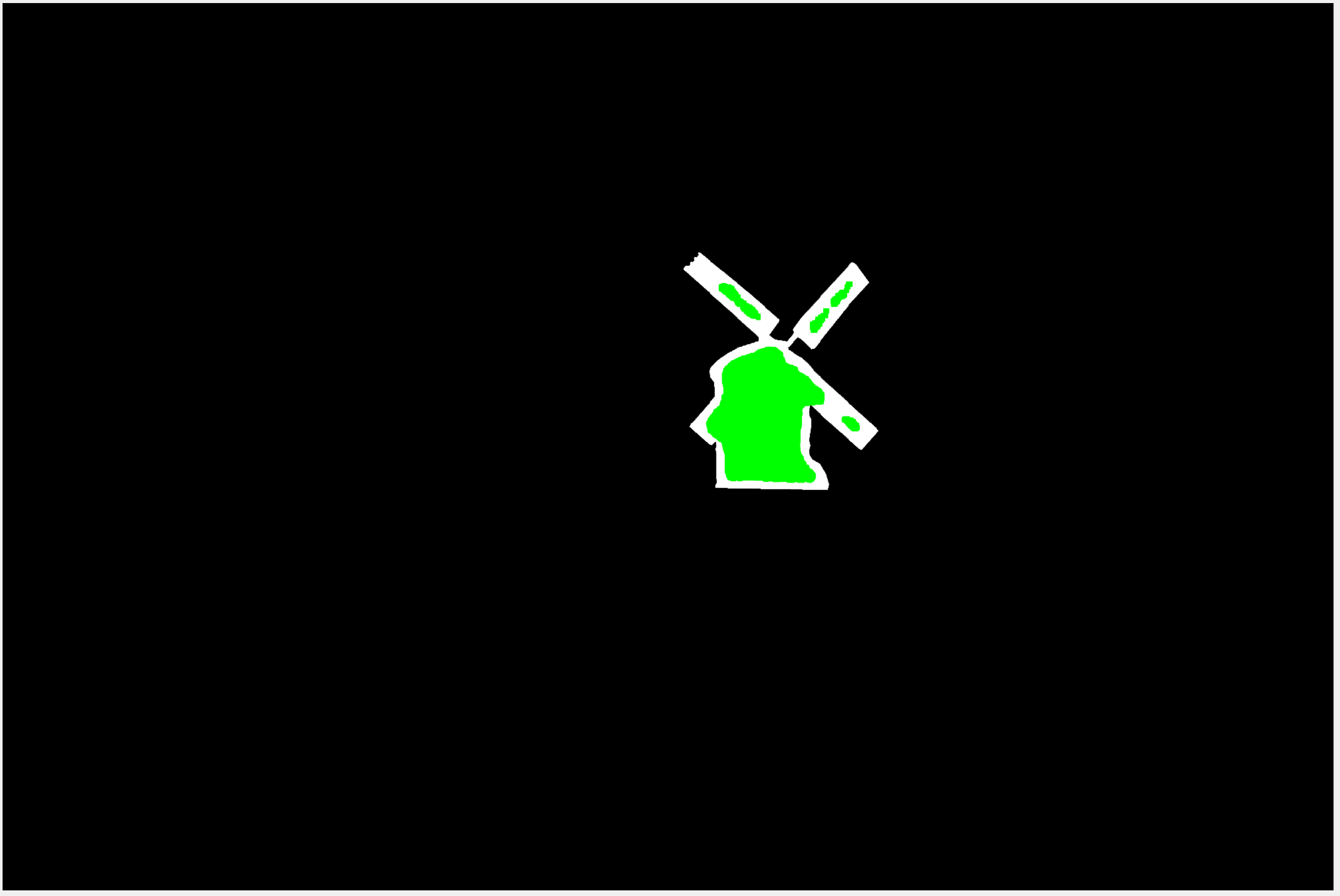}} & \raisebox{-.5\height}{\includegraphics[width=0.11\textwidth]{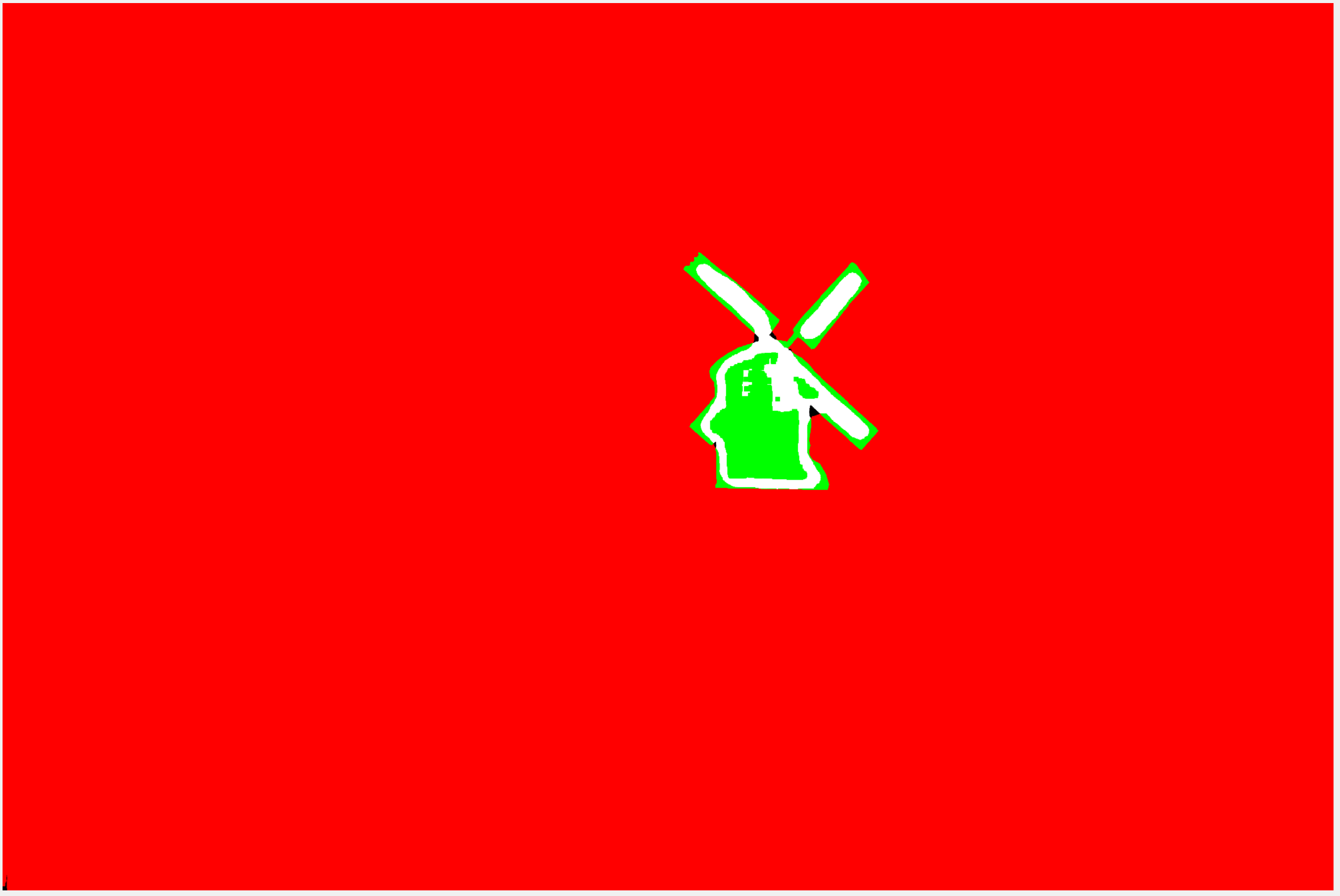}}\\ 
\\
\makecell{ISO 800 \\ forgery image} & \raisebox{-.5\height}{\includegraphics[width=0.11\textwidth]{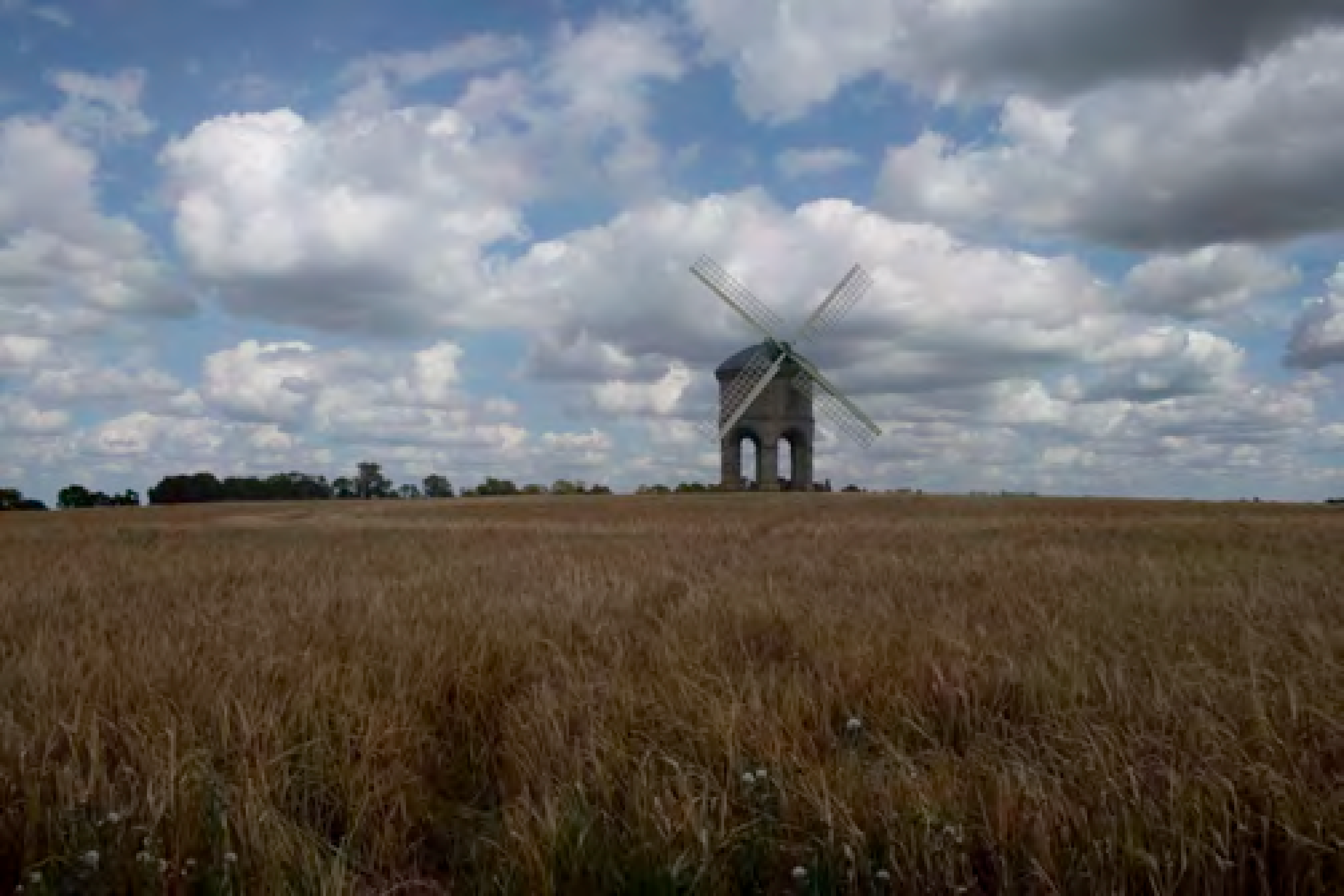}} &\raisebox{-.5\height}{ \includegraphics[width=0.11\textwidth]{detection_map/Canonm6/1/ISO800/forgery.pdf}} & \raisebox{-.5\height}{\includegraphics[width=0.11\textwidth]{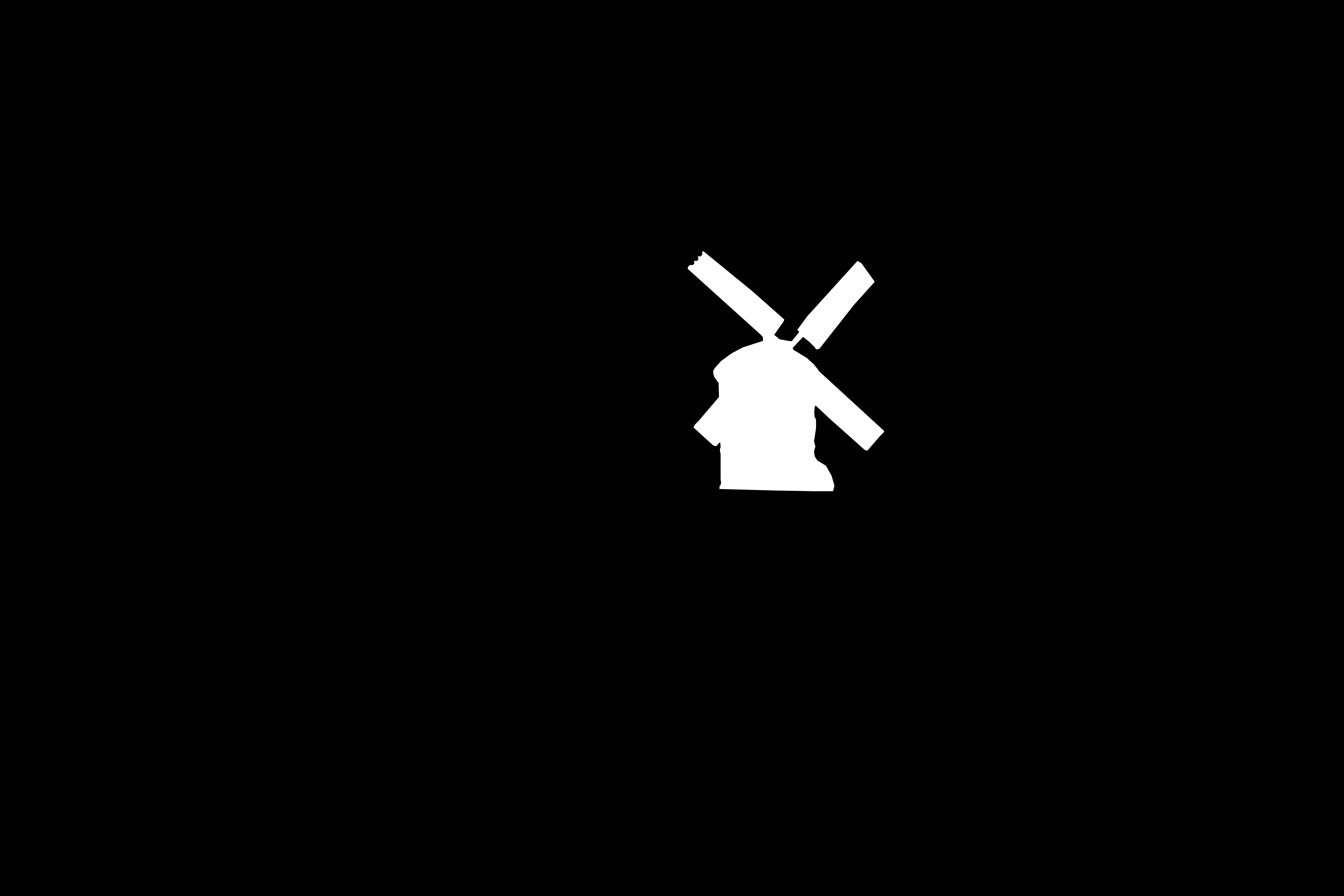}} & \raisebox{-.5\height}{\includegraphics[width=0.11\textwidth]{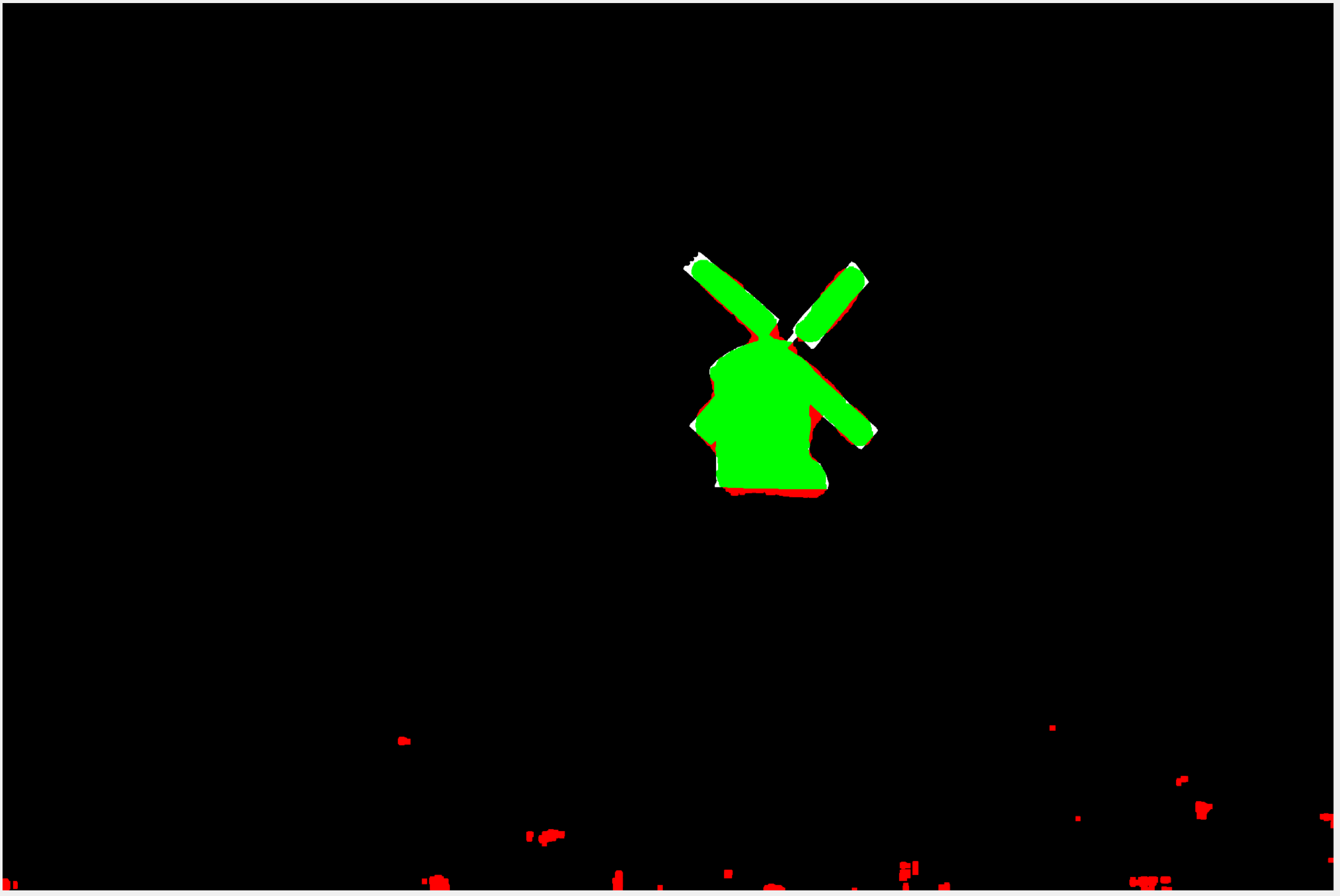}} & \raisebox{-.5\height}{\includegraphics[width=0.11\textwidth]{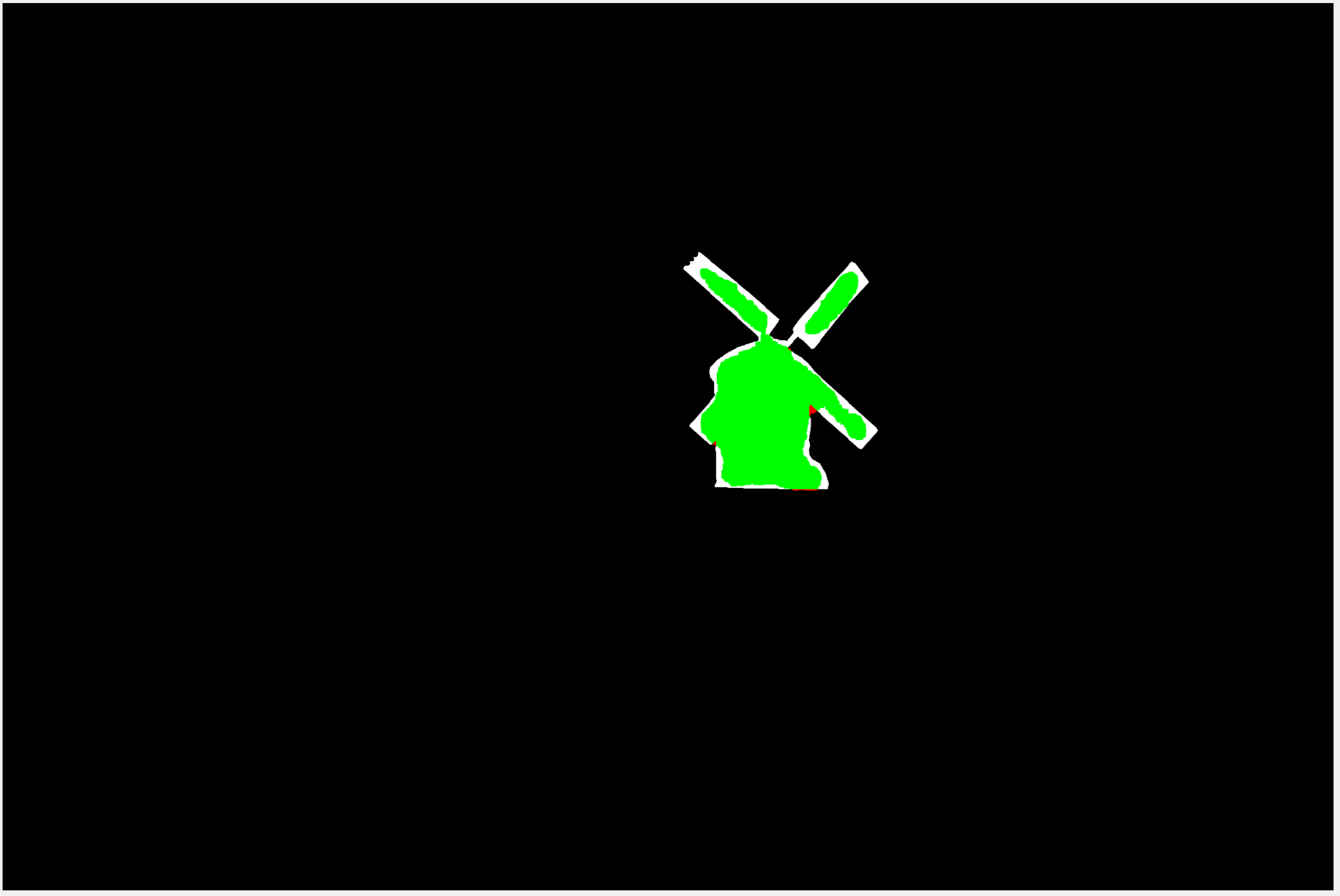}} & \raisebox{-.5\height}{\includegraphics[width=0.11\textwidth]{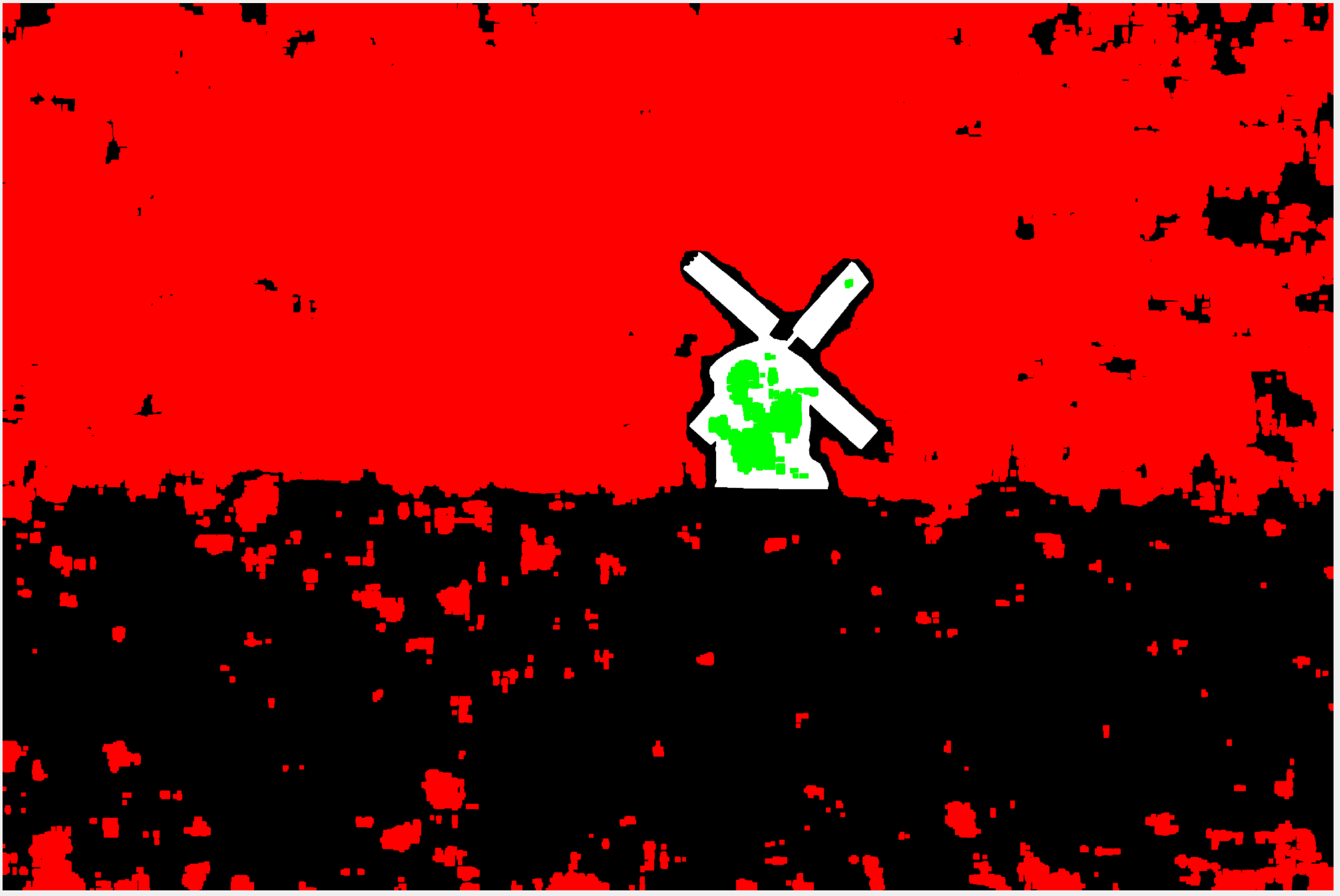}}\\
\\
  \makecell{ISO 6400 \\ forgery image} &\raisebox{-.5\height}{ \includegraphics[width=0.11\textwidth]{detection_map/Canonm6/1/ISO6400/original.pdf}} &\raisebox{-.5\height}{ \includegraphics[width=0.11\textwidth]{detection_map/Canonm6/1/ISO6400/forgery.pdf}} & \raisebox{-.5\height}{\includegraphics[width=0.11\textwidth]{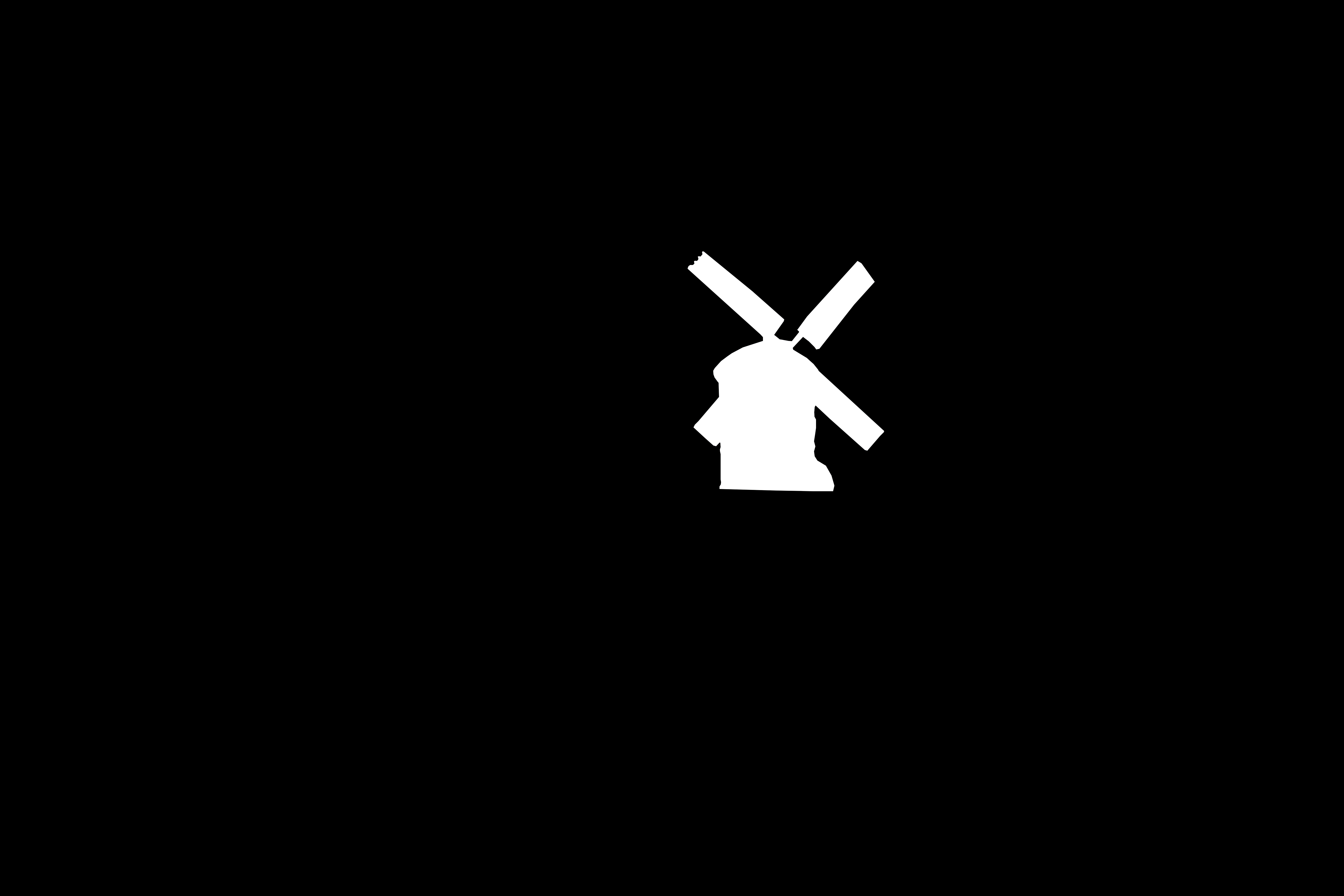}} & \raisebox{-.5\height}{\includegraphics[width=0.11\textwidth]{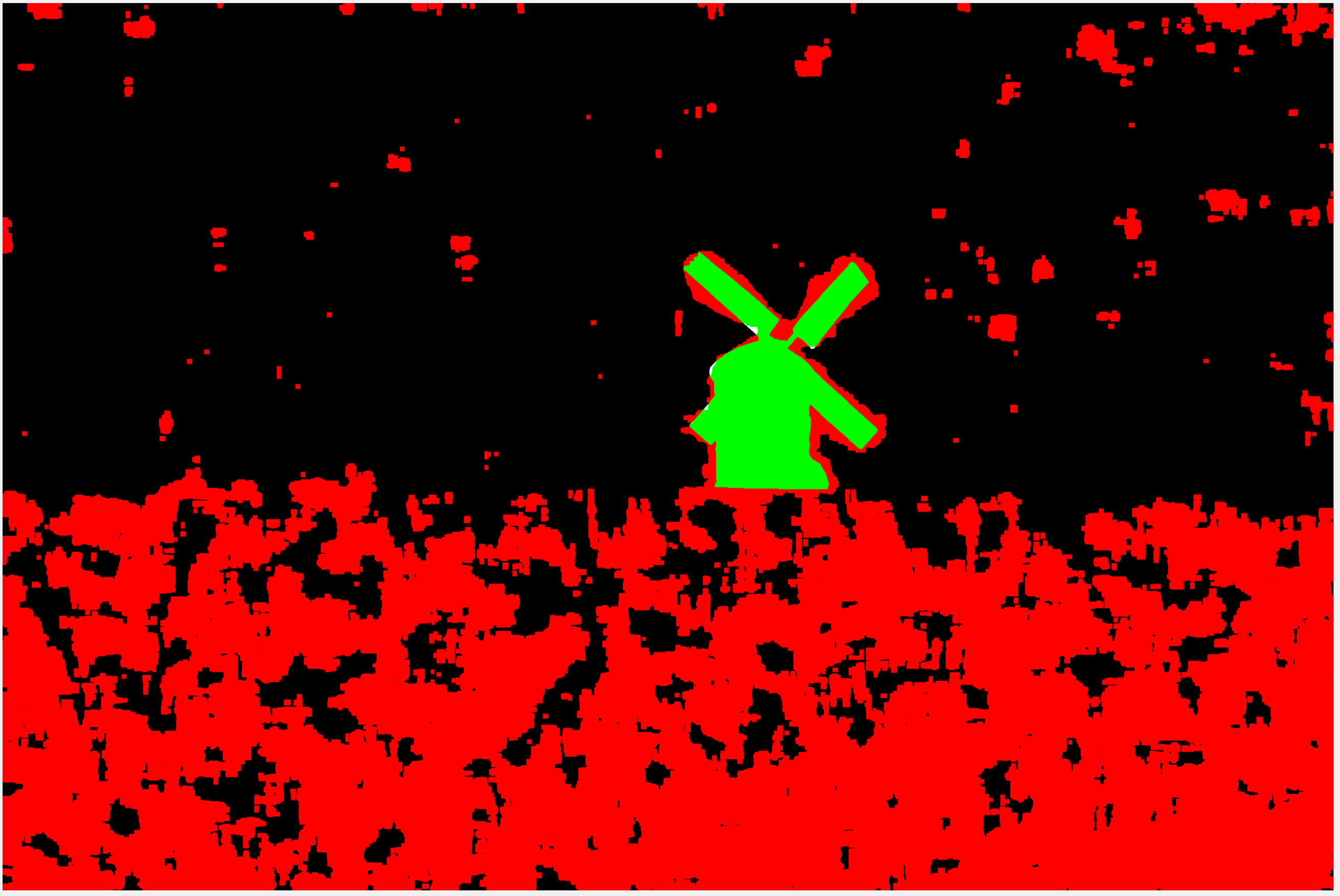}} & \raisebox{-.5\height}{\includegraphics[width=0.11\textwidth]{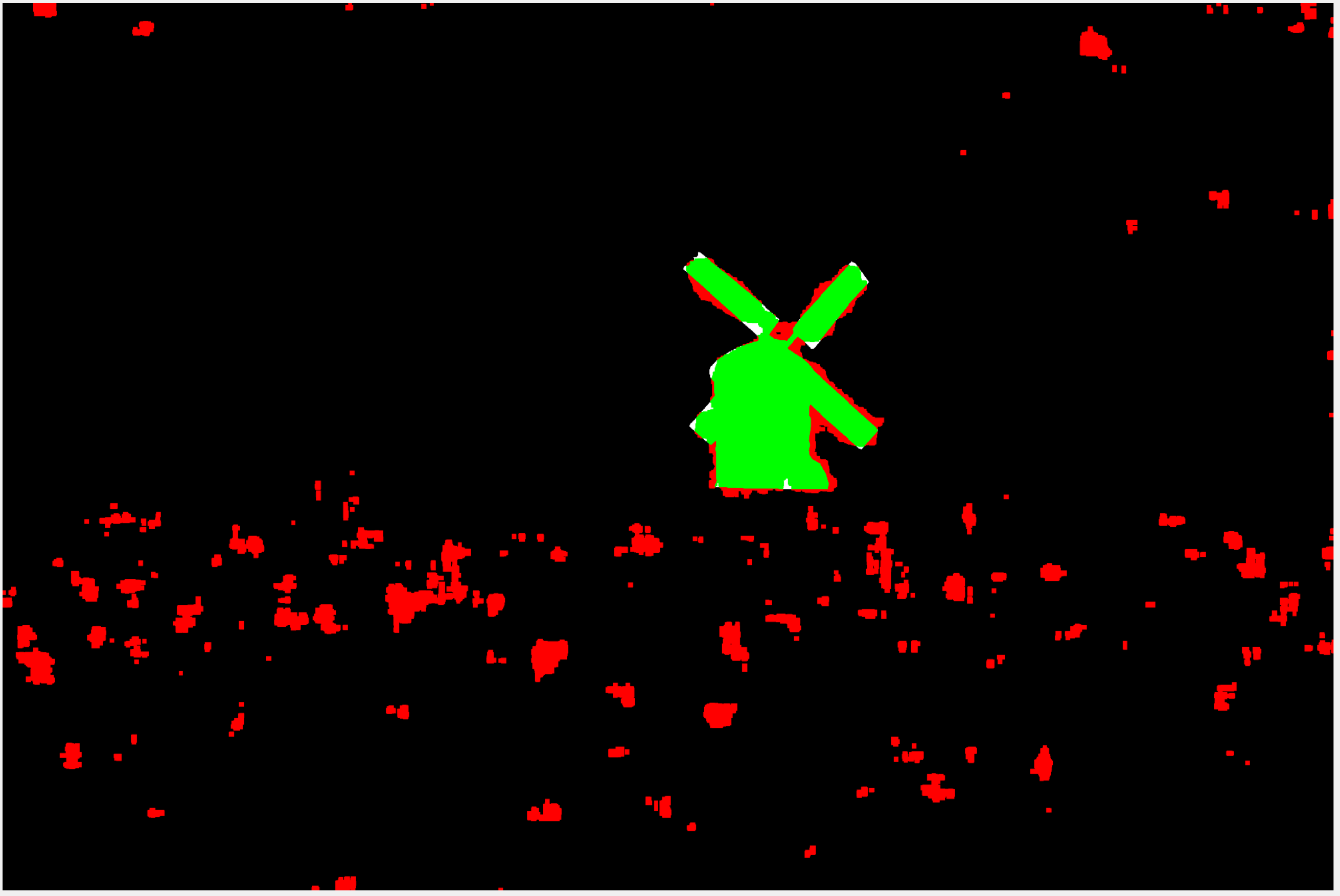}} & \raisebox{-.5\height}{\includegraphics[width=0.11\textwidth]{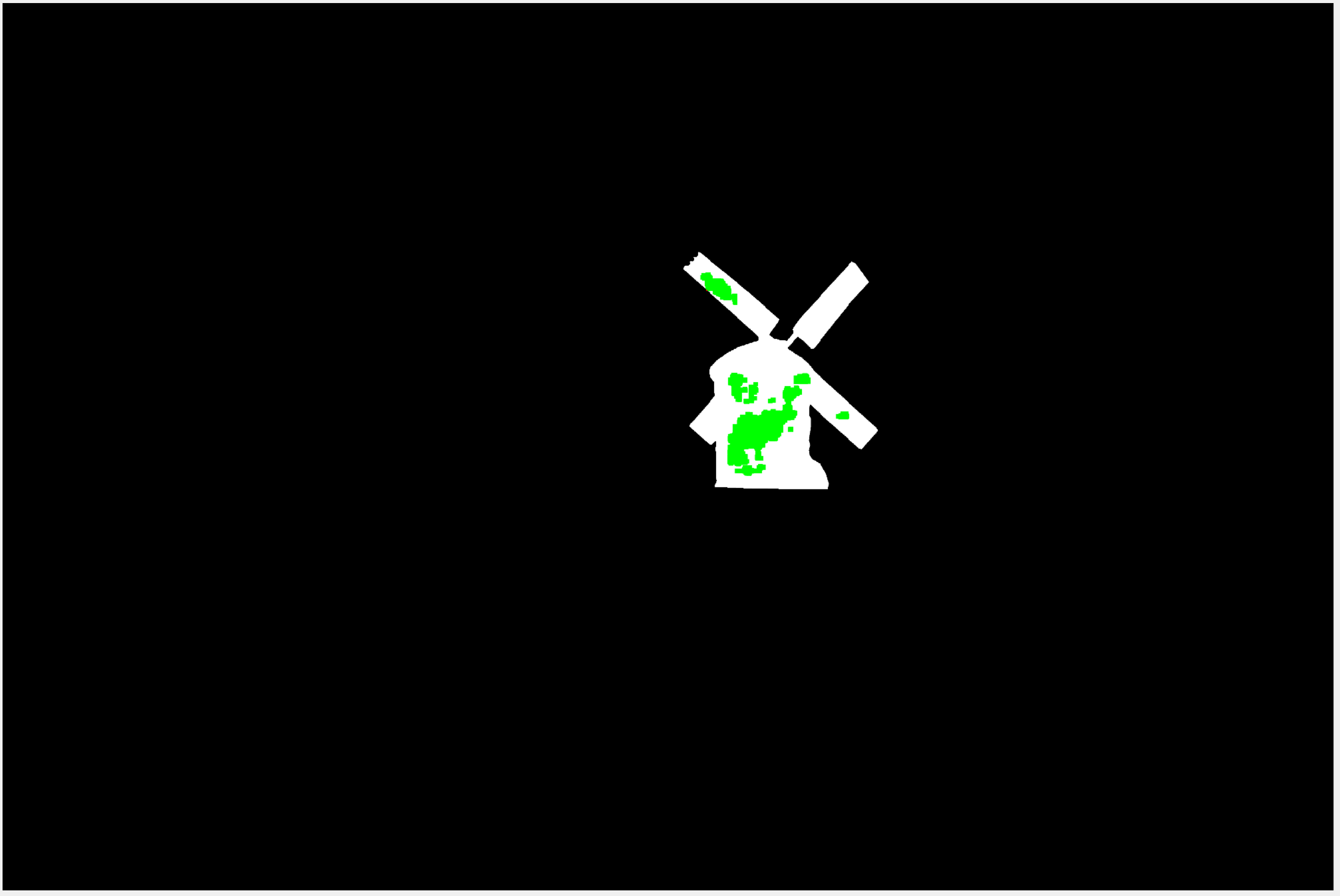}}
\end{tabular}
\caption{Forgery detection results on realistic forgeries from a Canon M6 with images of ISO speed 100, 800 and 6400. The images are taken with different exposure time to let them have similar exposure level. The Bayesian-MRF forgery detection algorithm is applied with the interaction parameter $\beta$ set to $10$ and probability prior $p_0$ set to $0.01$. The true detections are coloured with green and red for false detections. Missed tampered pixels are shown in white.}
\label{fig:forgery_detection}
\end{figure*}

Knowing that we cannot ignore the ISO speed in the correlation prediction training process, we also would like to investigate how mismatched ISO speeds of training and testing images would affect correlation prediction and subsequent forgery detection. In specific, we would like to investigate to what extent, a correlation predictor trained with images with a particular ISO speed can predict reliable correlation with images taken at other ISO speeds without significantly influencing the forgery detection results. We use Fig.\ref{fig:forgery_detection} to demonstrate the potential outcomes of forgery detection when the training image's ISO speed is significantly different from the test image's ISO speed.

Fig.\ref{fig:forgery_detection} shows the forgery detection results from tampered images with ISO speed 100, 800 and 6400 from a Canon M6. Images of the same scene taken at different ISO speeds are manipulated using Adobe Photoshop. For each image, the tampered region is replaced by using Photoshop's content-aware filling function, which leaves the tampered region at a similar noise level as its surrounding regions. We apply the Bayesian-MRF forgery detection algorithm from \cite{6719568} to the images. For all the images, we set the same parameters for the forgery detection algorithm: with the interaction parameter $\beta$ set to $10$ and probability prior $p_0$ set to $0.01$. The detection results show that the forgery detection algorithm works the best in terms of false detections when it is equipped with the matching ISO correlation predictor. We also notice that when we use ISO 100 correlation predictor for the forgery detection of the ISO 6400 forgery, despite the tampered region is correctly identified, there are a lot of false positives in the result. And when ISO 6400 correlation predictor is used for the detection of forgery in ISO 100 forgery image, while the entire authentic region is regarded as tampered, there are parts of the tampered region still undetected.

To explain these observations,  we have to consider the two potential outcomes of using images of different ISO speeds for the training of correlation predictors: the predicted correlation being either overestimated or underestimated.

Overestimation of the correlations (when correlation predictions are larger than the actual values) often occur when we use a correlation predictor trained with images of lower ISO speeds than the test image's ISO speed. As the actual intra-class correlations will be smaller than the predicted correlation, the corresponding pixels are more likely to be labeled as tampered, which results in an increased number of false detections as we have seen in Fig.\ref{fig:forgery_detection}. This is particularly harmful to real-life forensics. For most forgery detection algorithms, the authenticity of a pixel is checked by comparing its actual correlation with a threshold set with reference to the predicted correlations and expected inter-class correlation, which is expected to be zero. Though the actual algorithms can be different with more complexity by considering the distribution of the correlations from both inter- and intra-class as well as neighboring pixels' correlations, the comparison of whether the actual correlation sits closer to the predicted correlation or inter-class correlation when the correlation is overestimated can be a good indicator of how likely false detections can be introduced by a correlation predictor. Thus we would like to compare the two values: $d_1 = \rho-\bar{\rho}_\text{inter}$, which is the relative position from the inter-class correlation, $\bar{\rho}_\text{inter}$, to the actual computed correlation $\rho$ and $d_2 = \bar{\rho}_\text{intra}-\rho$, which is the relative position of the actual correlation, $\rho$, to the predicted intra-class correlation, $\bar{\rho}_\text{intra}$. Instead of comparing the $L_1$ distances, we compare these two values to focus more on the situation when the correlation is overestimated, which causes the actual correlation to be a value between the expected inter-class correlation and predicted correlation. We estimate $\bar{\rho}_\text{inter}$ as zero and use the predicted correlation to estimate $\bar{\rho}_\text{predict}$, and it gives $d_1 - d_2 \approx 2\rho - \bar{\rho}_\text{predict}$. When $d_1-d_2$ is negative, it indicates that the correlation has a large chance of being misidentified as an inter-class correlation.

\begin{figure}
\begin{center}
\includegraphics[width=0.45\textwidth]{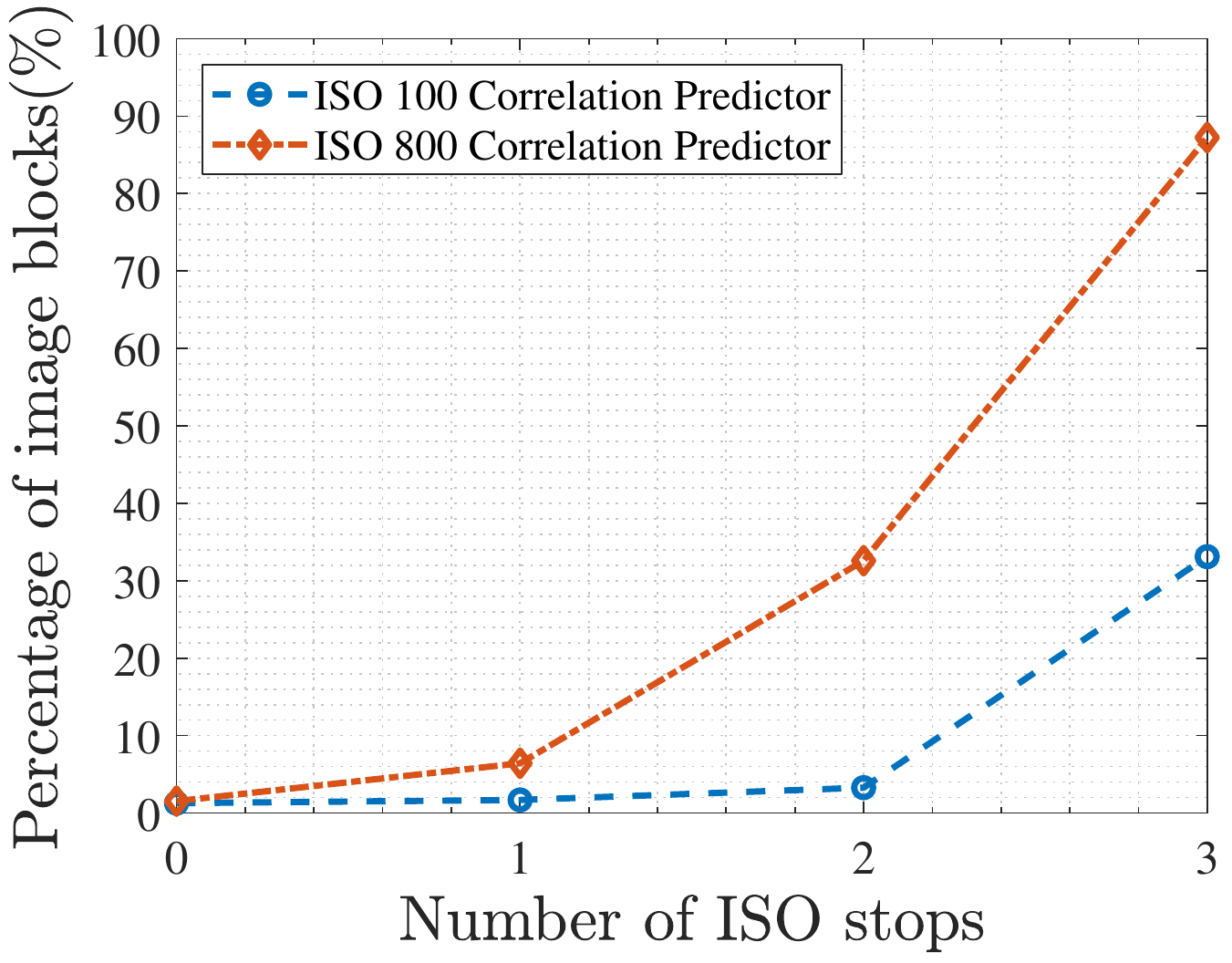}
\caption{A plot of the percentages of image blocks with $d_1-d_2$ smaller than 0 against the number of ISO stops the test image's ISO speed is above the ISO speed of the images used to train the correlation predictor for a Canon M6. The percentage indicates the portion of the authentic image blocks at risk of being misidentified as tampered blocks by forgery detection algorithms.}
\label{fig:percent}
\end{center}
\vspace{-.2cm}
\end{figure}

Again, use the camera Canon M6 as an example, we show the percentages of the image blocks with $d_1-d_2$ smaller than $0$ in Fig.\ref{fig:percent} when we use an ISO 100 and 800 correlation predictors to predict for test images with ISO speed number of stops above the training images. The plot shows that when the test images' ISO speeds are within the one-stop range of the training images' ISO speed, there is only a relatively small portion of blocks (i.e. less than $10\%$) with $d_1 - d_2$ smaller than 0 for both ISO 100 and ISO 800 correlation predictors. As the deviation from the test images' ISO speed to the training images' ISO speed increases, we start to see a higher percentage from Fig.\ref{fig:percent}, indicating an increased number of false detections could be introduced into forgery detection results. As we approximate $d_1-d_2$ as $2\rho-\bar{\rho}_\text{predict}$, it becomes an universal problem when $\rho < \frac{1}{2} \bar{\rho}_\text{predict}$.

\begin{figure*}
\centering
    \textbf{Canon M6}\par\medskip

\begin{subfigure}{.24\textwidth}
\includegraphics[width = \textwidth]{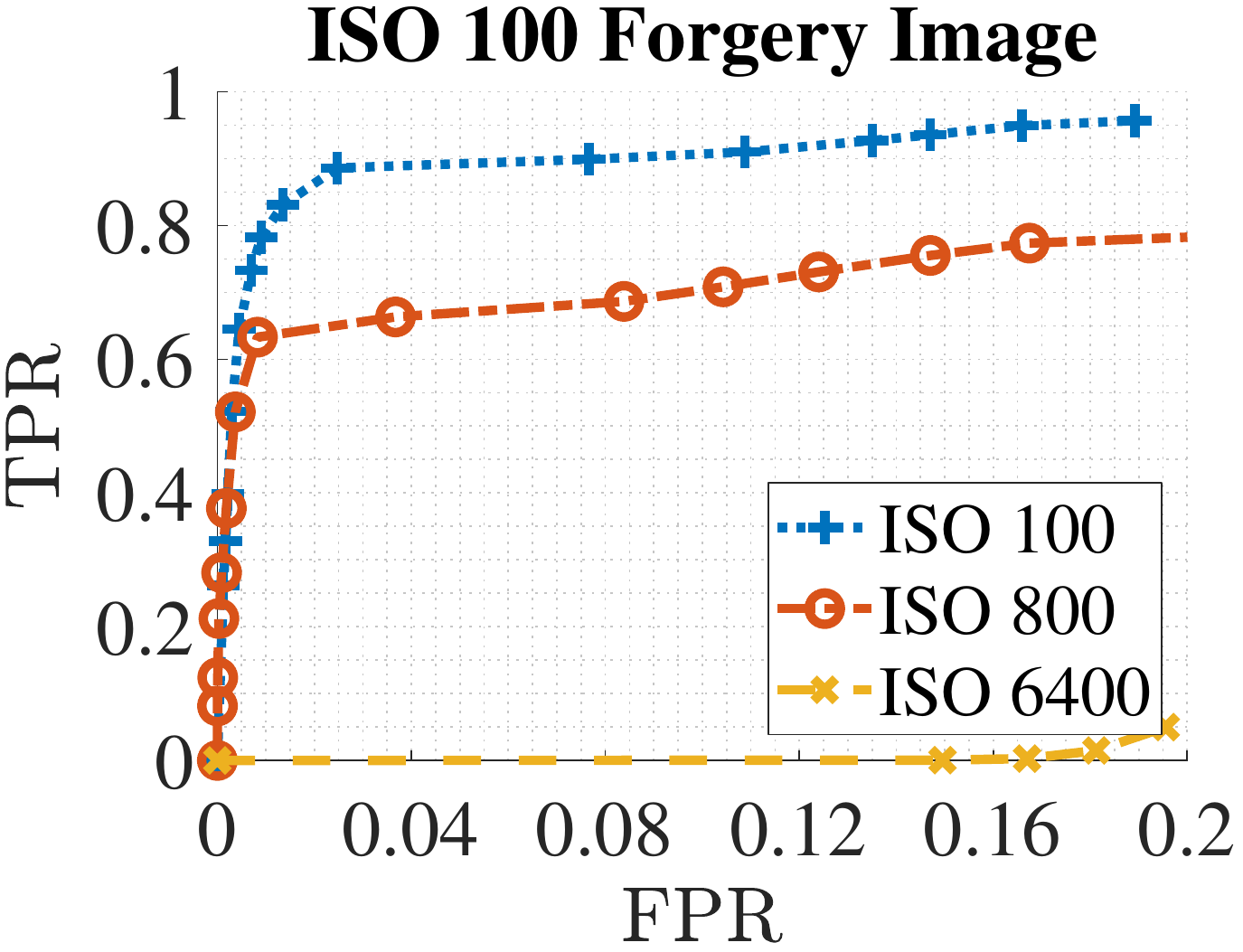}
\end{subfigure}
\begin{subfigure}{.24\textwidth}
\includegraphics[width = \textwidth]{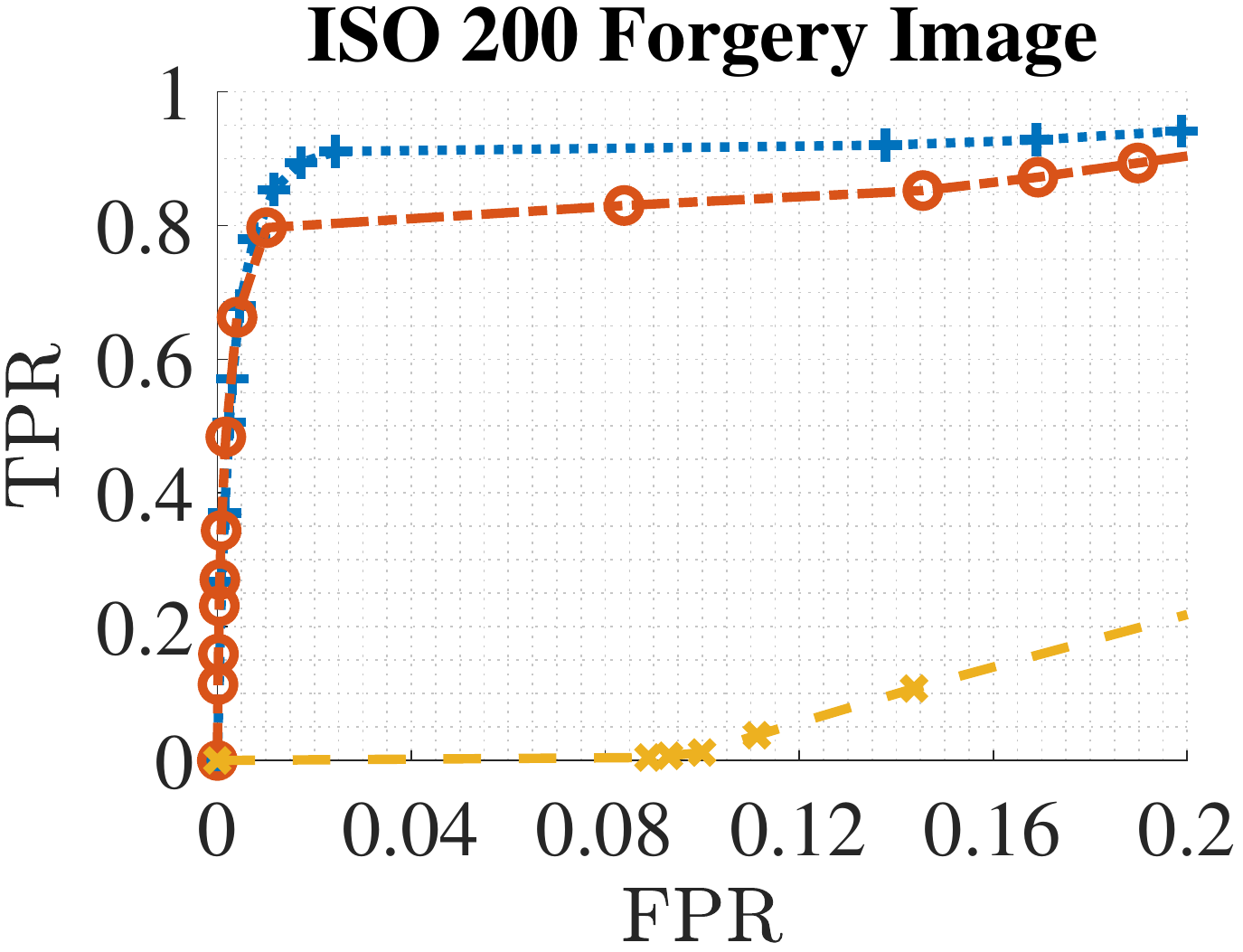}
\end{subfigure}
\begin{subfigure}{.24\textwidth}
\includegraphics[width = \textwidth]{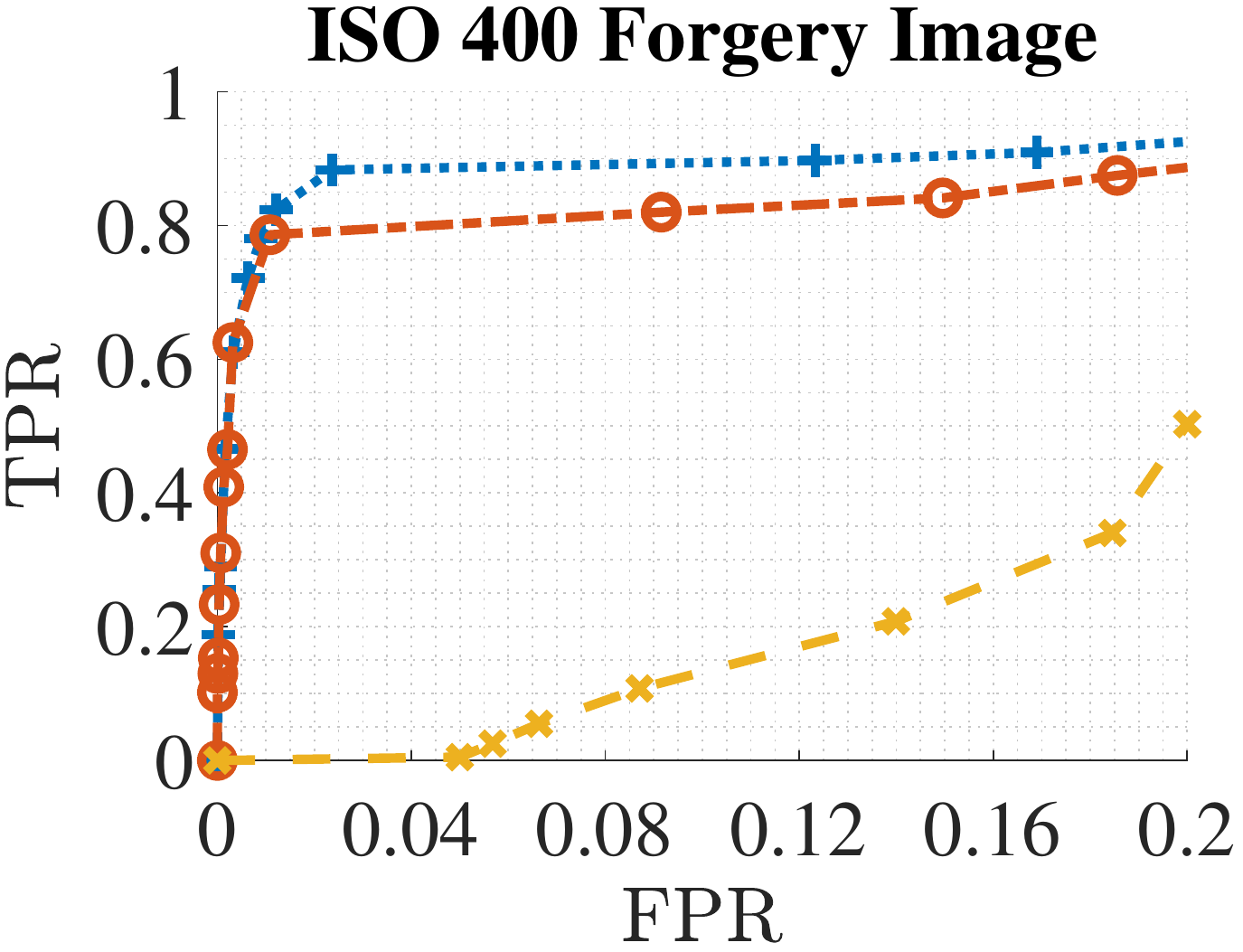}
\end{subfigure}
\begin{subfigure}{.24\textwidth}
\includegraphics[width = \textwidth]{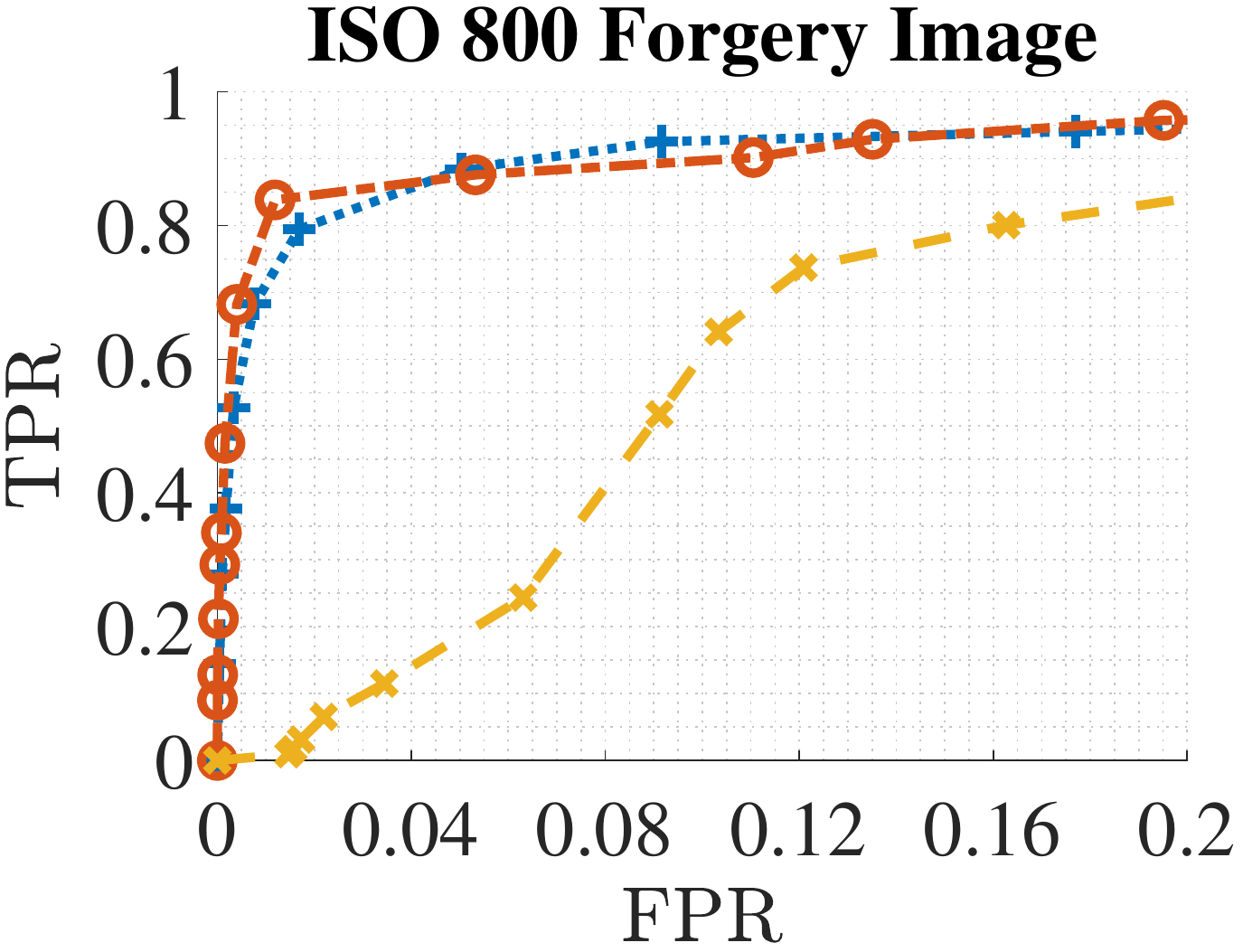}
\end{subfigure}
\begin{subfigure}{.24\textwidth}
\includegraphics[width = \textwidth]{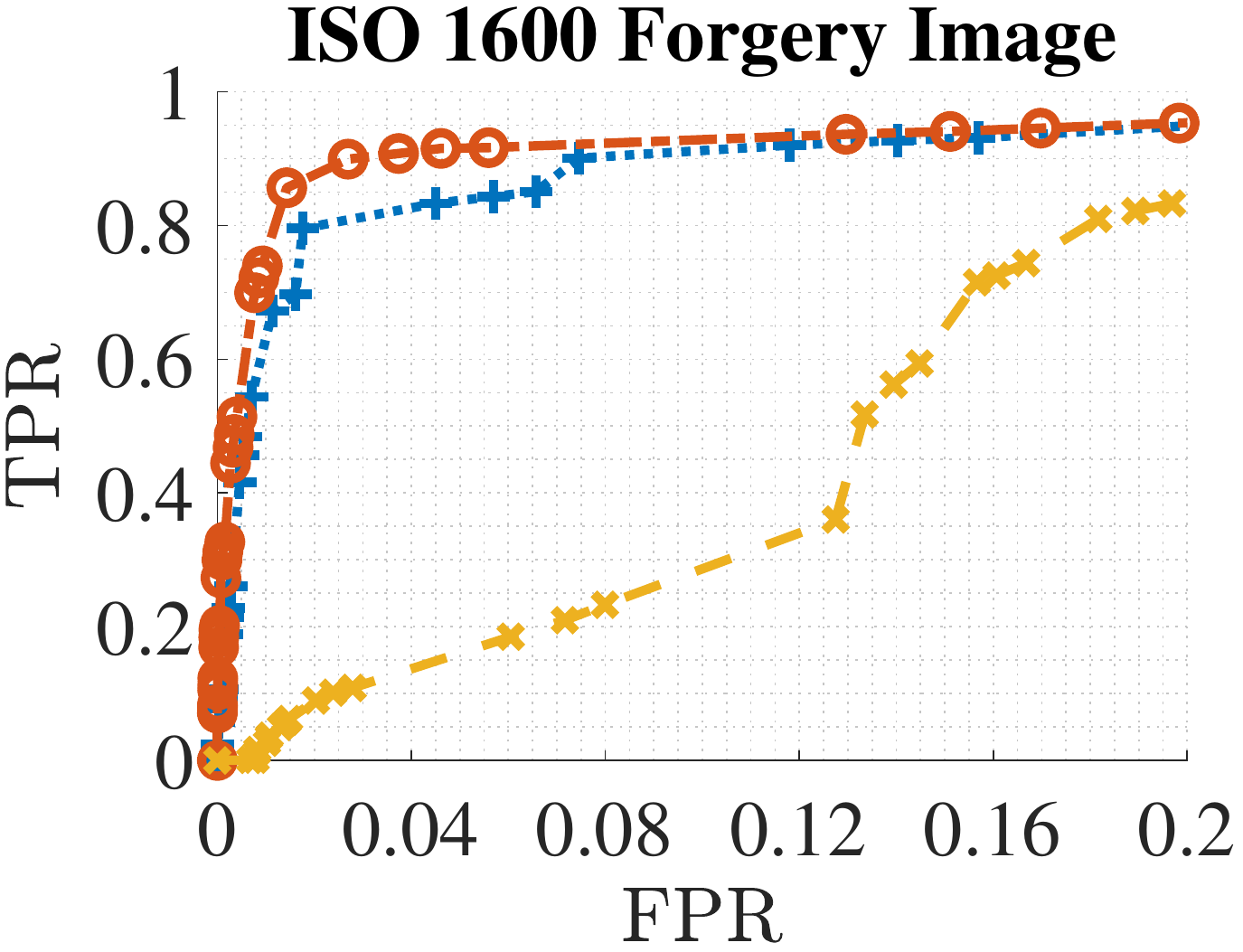}
\end{subfigure}
\begin{subfigure}{.24\textwidth}
\includegraphics[width = \textwidth]{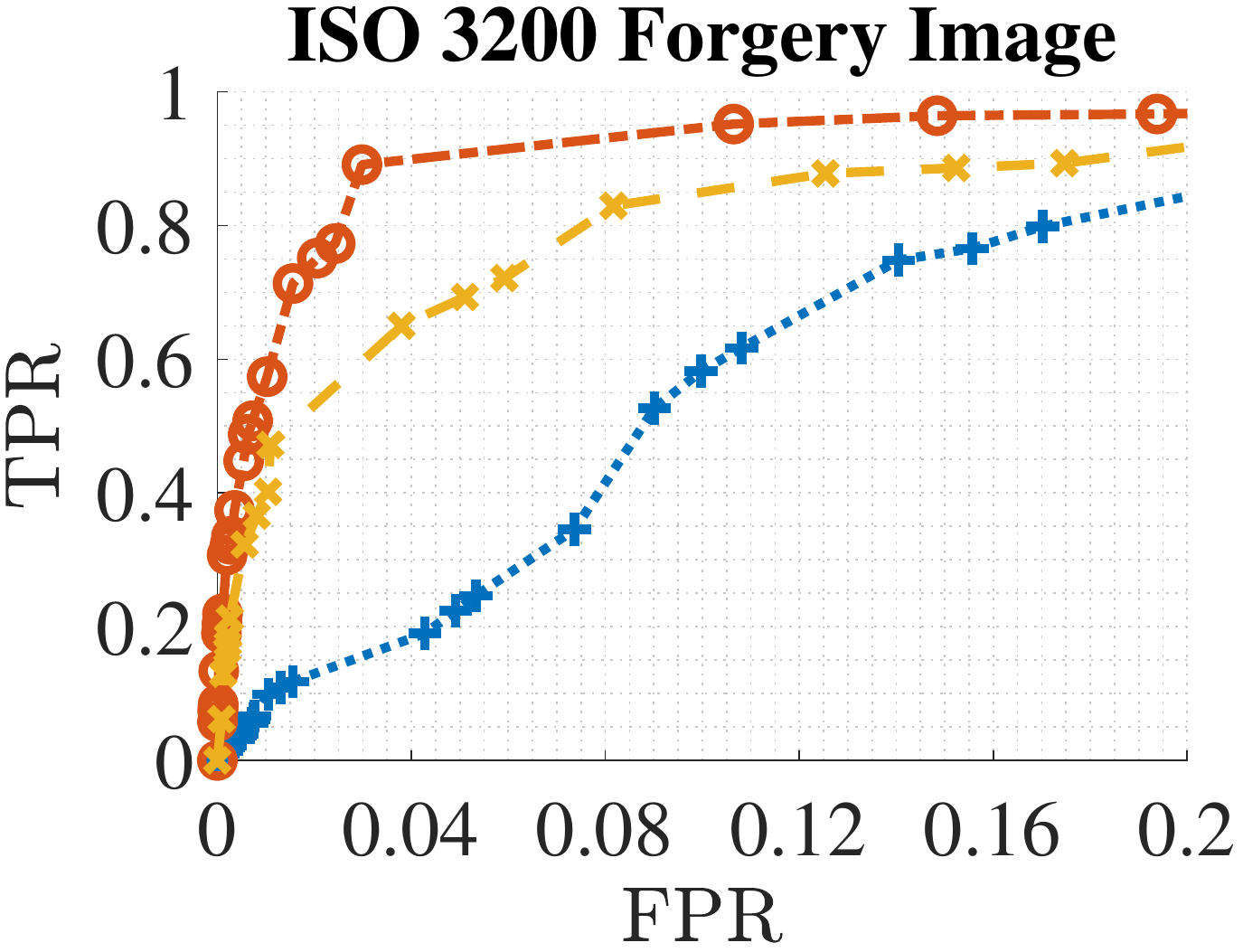}
\end{subfigure}
\begin{subfigure}{.24\textwidth}
\includegraphics[width = \textwidth]{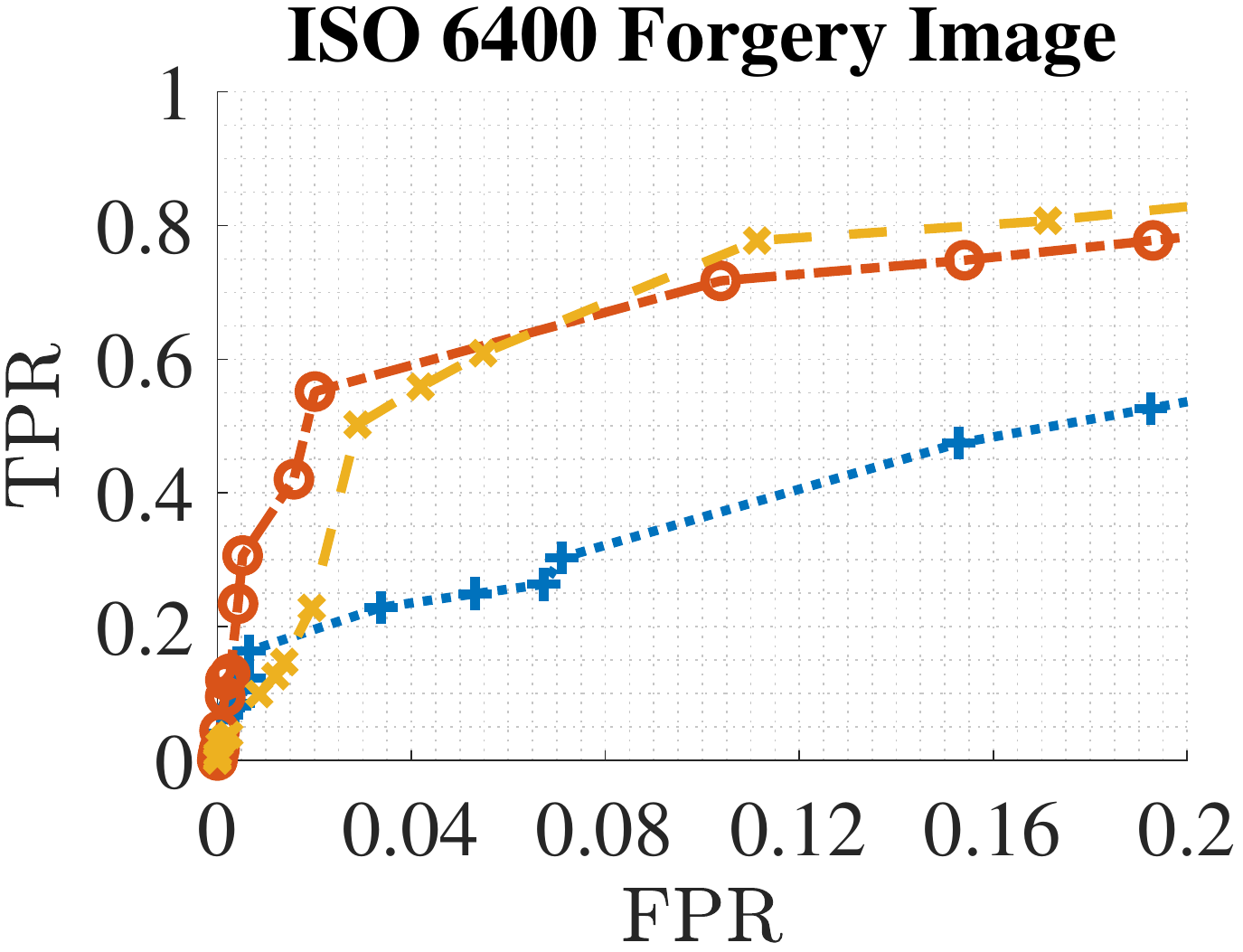}
\end{subfigure}
\par\medskip
    \textbf{Sigma SdQuattro}\par\medskip

\begin{subfigure}{.24\textwidth}
\includegraphics[width = \textwidth]{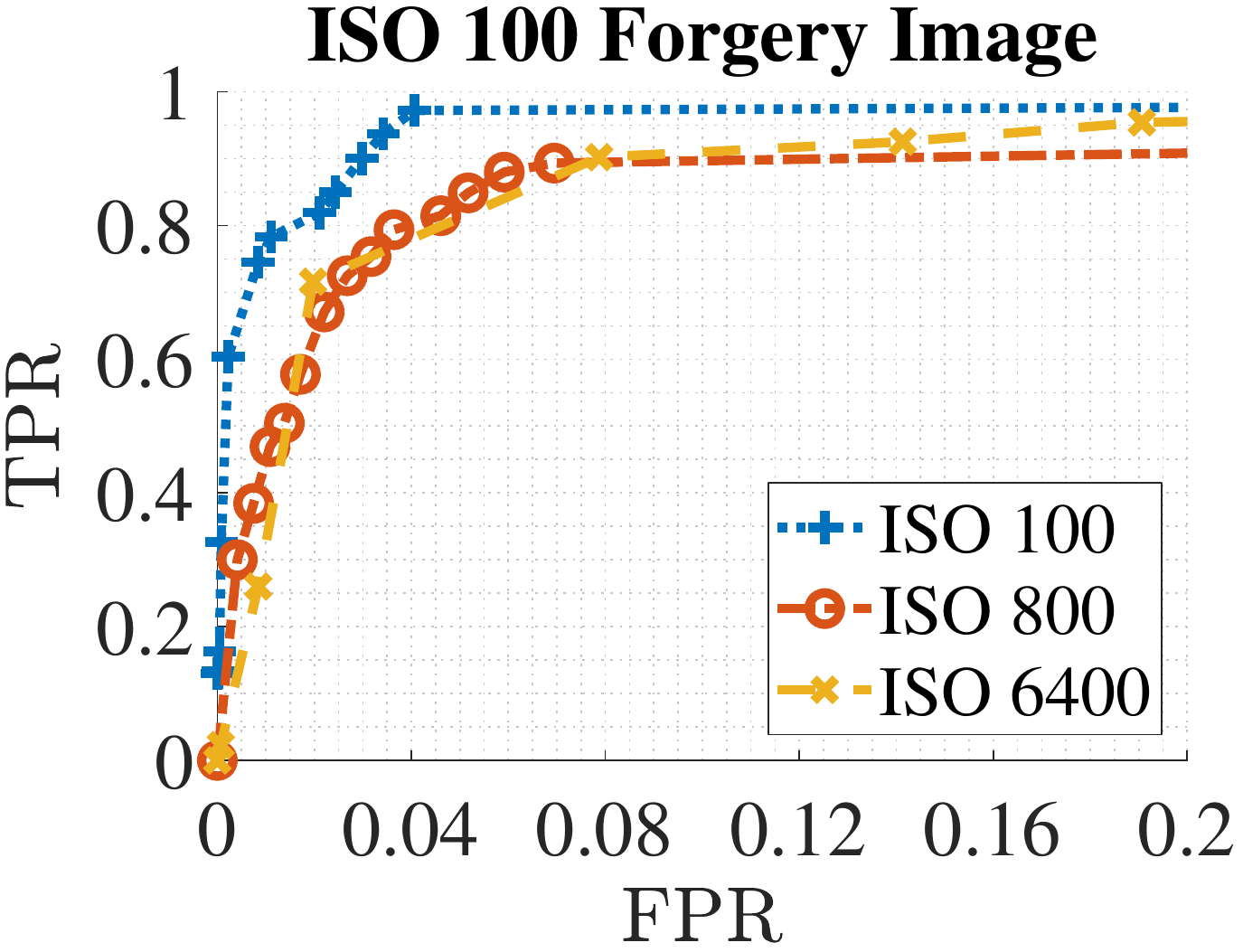}
\end{subfigure}
\begin{subfigure}{.24\textwidth}
\includegraphics[width = \textwidth]{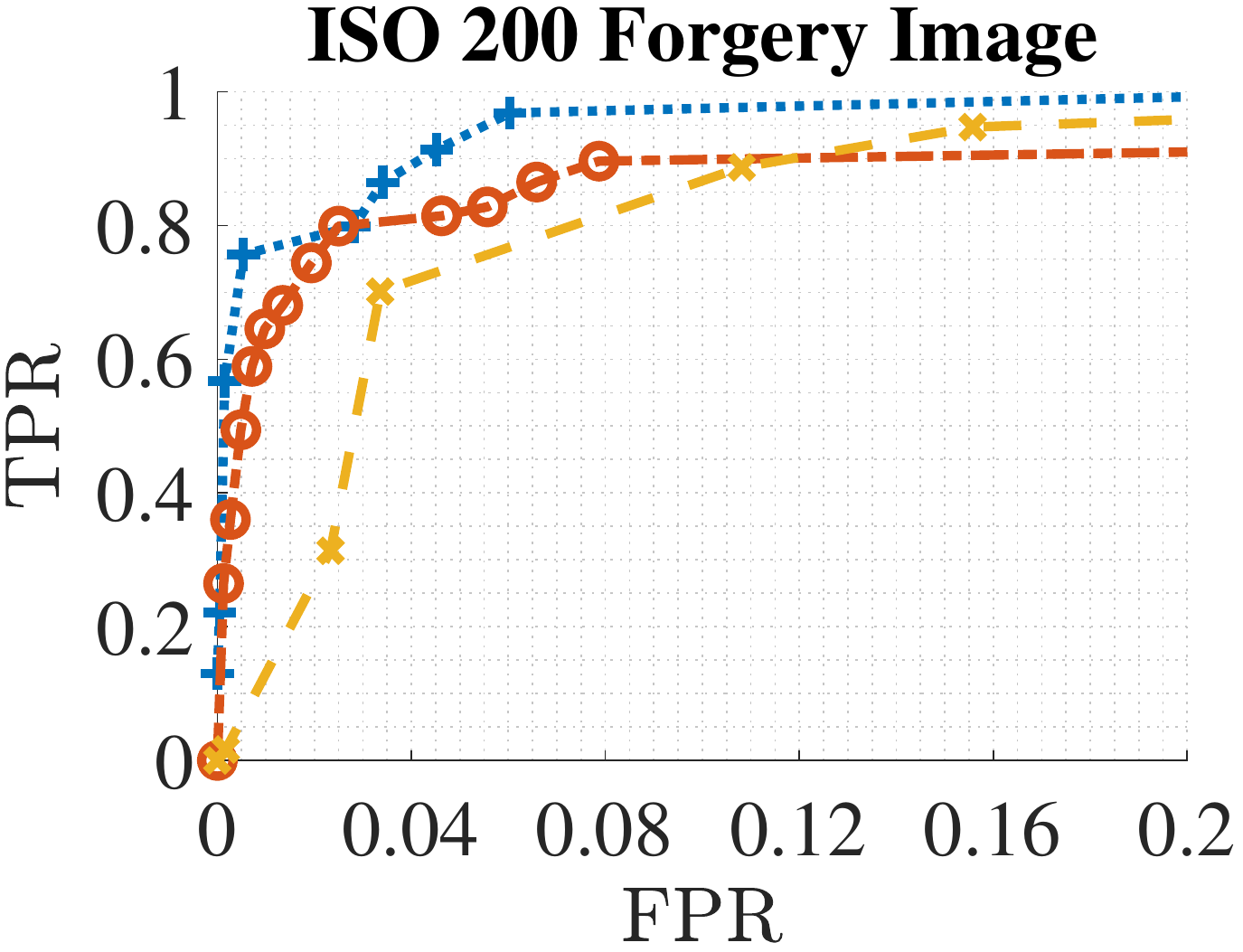}
\end{subfigure}
\begin{subfigure}{.24\textwidth}
\includegraphics[width = \textwidth]{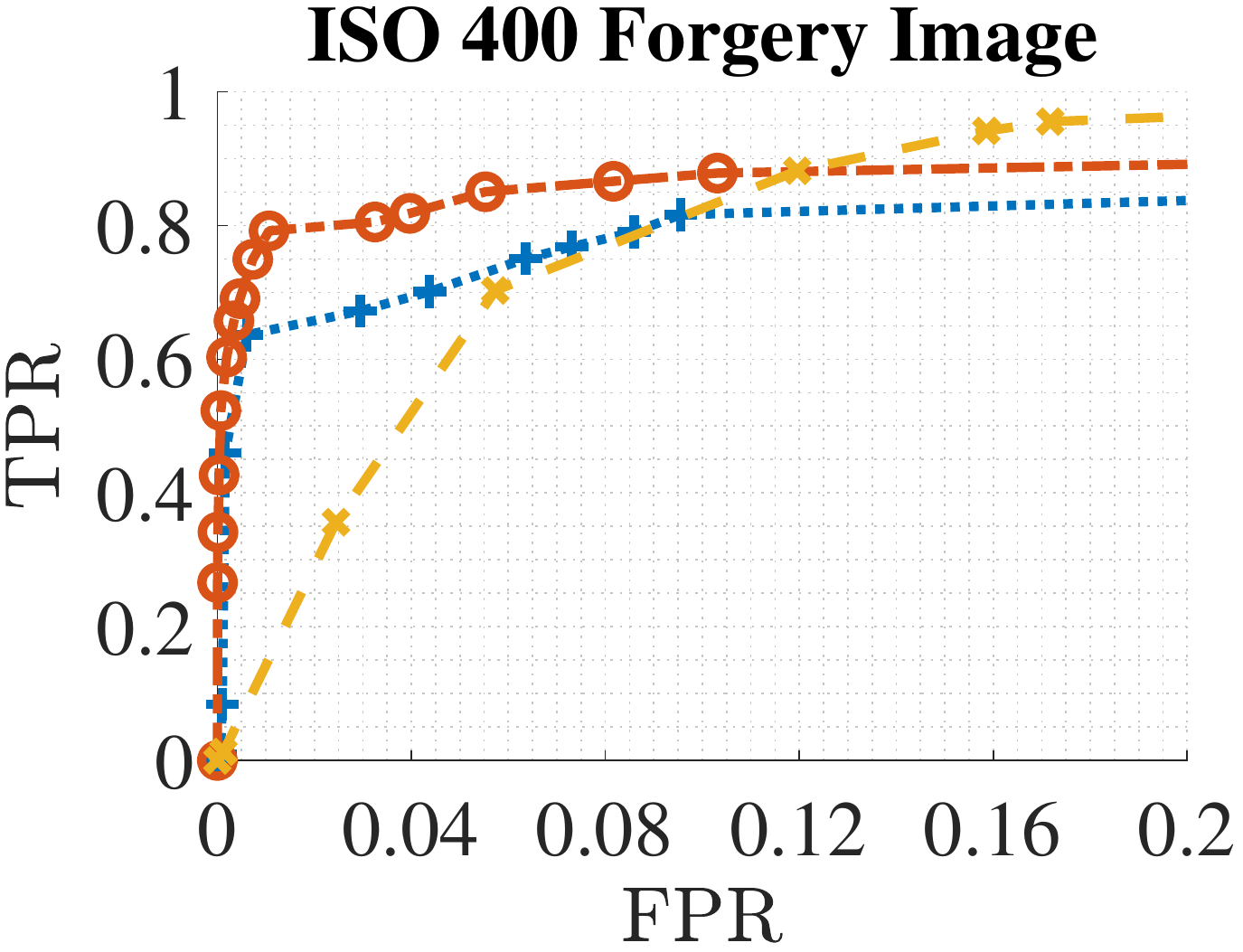}
\end{subfigure}
\begin{subfigure}{.24\textwidth}
\includegraphics[width = \textwidth]{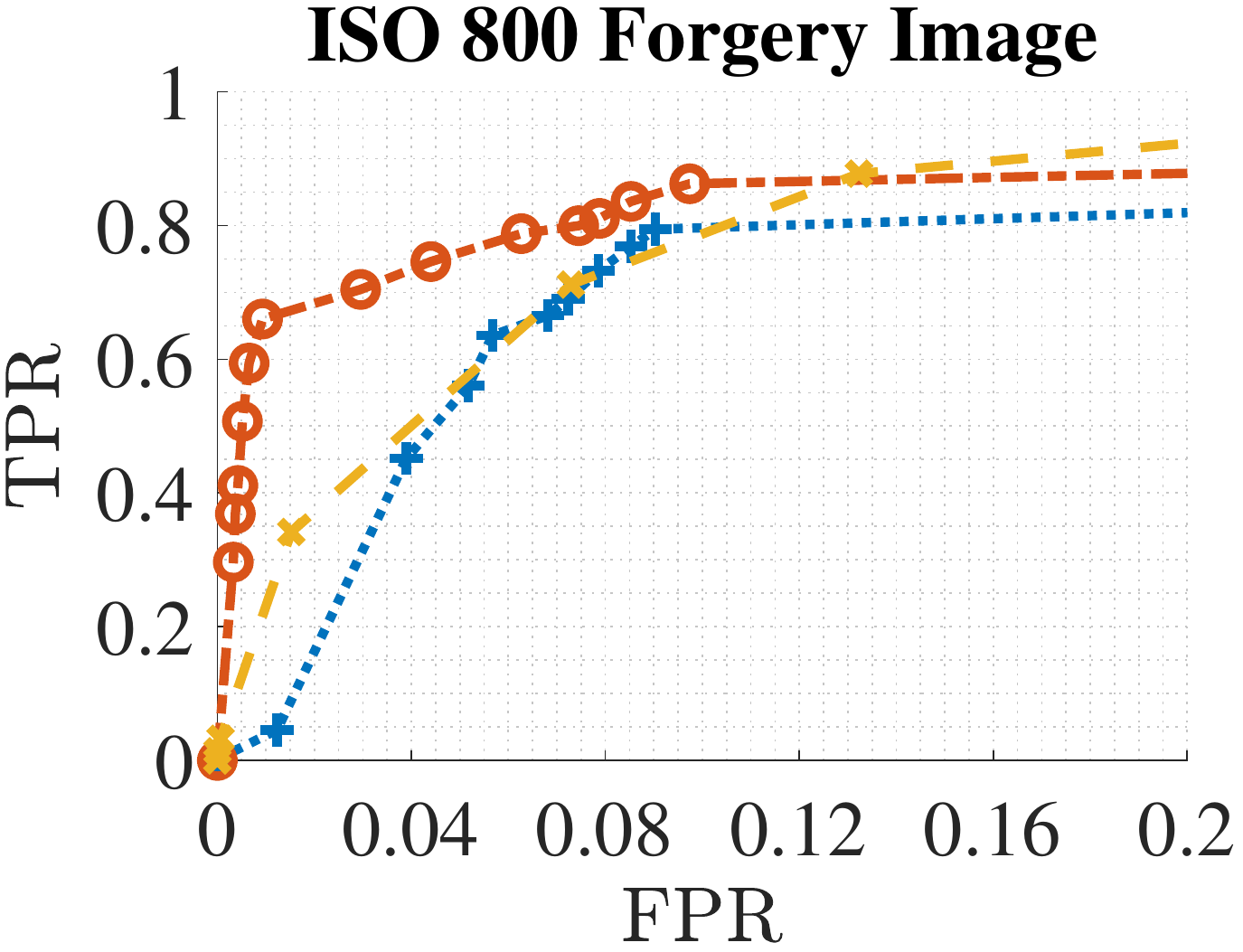}
\end{subfigure}
\begin{subfigure}{.24\textwidth}
\includegraphics[width = \textwidth]{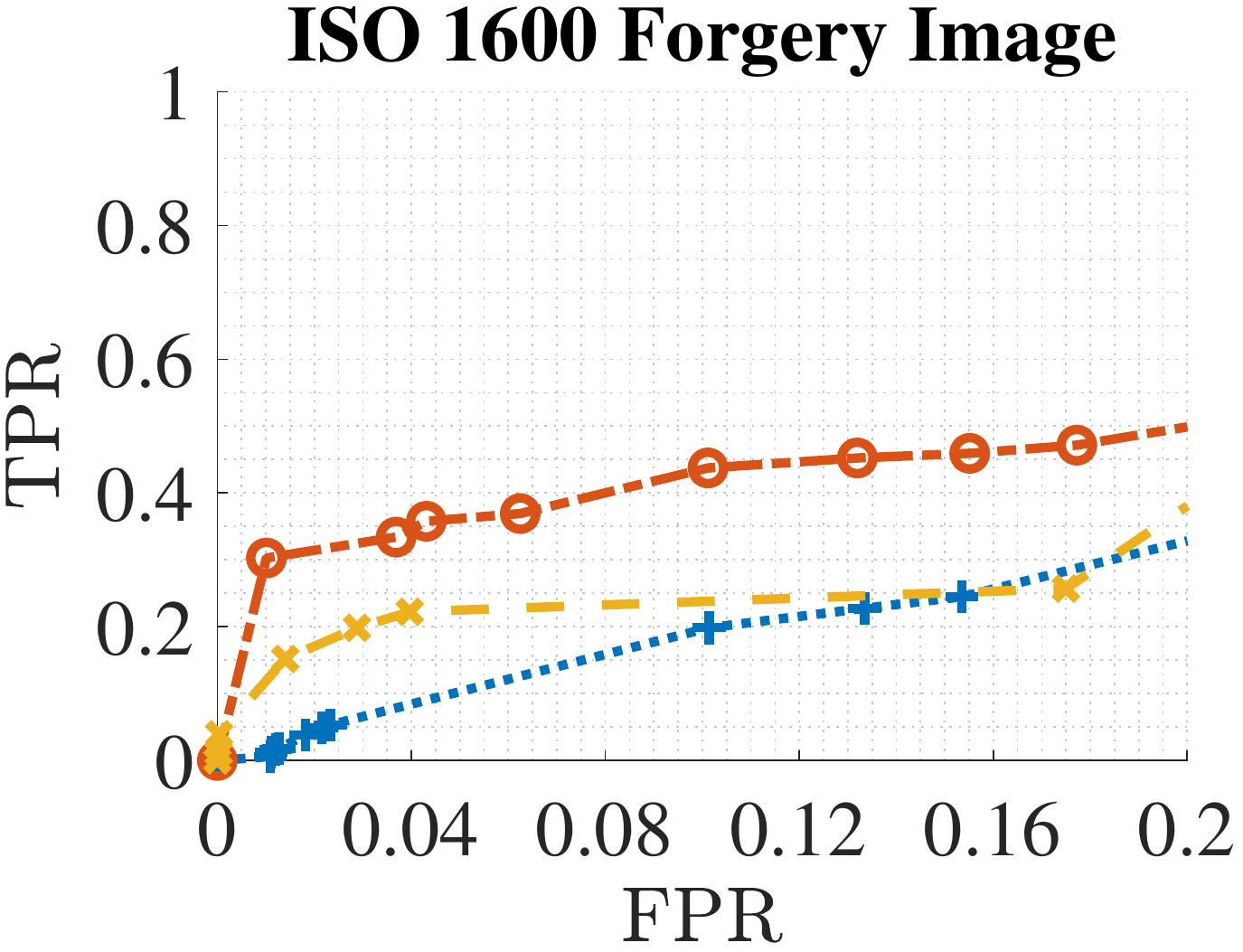}
\end{subfigure}
\begin{subfigure}{.24\textwidth}
\includegraphics[width = \textwidth]{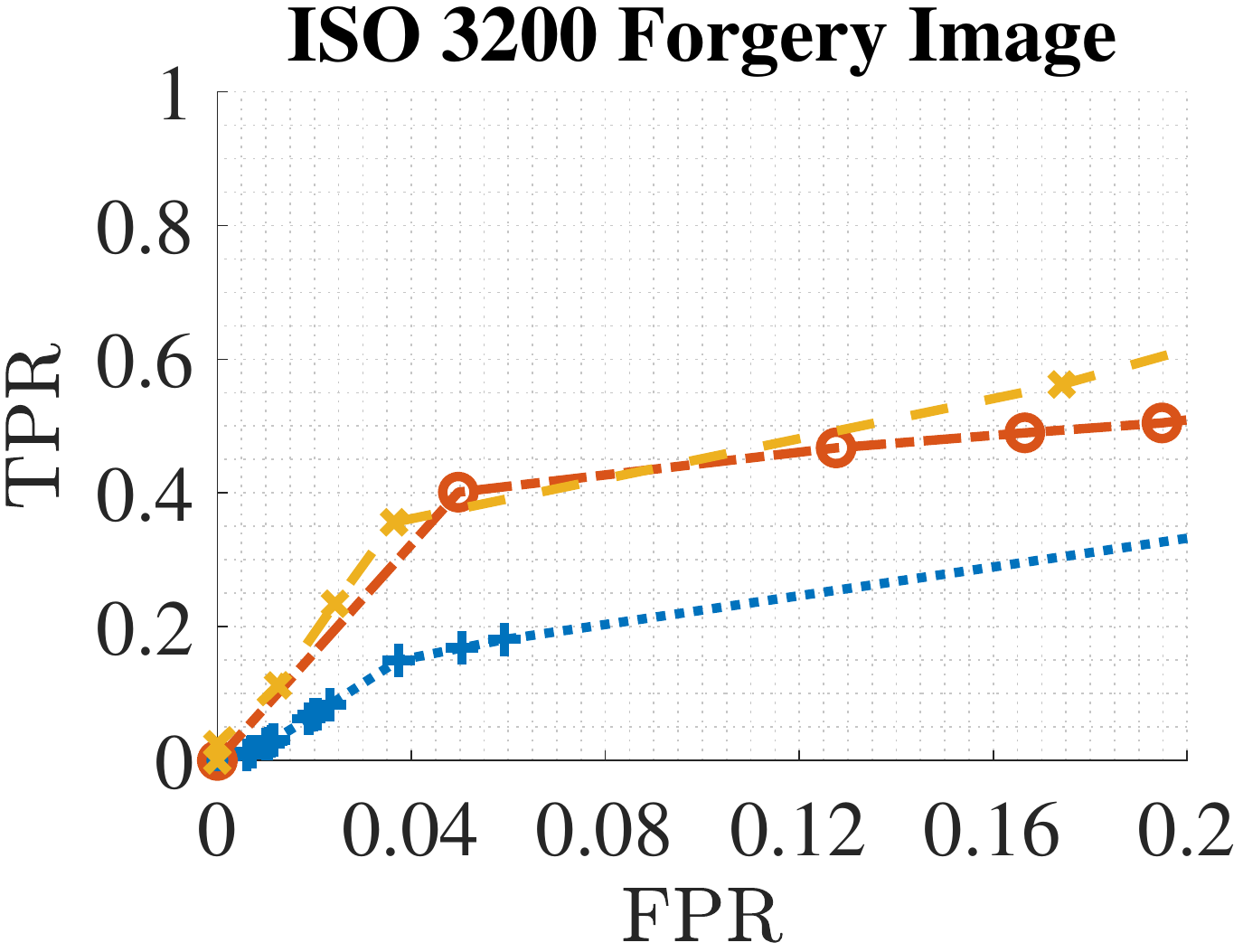}
\end{subfigure}
\begin{subfigure}{.24\textwidth}
\includegraphics[width = \textwidth]{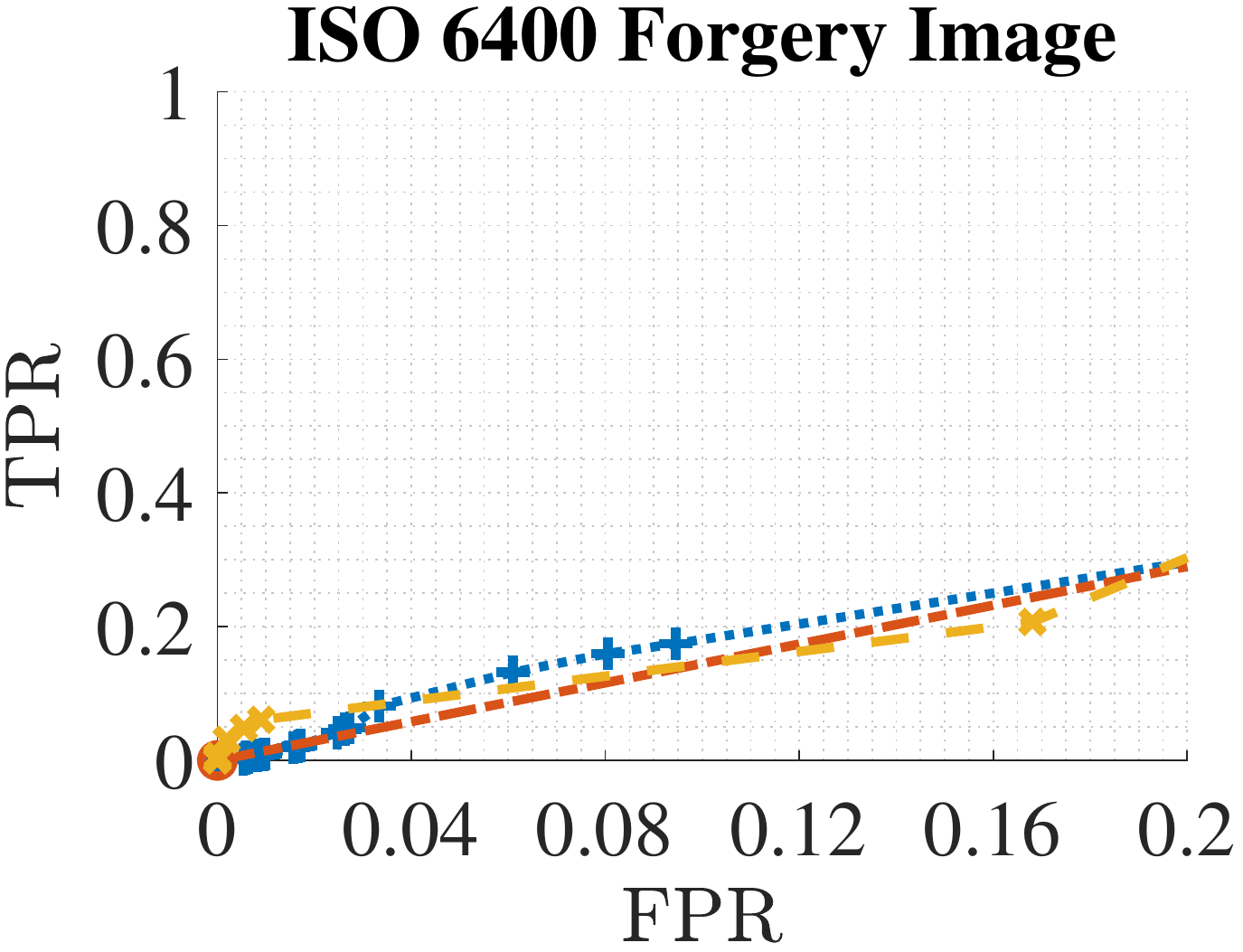}
\end{subfigure}
\caption{Receiver Operating Characteristic (ROC) curves of tampering localization using Bayesian-MRF forgery detection method \cite{6719568} on synthetic forgeries taken at different ISO speeds from a Canon M6 and Sigma SdQuattro. The legend shows the ISO speeds corresponding to the correlation predictors used to generate the ROC curves.}
\label{fig:ROCS}
\end{figure*}

Base on the correlation model derived from Equation (\ref{eqn:str}) and observations from experiments, we found that for image blocks of the same scene from images taken at different ISO speeds, it is generally true that the block-wise correlation in an image taken with ISO speed $G_1$ is twice larger than the correlation of the corresponding block from an image taken at ISO speed $G_2 = 2G_1$. Thus, we claim that $G_2 = 2 G_1$ is a safe choice to be set as the largest ISO speed a correlation predictor trained with images of ISO speed $G_1$ can reliably predict for. Similar behavior can be observed on other cameras as well and we show the receiver operating characteristic (ROC) curve for forgery detection in Fig.\ref{fig:ROCS} for further validation.\\

Each ROC curve in Fig.\ref{fig:ROCS} is plotted by running the Bayesian-Markov random field (MRF) based forgery detection algorithm from \cite{6719568} on 80 synthetic forgery images at each of the 7 presented ISO speeds. Three correlation predictors, each trained with 20 natural images taken at ISO speed 100, 800 and 6400, respectively, are used to predict the correlations for the forged images. We vary the interaction parameter $\beta$ in the range of $[1,1200]$ and the probability prior $p_0$ between $[0,1]$ to set different combinations of the parameters for the algorithm. This allows us to generate the enveloping curves for the ROCs to show the best performance. The 80 synthetic forged images are generated from 20 full-sized authentic images. From each full-sized image, we select 4 regions of $1024\times 1024$ pixels. We replace the center of each $1024 \times 1024$ pixel region's center with a tampered patch of $256\times 256$ pixels. The patch used to replace the center is cropped from the same original image but from a different position to ensure that it does not have the same PRNU. In fact, we fully facilitate the Warwick Image Forensics Dataset which provides images of the same content at different ISO speeds. This allows us to generate the synthetic forged images in the way that for one synthetic forged image at one ISO speed, we can find images of the same content at other ISO speeds as well. By doing this, Fig.\ref{fig:ROCS} not only allows us to compare the performance of different correlation predictors for forged images at one ISO speed but we can also systematically compare the performance of one correlation predictor for different ISO speeds.

We run the test on different cameras from Warwick Image Forensics Dataset. To save space, we only show the ROC curves of two most representative cameras, a Canon M6 and a Sigma SdQuattro in Fig.\ref{fig:ROCS}. Canon M6 represents the cameras that can generate relatively less noisy images (with a large peak to noise ratio (PSNR)) for most ISO speeds from the camera while Sigma SdQuattro represents the cameras whose image quality is highly dependent on the selected ISO speed. The false positive rate (FPR) and true positive rate (TPR) are computed at the pixel-level. As for real-life tampering localization application, we usually require the method to produce a small FPR, thus we focus on the range of $[0, 0.2]$ of FPR in the plots.

From Fig. \ref{fig:ROCS}, we first notice that for ISO 100, 800 and 6400 forgery images, the matching ISO correlation predictor works the best in both cameras in almost every case. The only exception is for Sigma SdQuattro ISO 6400 forgery images. In this case, despite the ISO 6400 correlation predictor can make predictions accurately as we have seen from Table \ref{tab:warwickr2}, none of the three correlation predictors can produce accurate detections. This is because, for high ISO images from this camera, the images' intra-class correlations are generally very close to zero and hard to be separated from inter-class correlations. For such images, PRNU-based methods may not be the best tool to perform forgery localization. However, the ISO specific correlation predictor can still be helpful in such a scenario as it will be able to accurately predict the correlations close to zero. Thus, the users can be warned that the PRNU based methods may not be suitable under such a scenario. Overall, the results show the benefit of using a matching ISO correlation predictor for forgery detection.

For both cameras, we observe that the detection results of using the ISO 100 correlation predictors (i.e. predictors trained with images taken at ISO speed 100) are better when the forged image's ISO speed is smaller than 400. While the Canon M6's relatively good high PSNR at higher ISO speeds allows the ISO 100 correlation predictor to perform reasonably well for a forged image with ISO speed up to 1600, it is not the case for the Sigma SdQuattro camera. From ISO 400 and above, the ISO 100 correlation predictor for the Sigma SdQuattro starts to struggle. And the similar effect can be observed for ISO 800 correlation predictors when they are used to predict for images with ISO speed much higher than 800. Thus, it conforms to our argument that a predictor trained with images taken at ISO speed $G_1$ can perform reliably on the images taken at an ISO speed $G_2$ that is lower than or equal to $2 G_1$. While depending on the camera, some correlation predictors may perform when the test image's ISO speed is above the range, the above argument provides a safe range for the choice of correlation predictor's training ISO speed without risking too many false detections.

Fig.\ref{fig:ROCS} also shows the situation when the correlation predictors underestimate the test image's correlations. Underestimation often occurs when we use a correlation predictor trained with images of a much higher ISO speed than the test image's ISO speed. In the plots, we noticed that the ISO 6400 correlation predictors, especially for the Canon M6 camera, appear to have difficulty in correctly localizing the forgery for images with low ISO speed. This is because when the correlation predictor underestimates the correlations, it eventually reduces the forgery detection algorithm's capability of correctly identifying tampered pixels. Thus, to avoid the underestimation but still provides a practical range from which a training ISO speed can be conveniently selected, we empirically set the lower bound of the ISO speed a correlation predictor can be used for to half of the ISO speed of its training images. From the plots, we see by using this range, the corresponding detection results either outperform other correlation predictors or are on par with the best performance. Altogether, we conclude that for a test image taken at ISO speed $G_1$, using correlation predictors trained with images of ISO speed, $G_2$, which is in the one-stop range of $G_1$ ($G_2 \in [G_1/2,2G_1]$) can produce forgery detection result without risking false detections being excessively introduced due to the correlation predictor.
\section{ISO specific correlation prediction process}
\label{sec:infer}
Observing the ISO speed's impact on correlation prediction, we concluded that reliable correlation predictions should be made in an ISO specific way. Thus, we propose an ISO specific correlation prediction process. To predict correlations for an image of ISO speed $G_1$, we have to use a correlation predictor, preferably trained with images of the same ISO speed at $G_1$, or similar to $G_1$. An ISO speed $G_2$ is considered as similar to $G_1$ if $G_2$ is in the one-stop range of $G_1$. The images used for the training of the correlation predictor should cover diverse image feature settings: including both bright and dark scenes, highly textured and flat patterns, etc. To cover such a diverse set of image features, it usually requires a large number of images. Thus, a good correlation predictor should be trained with no less than 20 full-sized images. With a relatively large collection of images of good feature diversity taken at an ISO speed similar to the test image, the weight for each defined feature can be learned following the process presented in \cite{chen2008determining} for the correlation predictor.

In order to complete the correlation prediction process, we need to have the knowledge of the ISO speed $G_1$ to find images of the same or similar ISO speeds to form the training set. However, as the image in question may have undergone some unknown manipulations, either on its image content or metadata, the ISO speed information presented in the metadata can be unreliable or even unavailable. Thus, we can often face the problem when we have an image of unknown ISO speed and we would like to select images with the closest ISO speed to the image to train a correlation predictor.

As a known factor, for the same camera, the higher the ISO speed is, the higher the level of noise is introduced to the content of images. Thus, it is intuitive to infer an image's ISO speed by exploiting its noise characteristics in the content. Based on the Poissonian-Gaussian noise model \cite{radioCCD}, methods are proposed in \cite{4623175,foi2009clipped, thai2014camera} to infer the camera gain, $g$, from a RAW image, which then can be directly related to the camera's ISO speed. Despite these methods showing promising performance on RAW images, as the noise model generally cannot be applied directly to non-RAW image formats, their performance is suboptimal and cannot be practically used to infer a JPEG image's ISO speed. Furthermore, for similar reasons, though many noise level estimation algorithms \cite{ liu2006noise, 6607209, 5459476, nam2016holistic} may work well on RAW images to give clues about an image's ISO speed, JPEG images still pose challenges. As JPEG is one of the most common image formats, being able to identify a JPEG image's ISO speed is a prerequisite for ISO specific correlation prediction.
\begin{figure}[]
\centering
\includegraphics[width=.48\textwidth]{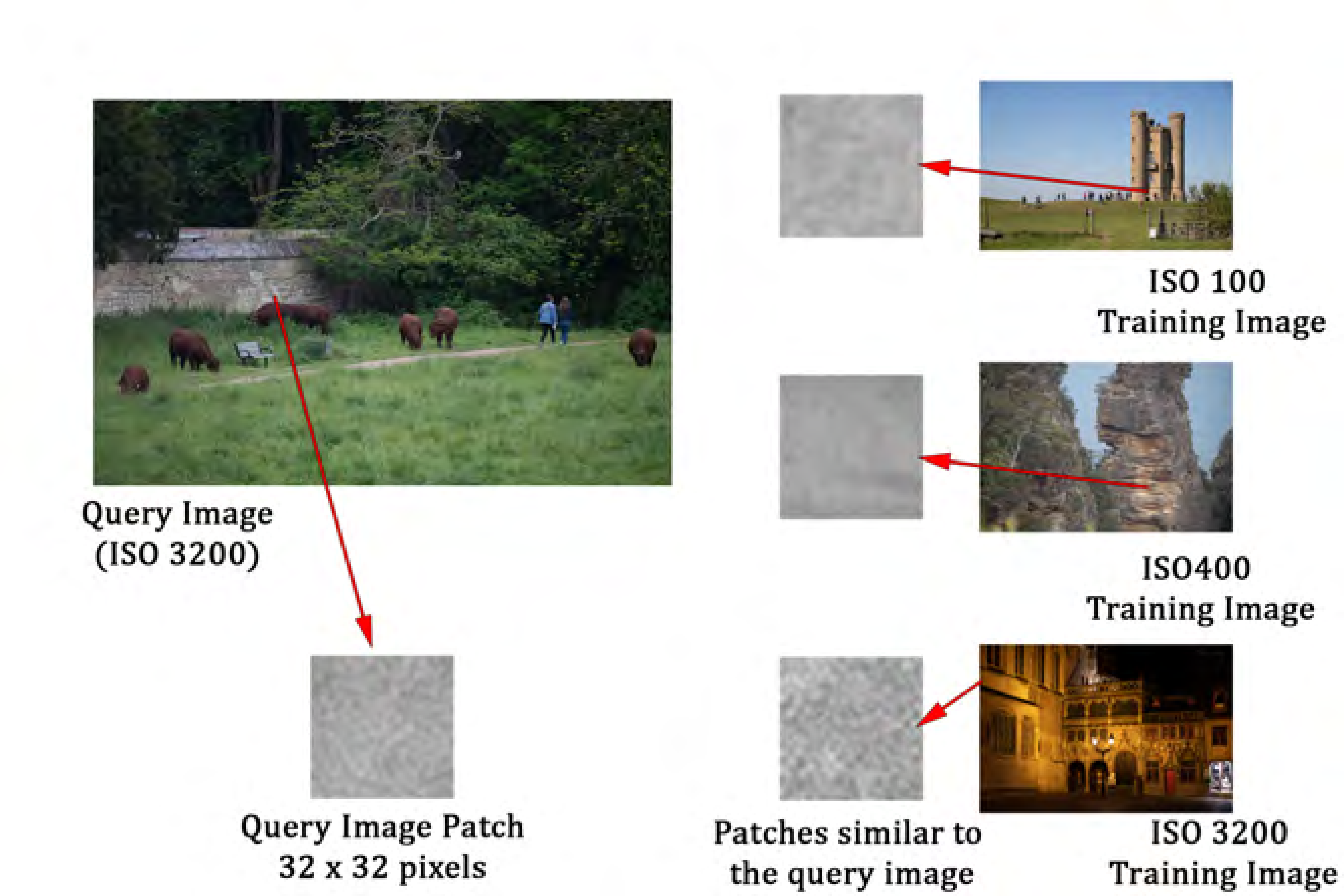}
\caption{A demonstration of the idea behind the proposed ISO speed inferring  method. We expect patches from different images to show similar noise characteristics if they have similar content and the same ISO speed. The example shows a patch from an ISO 3200 query image. It shows similar noise characteristics with a patch of similar content from an ISO 3200 training image.}
\label{fig:ISO_determine}
\end{figure}

Though finding an accurate noise model for a JPEG image can be of great complexity, we can simplify this problem by making the following assumption: \textit{image patches from the same camera with similar content and JPEG quality factor should show similar noise characteristics if they are of the same ISO speed, and vice versa} as shown in Fig.\ref{fig:ISO_determine}. Thus, we propose a method called Content-based Inference of ISO Speed (CINFISOS, pronounced as /\textipa{'sin.f@.s@s}/) to determine an image's ISO speed by doing patch-wise noise comparison with patches of similar content from images taken with the same camera at different ISO speeds.

Consider the case when we have a query image, $\mathcal{Q}$, and $t$ candidate training sets, $\mathcal{S} = \{ S_1, ..., S_t\}$, each consists of multiple images and the sets are with different ISO speeds. We would like to find the set with the ISO speed closest to the query image $\mathcal{Q}$. The query image is first partitioned into a set of non-overlapping patches, $\mathcal{P} =\{p_i\}$, each patch of size $d \times d$ pixels. As we would like to use the patches to best represent the image's noise characteristics, patches with too many dark and saturated pixels in any color channel should be removed. We consider the patches in the RGB color space. For each pixel $q$ in the $j$th channel of the patch, $p_i^j$, the pixel is considered as dark or saturated if its pixel value $I(q)$ is not in the range $[\lambda_1, \lambda_2]$:
\begin{equation}
U(q) = 
\begin{cases}
1, \text{if } I(q) < \lambda_1 \text{ or } I(q) > \lambda_2 \\
0, \text{otherwise}
\end{cases}
\end{equation}
The $i$th patch is excluded from $\hat{\mathcal{P}}$ if $ \forall j  (\sum\limits_{q \in p_i^j} U(q) > \lambda_\tau d^2)$, when the ratio of the dark or saturated pixels in every channel of the patch is over a limit $\lambda_\tau$. In addition to removing the dark and saturated pixels, the image's noise characteristics can be better revealed by including only the less textured patches. Thus, we only keep $m$ least textured patches in $\mathcal{P}_\mathcal{Q}$, the set of patches that we believe can best represent the query image's noise characteristics. To evaluate how textured a patch is, we use the texture feature definition from \cite{chen2008determining} but extends its definition to patches of three color channels by a simple summation:
\begin{equation}
f_T(p_i) = \sum_{j=1}^3 (\frac{1}{d^2} \sum\limits_{q \in p_i^j} \frac{1}{1 + \text{var}_5(\textbf{F}(q))})
\end{equation}
where $\textbf{F}( )$ is the high-pass filter and $\text{var}_5()$ measures the variance of $5\times 5$ neighbourhood. The feature $f_T$ is defined in the range $[0,1]$ with lower values for more texture patches. We select $m$ least textured patches from $\hat{\mathcal{P}}$ to form the set of qualified query image patches $\mathcal{P}_\mathcal{Q}$:
\begin{equation}
\mathcal{P}_\mathcal{Q} = \{p_i | (p_i \in \hat{\mathcal{P}}) \land (f_T(p_i) > f_T{_{m+1}})\}
\end{equation}
$f_T{_{m+1}}$ is the texture feature of the $m+1$th least textured patch from $\hat{\mathcal{P}}$. As $\mathcal{P}_\mathcal{Q}$ only contains patches with relatively smooth texture, we can approximate their image content by applying a low pass filter. We implement the method of finding patches with similar content using a block-matching method similar to \cite{dabov2007image}. The distance between two patches in each color channel is measured as the Euclidean distance between the discrete cosine transforms (DCT) of the two with hard thresholding applied. And the overall distance between two patches is the summation of the distances in the three color channels:
\begin{equation}
\Delta(p_i, p_k) = \sum_{j=1}^3 \|\Gamma(\text{DCT}(p_i^j), \lambda_\text{DCT})- \Gamma(\text{DCT}(p_k^j),\lambda_\text{DCT}) \|_2
\end{equation}
where $\Gamma(x, \lambda_\text{DCT})$ is the hard thresholding operation:
\begin{equation}
\Gamma(x, \lambda_\text{DCT}) = 
\begin{cases}
 x, \text{if } x > \lambda_\text{DCT},\\
 0, \text{otherwise}	
\end{cases}
\end{equation}
For each patch $p_i$ in $\mathcal{P}_\mathcal{Q}$, from each candidate training set $S_k$, $n$ patches with the least distance to $p_i$ will be selected. Though the exhaustive search for the patches with the shortest distance is computationally expensive, this step can be easily parallelized. We call this set of selected patches as $\mathcal{P}_k^i$. We define the distance, which measures the sum of the absolute differences in noise characteristics in all three color channels from each patch $p_i$ in $\mathcal{P}_\mathcal{Q}$ to each candidate training image set $S_k$, as:
\begin{equation}
D(p_i, S_k) = \sum_{j=1}^3 (|\text{var}(p_i^j - \tilde{p}_i^j)  - \frac{1}{n}\sum \limits_{p_l \in \mathcal{P}_k^i} \text{var} (p_l^j - \tilde{p}_l^j)|)
\label{eqn:infer_distance}
\end{equation}
where $\tilde{p}_l^j$ is the low-pass filtered version of the patch $p_l$ of the $j$th channel:
\begin{equation}
\tilde{p}_l^j = \text{IDCT}(\Gamma(\text{DCT}(p_l^j), \lambda_\text{DCT}))
\end{equation}
For each patch $p_i$ in $\mathcal{P}_\mathcal{Q}$, it will have a vote for a candidate training set, $S_k$, who has the smallest $D(p_i, S_k)$. The candidate training set with the closest ISO speed to the query image will be determined by a simple majority vote from all the patches in $\mathcal{P}_\mathcal{Q}$. The ISO speed that receives the majority votes will be deemed as the ISO speed of the query image and the correlation predictor can be trained with the corresponding images.
\section{Experiments}
\subsection{Inferring ISO speed with CINFISOS}
\begin{table*}
\caption{Patch level accuracy of the proposed ISO speed inferring method on images from Warwick Image Forensics Dataset}
\begin{center}
\begin{tabular}{c|c|c|c|c|c|c|c}
 		      			 & ISO 100 & ISO 200 & ISO 400 & ISO 800 & ISO 1600 & ISO 3200 & ISO 6400 \\ \hline
 Canon 6D     		      &	0.954 &  0.843  &  0.619  &  0.740  &  0.637   &   0.755  &  0.806  \\ 
 Canon 6D MKII		      &	0.999 &  0.952  &  0.593  &  0.795  &  0.764   &   0.723  &  0.744   \\
 Canon 80D			      &	0.990 &  0.893  &  0.789  &  0.882  &  0.851   &   0.879  &  0.997  \\
 Canon M6	 		      &	1.000 &  0.937  &  0.682  &  0.869  &  0.836   &   0.911  &  0.983   \\		
 Fujifilm XA\_10\_1	      & 0.725 &  0.574  &  0.543  &  0.666  &  0.612   &   0.704  &  0.668 	\\	
 Fujifilm XA\_10\_2       & 0.699 &  0.602  &  0.587  &  0.673  &  0.578   &   0.625  &  0.654  \\ 
 Nikon D7200     	      &	0.998 &  0.891  &  0.734  &  0.859  &  0.800   &   0.860  &  0.918  \\ 
 Olympus EM10 MKII        &	0.989 &  0.928  &  0.631  &  0.694  &  0.712   &   0.697  &  0.731  \\
 Panasonic Lumix TZ90\_1  &	0.961 &  0.802  &  0.554  &  0.581  &   N.A.   &   0.720  &  N.A. 	\\ 
 Panasonic Lumix TZ90\_2  &	0.908 &  0.769  &  0.580  &  0.576  &   N.A.   &   0.708  &  N.A.	\\  
 Sigma SdQuttro           &	0.881 &  0.825  &  0.512  &  0.716  &   0.565  &   0.601  &  0.642  \\
 Sony Alpha68     		  &	0.948 &  0.883  &  0.714  &  0.850  &   0.748  &   0.863  &  0.993  \\ 
 Sony RX100\_1     		  &	0.913 &  0.856  &  0.741  &  0.869  &   0.677  &   0.549       &  0.648  \\ 
 Sony RX100\_2     		  &	0.998 &  0.915  &  0.791  &  0.837  &   0.625  &   0.610       &  0.763  
\end{tabular}
\label{tab:final}
\end{center}
\vspace{-.2cm}
\end{table*}
To test the performance of the proposed CINFISOS, we conduct experiments on our Warwick Image Forensics Dataset. In the previous section, we concluded that for a correlation predictor trained with ISO speed $G_1$, reliable correlation predictions can be made for images taken with ISO speed in the range of $[G_1/2, 2G_1]$. Therefore, to select a correlation predictor trained with images of an ISO speed suitable for the image in question, the inferred ISO speed only needs to be within the one-stop range of the real value. As a result, we only need a few candidate training sets, $S_k$, to cover a broad range of ISO speeds to give reliable correlation predictions. 

In our experiments, for each camera in the Warwick Image Forensics Dataset, we have three candidate training sets with images of ISO speed 100, 800 and 6400, respectively (with the exception for the two Panasonic Lumix TZ90, of we select the ISO 3200 candidate training set instead of the ISO 6400 training set). These three ISO speeds are selected as they cover a broad range of commonly used ISO speeds. Besides, we deliberately avoid overlapping between the one-stop range of the ISO speeds, each of the three candidate ISO speed can predict for, to make it easier for the performance evaluation.

To apply CINFISOS, we set the following parameters. The size of each query image patch is $32\times 32$ pixels. $m = 50$ is the number of patches in the qualified query set $\mathcal{P}_\mathcal{Q}$. $\lambda_\text{DCT}$ is set to $13.0315$ in a similar manner as how it is set in \cite{dabov2007image}. For each query patch, we find $5$ similar patches from each candidate set. For each camera in the Warwick Image Forensics Dataset apart from the two Panasonic Lumix TZ90, we have 20 query images, each with ISO speed 100, 200, 400, 800, 1600, 3200 and 6400 in the JPEG format. Each candidate training set consists of 20 images. For the two Panasonic Lumix TZ90, in addition to the fact that ISO 6400 images are unavailable, we also excluded ISO 1600 query images as both ISO 800 and 3200 can be considered as inferred correctly.

We run the experiment with a desktop equipped with an Intel Core i7-9700K CPU. With the afore-mentioned setup, it takes around 130 seconds for CINFISOS to run on a full-resolution query image (e.g. $4160\times 6240$ pixels for an image from a Canon 6D MKII), including the exhaustive search for similar patches among 60 full-resolution training images. The patch-level accuracy, which measures the percentage of patches voting correctly for the inferred ISO speed, is reported in Table \ref{tab:final}. We notice that the accuracy varies greatly between cameras at different ISO speeds but the accuracy is above 0.5 in every case. It means that overall, every single patch is more likely to vote correctly. Given this patch-level accuracy, a $99.52\%$ accuracy at the image-level is observed with only 9 out of 1880 test images wrongly inferred.
\subsection{Forgery detection with ISO specific correlation prediction}
\begin{figure}
\center
\begin{tabular}{c}
(a) Canon M6 \\
   \includegraphics[width=.43\textwidth]{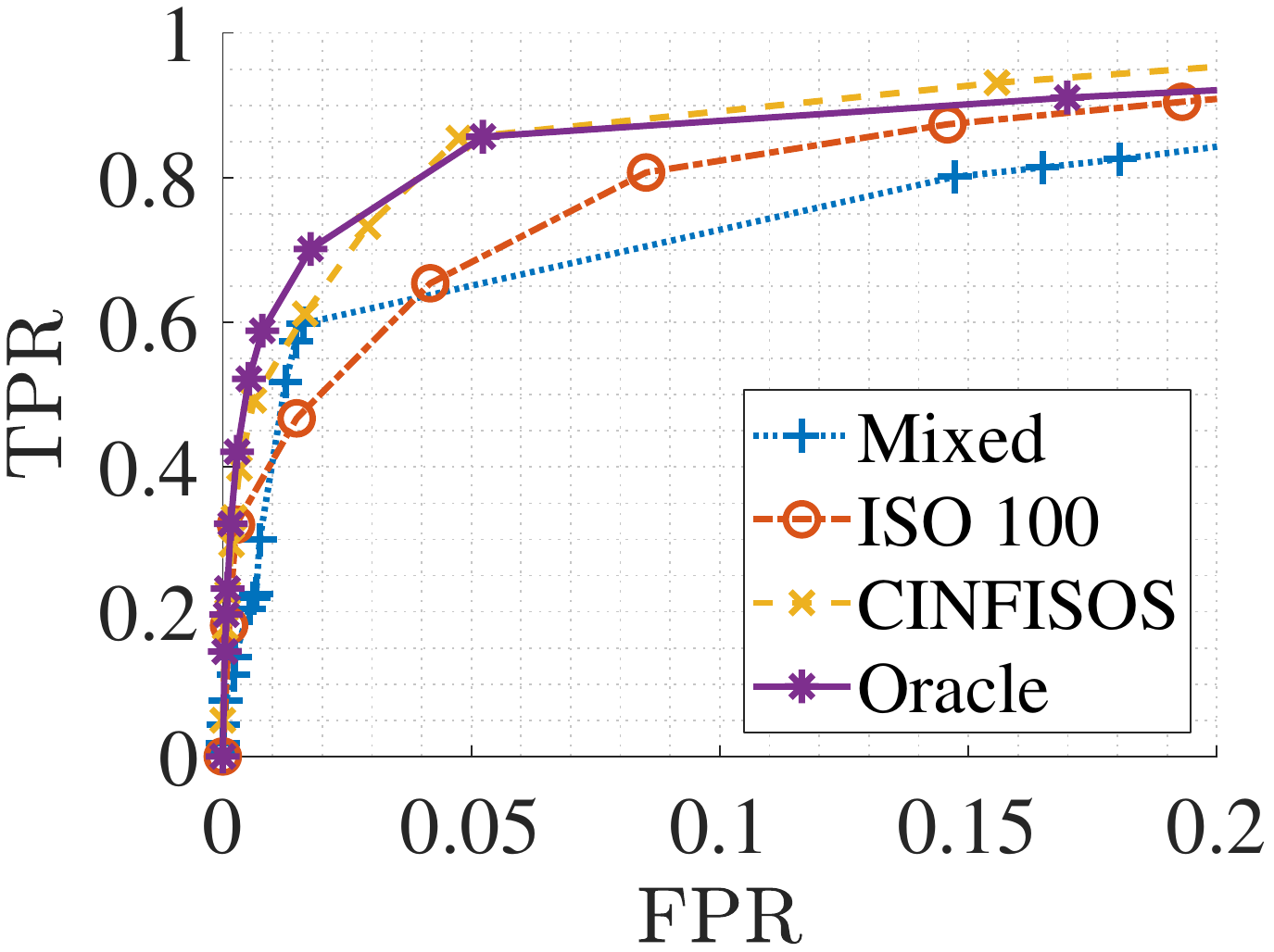}\\   (b) Sigma SdQuattro  \\ \includegraphics[width=.43\textwidth]{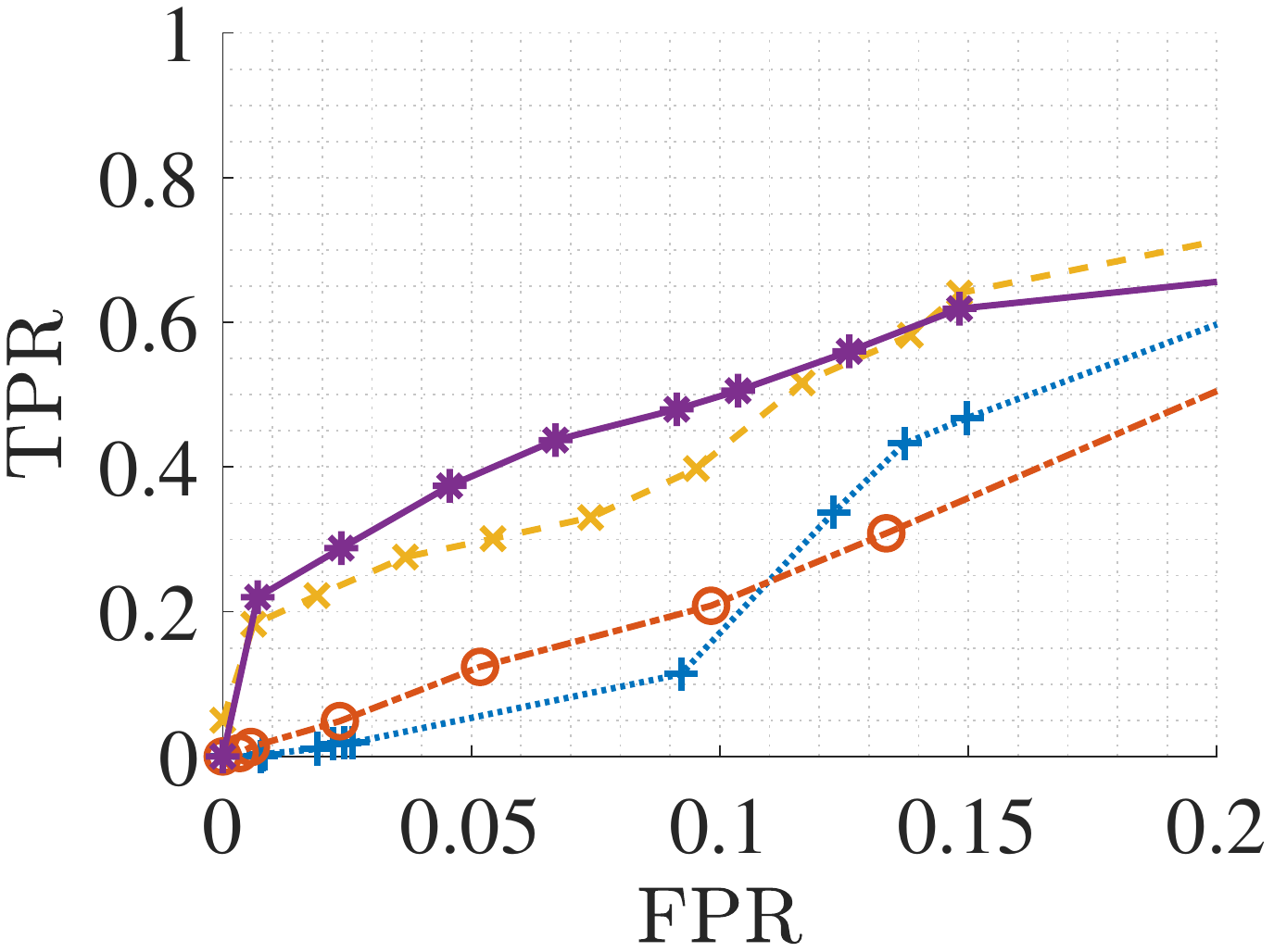}
   \end{tabular}
    \caption{ The ROC curves depicting the performance of detector with various correlation predictors tested on 560 synthetic forgery images of 7 different ISO speeds for two cameras (a) a Canon M6 and (b) a Sigma SdQuattro. Forgery detections are carried out with the Bayesian-MRF forgery detection algorithm from \cite{6719568} with correlation predictions generated from (i) a mixed ISO correlation predictor (ii) an ISO 100 correlation predictor (iii) the proposed ISO specific correlation prediction process with CINFISOS and (iv) the proposed ISO specific correlation prediction process with an oracle correlation predictor.}
    \label{fig:final_ROC}
\end{figure}
The high accuracy of CINFISOS in identifying the ISO speed of an image within its one-stop range allows us to conduct the proposed ISO specific correlation prediction process even when we do not know the test image's ISO speed. Thus, we would like to test the performance of the proposed ISO specific correlation prediction process in terms of forgery detection.

We apply the Bayesian-MRF forgery detection algorithm\cite{6719568} on the synthetic forgery images from two cameras: a Canon M6 and a Sigma SdQuattro for the test. The images are the same as the ones used in Section \ref{sec:set}. There are 560 synthetic images from each camera and they are equally distributed over 7 different ISO speeds (namely ISO speed 100, 200, 400, 800, 1600, 3200 and 6400). We carry out the proposed ISO specific correlation prediction process in two ways: (a) using the proposed CINFISOS to determine whether a correlation predictor is suitable for the test image, and (b) with an oracle correlation predictor. With the aforementioned one-stop range setting, we only need three correlation predictors, namely an ISO 100, an ISO 800 and an ISO 6400 correlation predictor to cover the whole range of the ISO speeds we need to predict for with CINFISOS. We apply CINFISOS on each synthetic image to determine which of the three correlation predictors should be used to produce the predictions of each image. The oracle correlation predictor uses a matching-ISO correlation predictor for each image according to its ISO speed information. We trained 7 different correlation predictors for the 7 different ISO speeds presented in this test, each with 20 natural images, to realize the oracle correlation predictor.

We compare the forgery detection results by our proposed ISO specific correlation prediction process against the results by using correlation predictions with a mixed ISO correlation predictor and an ISO 100 correlation predictor. Mixed ISO correlation predictors represent the situation when we select training images randomly without considering the images' ISO speeds. Thus, the mixed ISO correlation predictors' performance can be viewed as the baseline for the forgery detection results when we disregard the impact from ISO speed on correlation prediction completely. For each camera, the mixed ISO correlation predictor is trained with 20 training images randomly selected from the 60 images of three different ISO speeds. The ISO 100 correlation predictor is the same as the one used in our proposed ISO specific correlation prediction process. We vary the interaction parameter $\beta$ and the probability prior $p_0$ for the Bayesian-MRF forgery detection method to generate the enveloping ROC curves. Each data point on the curve is generated by summing the detection results of the 560 synthetic images from each camera. The ROC curves for the detection results are shown in Fig.\ref{fig:final_ROC}. We focus on the low false positive rate range of $[0,0.2]$.

Unsurprisingly, the detection result from the oracle correlation predictor comes as the best above all the predictors for both cameras. However, the detection results based on the proposed CINFISOS are comparable to the oracle correlation predictor's ones. It shows the effectiveness of the proposed CINFISOS and validates that the one-stop range for ISO speed prediction is a feasible choice without significantly sacrificing the forgery detection performance. In comparison, the mixed ISO and ISO 100 correlation predictors have worse performance. Though in Fig.\ref{fig:ROCS}, we have noticed that the ISO 100 correlation predictor can predict well for images with ISO speed up to 1600, its poor performance on images of higher ISO speed is evident. Thus, it is not a good choice to use a correlation predictor trained with low ISO speed for all the images. To conclude, the proposed ISO specific correlation prediction process shows superior performance in terms of forgery detection.

\section{Conclusion}
In this work, we did both analytical and empirical studies on the impact of different camera sensitivity (ISO speed) settings on PRNU-based digital forensics. First, we show how the correlation between an image's noise residual with the device's reference PRNU can be dependent on the image's ISO speed. With this dependency in mind, we empirically show how mismatched ISO speeds may influence the correlation prediction process. Thus, we proposed an ISO-specific correlation prediction process to be used in PRNU-based forgery detection. To address the problem that the information about the ISO speed of an image may not be available, a method called Content-based Inference of ISO Speed (CINFISOS) is proposed to infer the image's ISO speed from its content. Clear improvements are observed in correlation predictions and forgery detection results by applying our proposed ISO specific correlation prediction process with CINFISOS. By pointing out the influence of camera sensitivity setting on PRNU-based forensic methods, the provided solutions from this work can make the forensic analysis more reliable and trustworthy.
\section*{Acknowledgment}
This work is supported by the EU Horizon 2020 Marie Sklodowska-Curie Actions through the project entitled Computer Vision Enabled Multimedia Forensics and People Identification (Project No. 690907, Acronym: IDENTITY)
\bibliography{refs}
\begin{IEEEbiography}[{\includegraphics[width=1in,height=1.25in,clip,keepaspectratio]{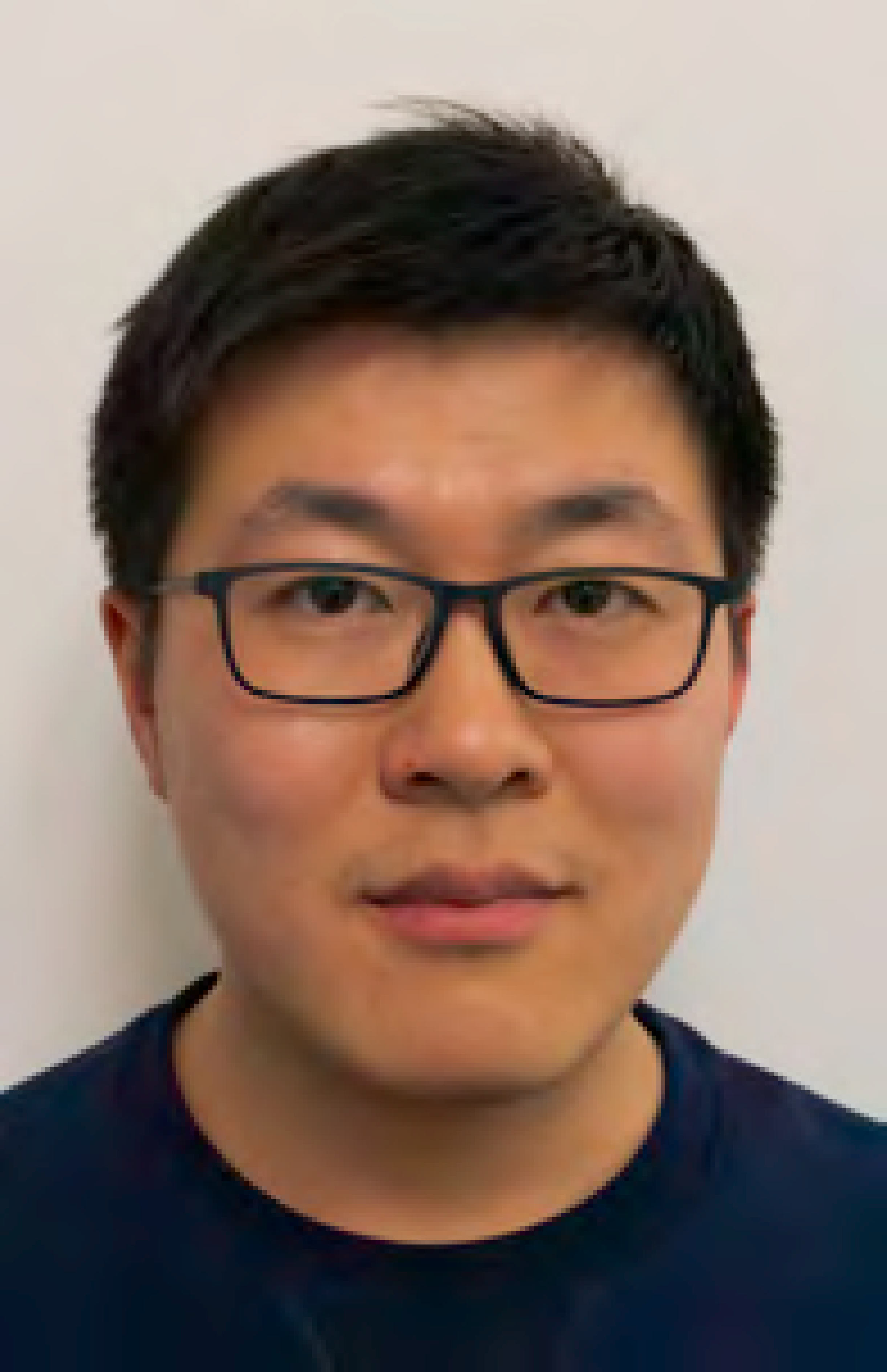}}]{Yijun Quan}
received the BA degree in Natural Science from Trinity College, University of Cambridge, UK, in 2015, the MSc degree in Computer Science from University of Warwick, UK, in 2016. He is currently a Ph.D. Candidate at University of Warwick. He was a visiting scholar to South China University of Technology (SCUT) under Marie Sklodowska-Curie fellowship in 2018. His research interests include multimedia forensics and security, machine learning, image processing and computational photography.
\end{IEEEbiography}
\begin{IEEEbiography} [{\includegraphics[width=1in,height=1.25in,clip,keepaspectratio]{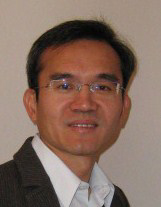}}]{Chang-Tsun Li} received the BSc degree in electrical engineering from National Defence University (NDU), Taiwan, in 1987, the MSc degree in computer science from U.S. Naval Postgraduate School, USA, in 1992, and the PhD degree in computer science from the University of Warwick, UK, in 1998. He was an associate professor of the Department of Electrical Engineering at NDU during 1998-2002 and a visiting professor of the Department of Computer Science at U.S. Naval Postgraduate School in the second half of 2001. He was a professor of the Department of Computer Science at the University of Warwick (UK) until January 2017 and a professor of Charles Sturt University (Australia) from January 2017 to February 2019. He is currently a professor of the School of Information Technology at Deakin University, Australia. His research interests include multimedia forensics and security, biometrics, data mining, machine learning, data analytics, computer vision, image processing, pattern recognition, bioinformatics, and content-based image retrieval.  The outcomes of his multimedia forensics research have been translated into award-winning commercial products protected by a series of international patents and have been used by a number of police forces and courts of law around the world. He is currently the EURASIP Journal of Image and Video Processing (JIVP) and Associate Editor of IET Biometrics. He involved in the organisation of many international conferences and workshops and also served as member of the international program committees for several international conferences. He is also actively contributing keynote speeches and talks at various international events.
\end{IEEEbiography}

\end{document}


\title{Supplementary Material for `\textit{On Addressing the Impact of ISO Speed upon PRNU and Forgery Detection}'}

\author{Yijun Quan, Chang-Tsun Li~\IEEEmembership{Senior Member,~IEEE}}

\maketitle

\section*{A case study on JPEG compression's impact images of different ISO speeds}
\begin{figure*}
\begin{center}
\begin{tabular}{ccc}
(a) Canon 6D MKII &  (b) Nikon D7200 & (c)  Sigma SdQuattro \\ 
\includegraphics[width = .3\textwidth]{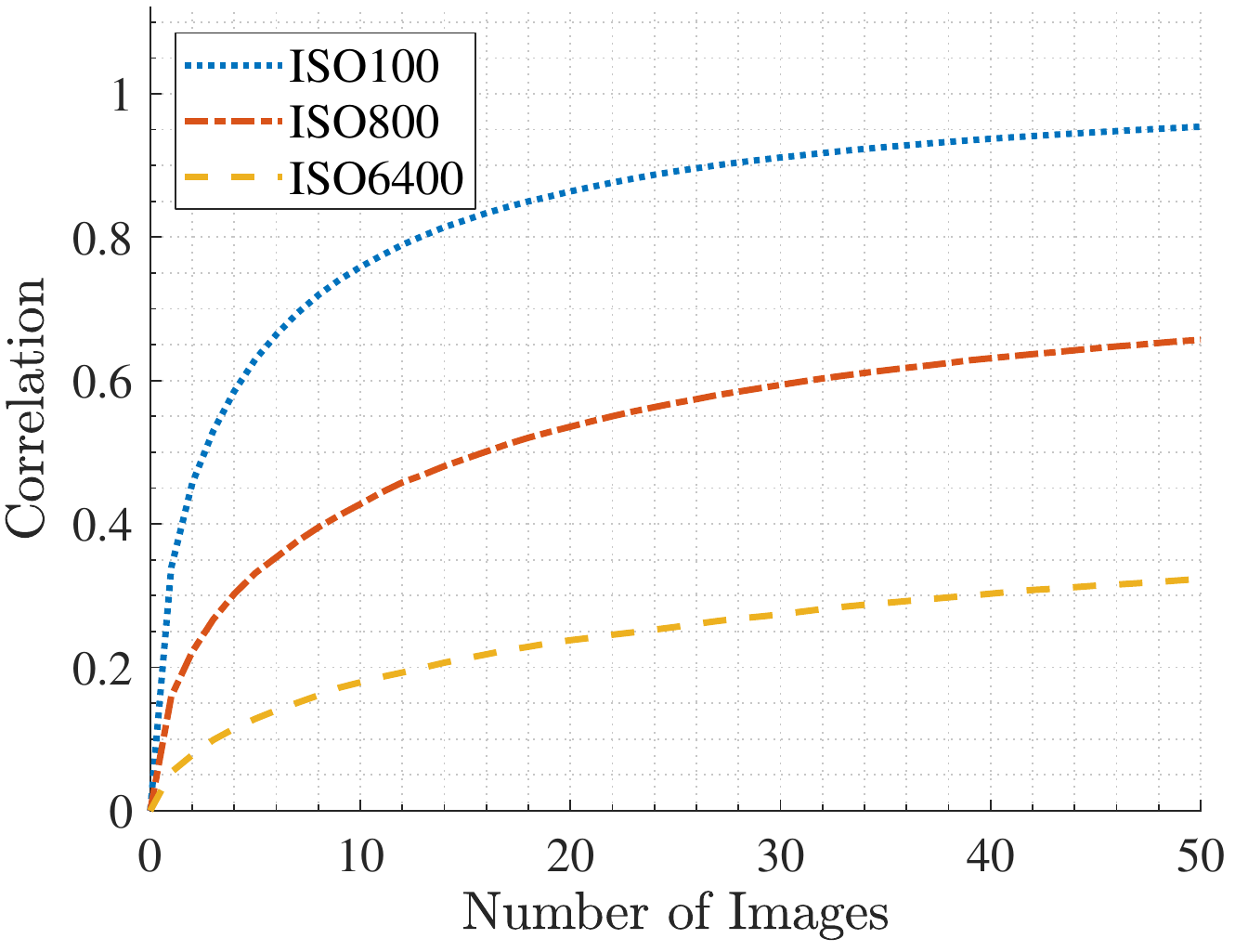} &
 \includegraphics[width = .3\textwidth]{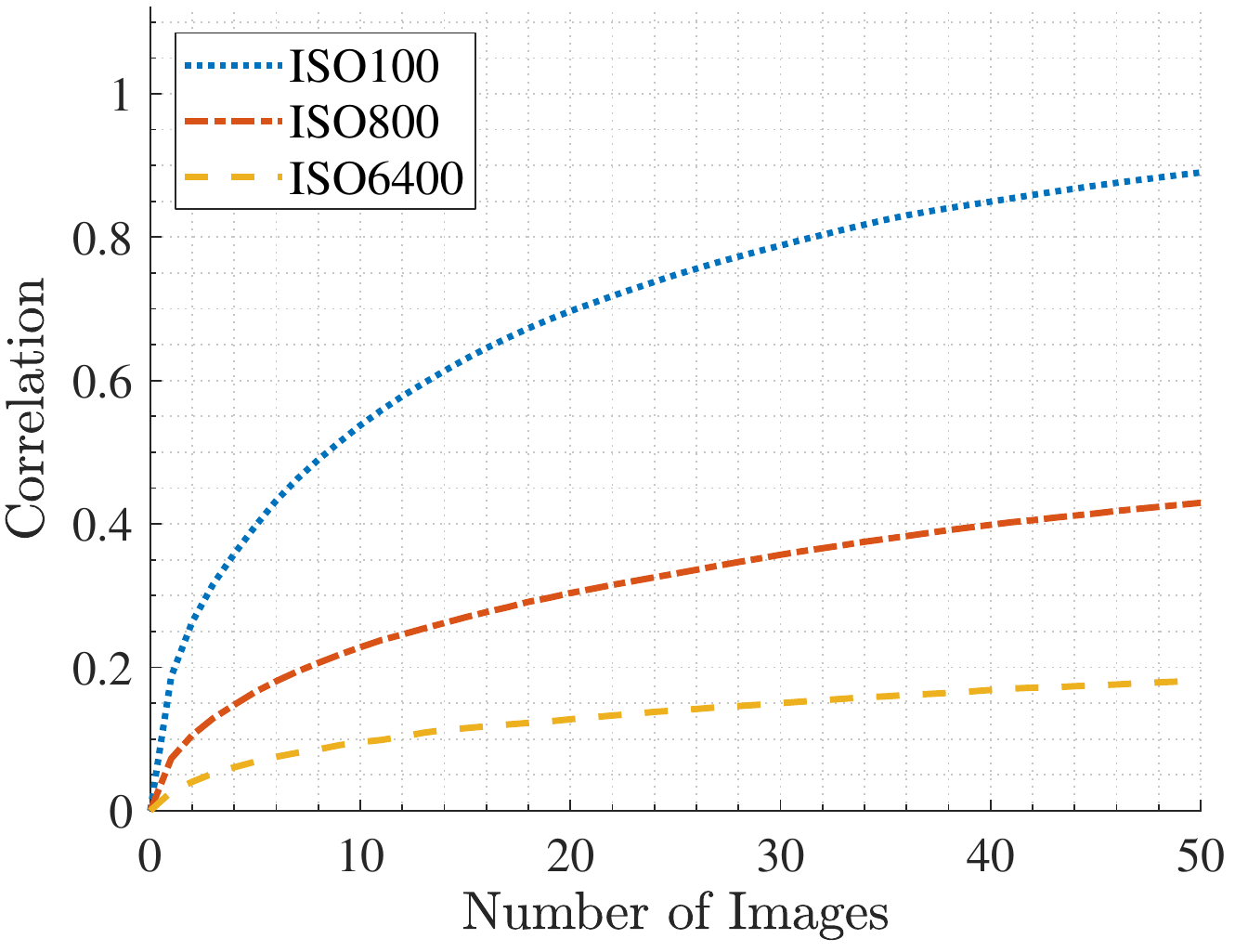} &
 \includegraphics[width = .3\textwidth]{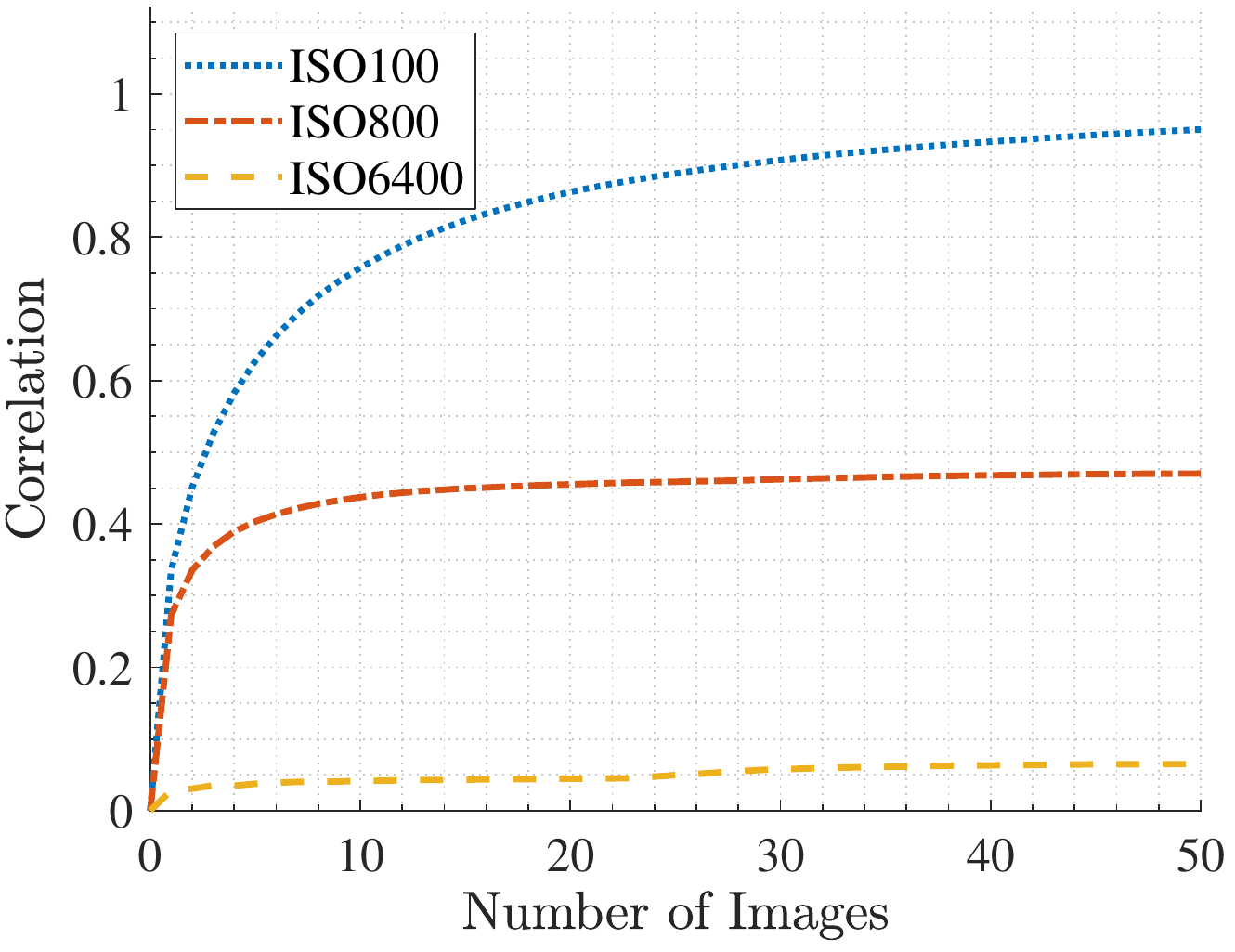} \\
\end{tabular}
\caption{The plots show how the number of JPEG images used for reference PRNU extraction may affect the quality of the extracted reference PRNU from three cameras: (a) Canon 6D MKII, (b) Nikon D7200 and (c) Sigma SdQuattro. We use the correlation between the extracted reference PRNU with another reference PRNU extracted from 100 flat-field images of ISO speed 100 to indicate the quality of the extracted reference PRNU.}
\label{fig:converge}
\end{center}
\vspace{-.2cm}
\end{figure*}
\begin{figure*}
\begin{center}
\begin{tabular}{ccc}
\multicolumn{3}{c}{Canon 6D MKII}\\
\includegraphics[width = .28\textwidth]{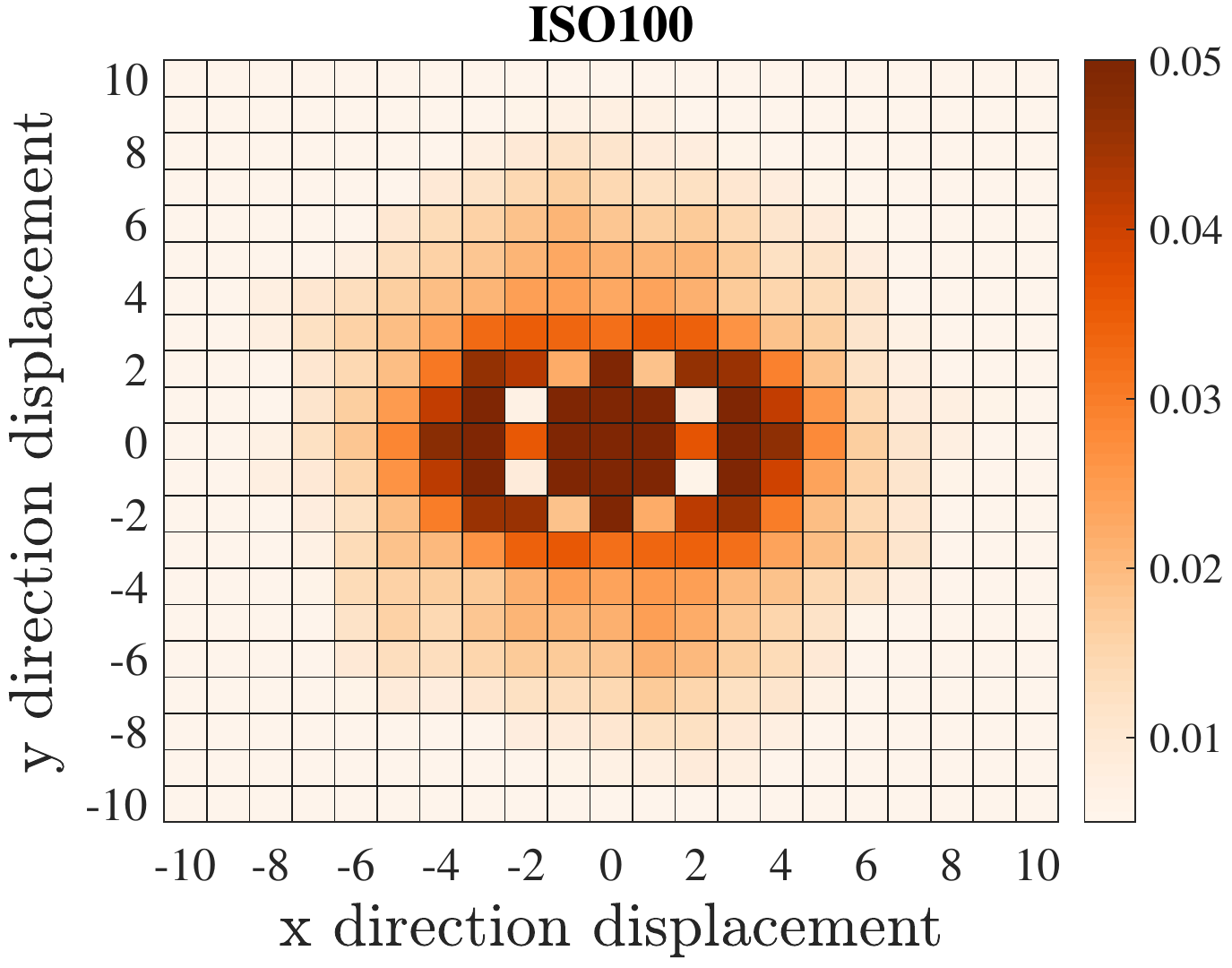} & \includegraphics[width = .28\textwidth]{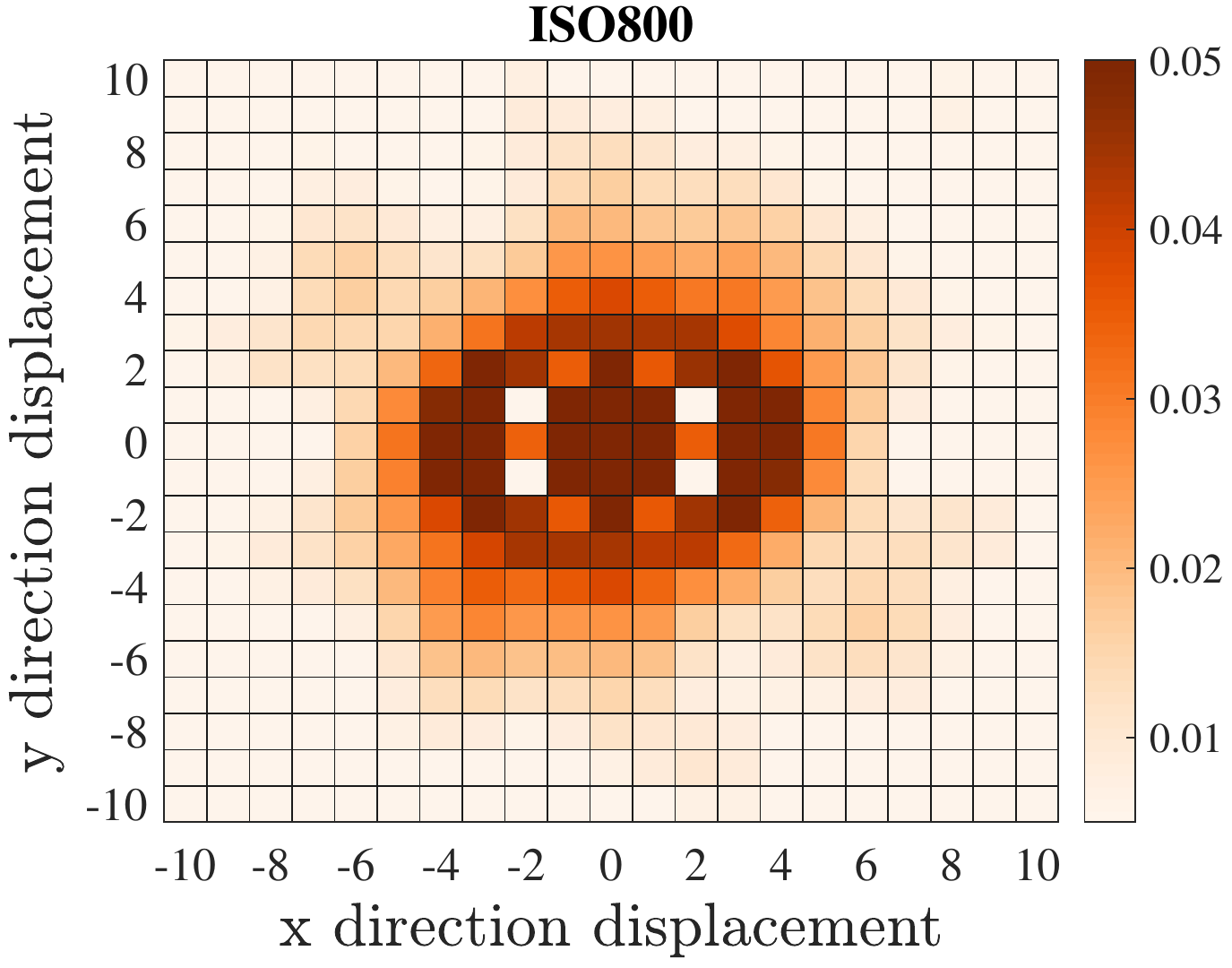} & \includegraphics[width = .28\textwidth]{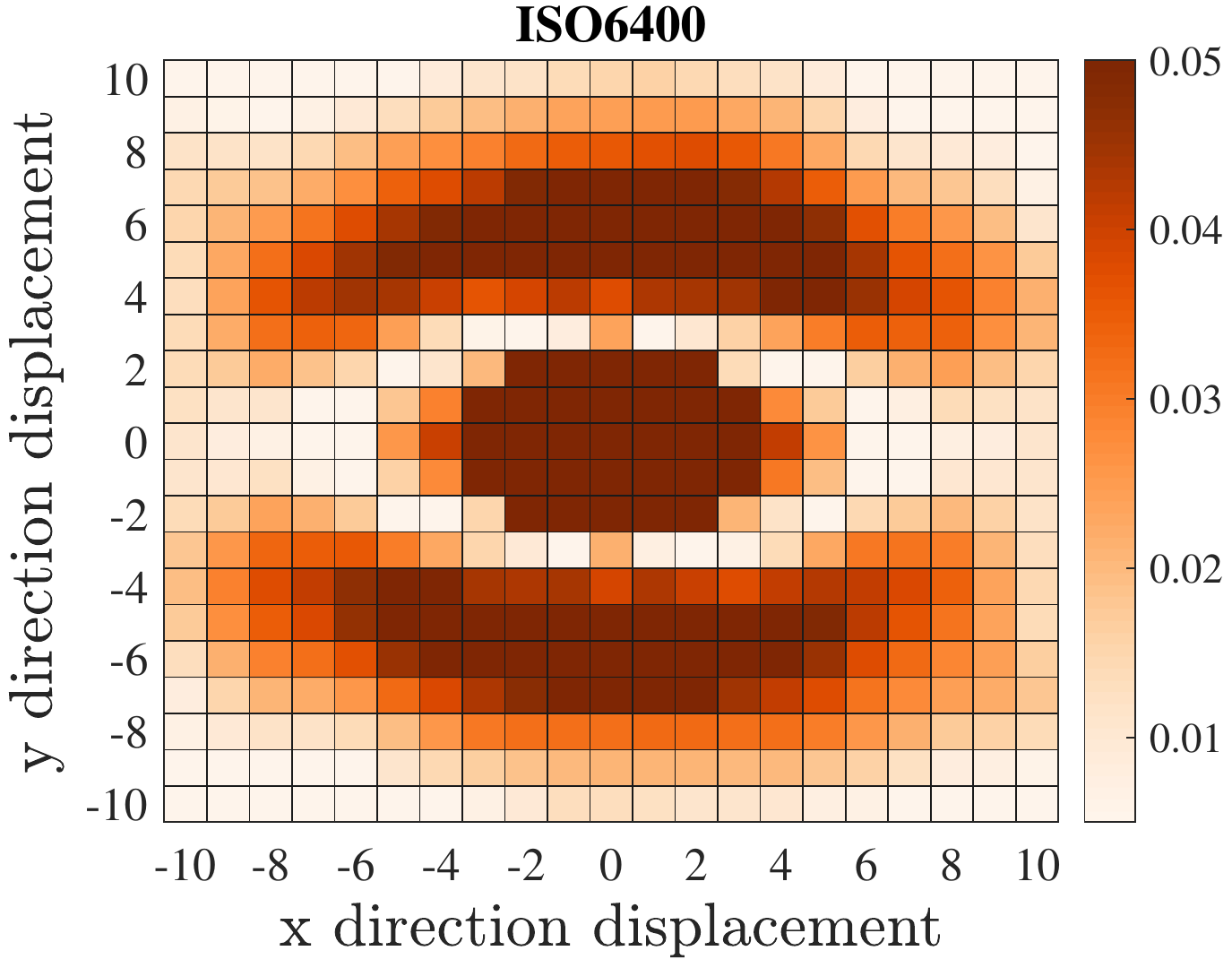}\\
\multicolumn{3}{c}{Nikon D7200}\\
\includegraphics[width = .28\textwidth]{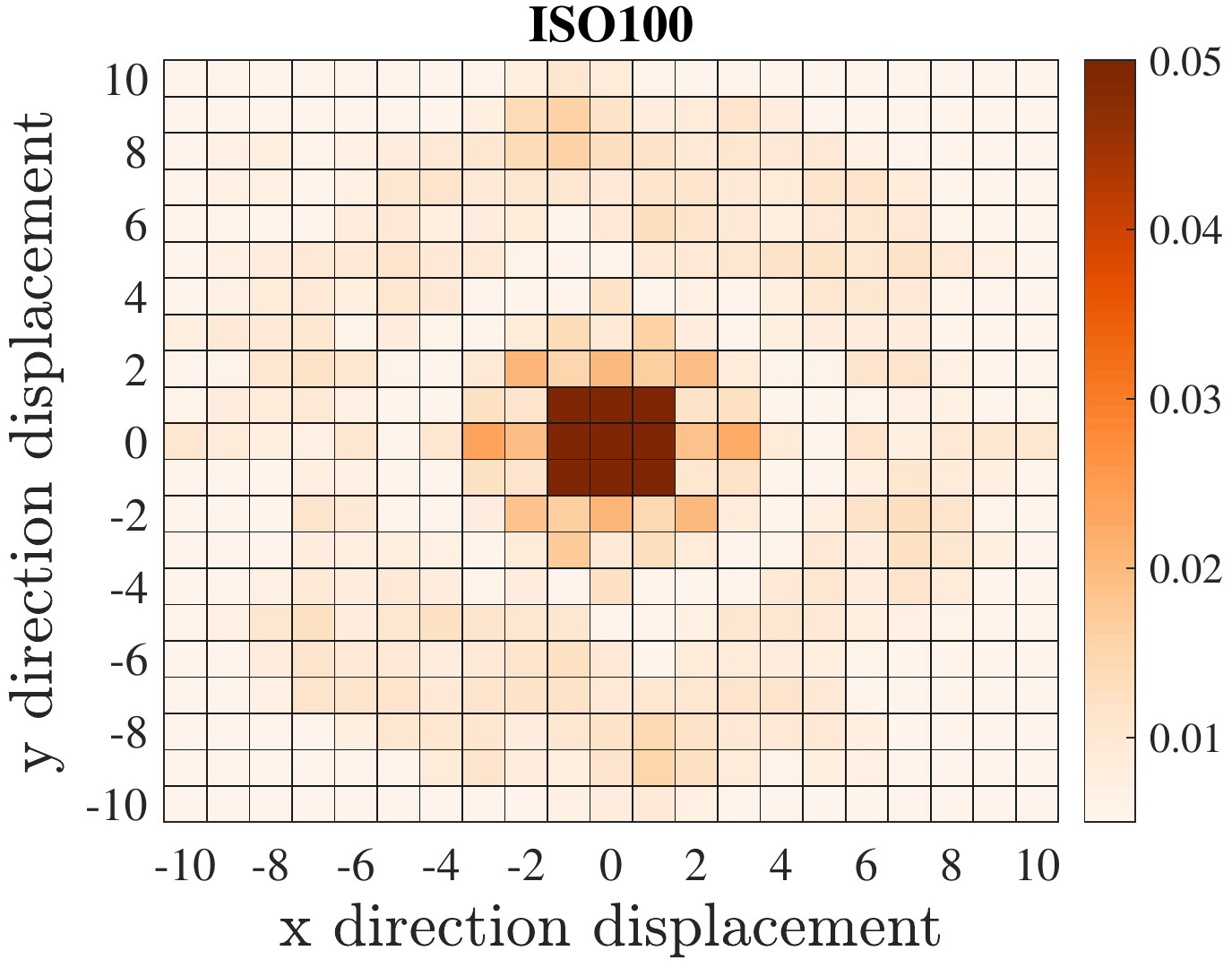} & \includegraphics[width = .28\textwidth]{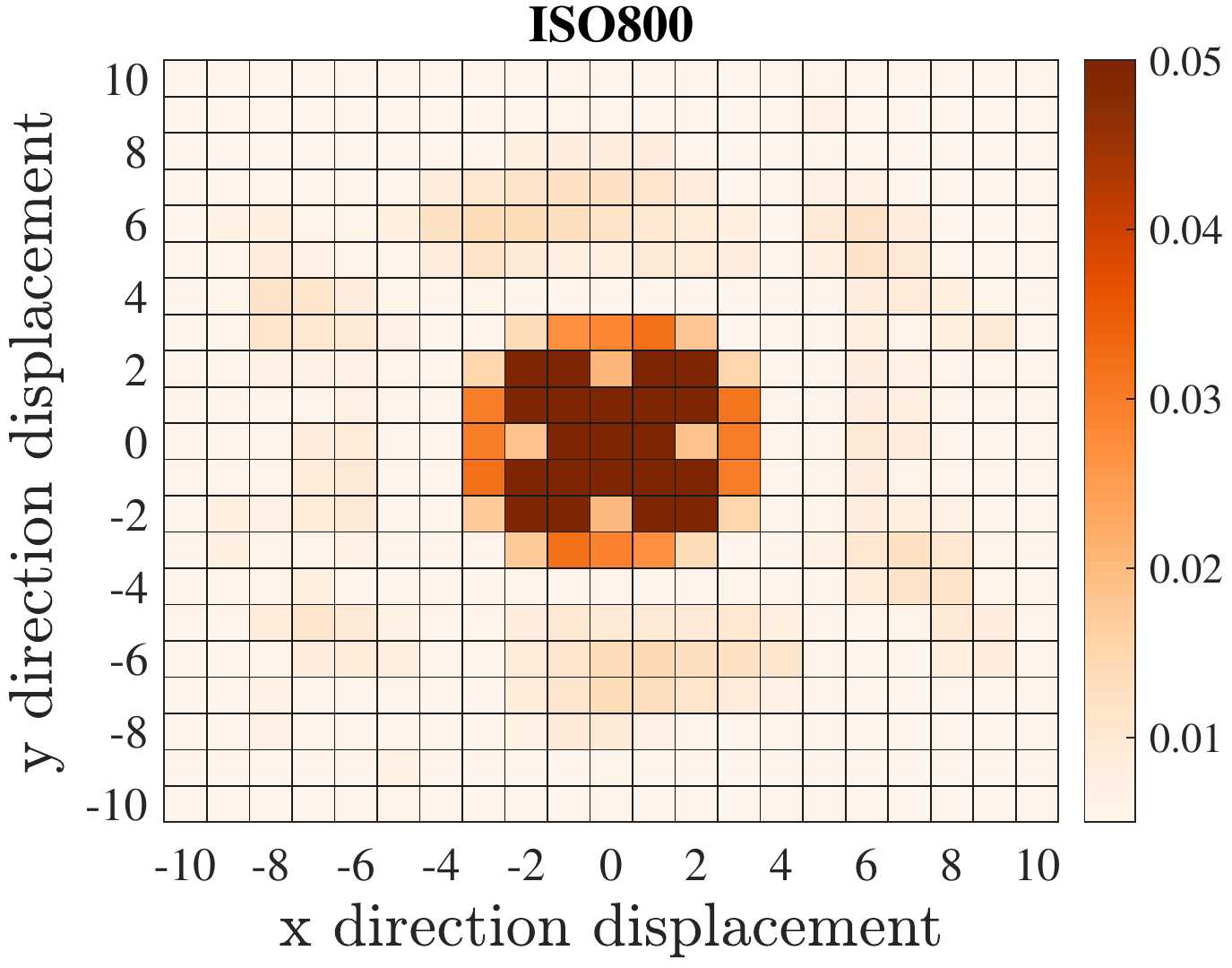} & \includegraphics[width = .28\textwidth]{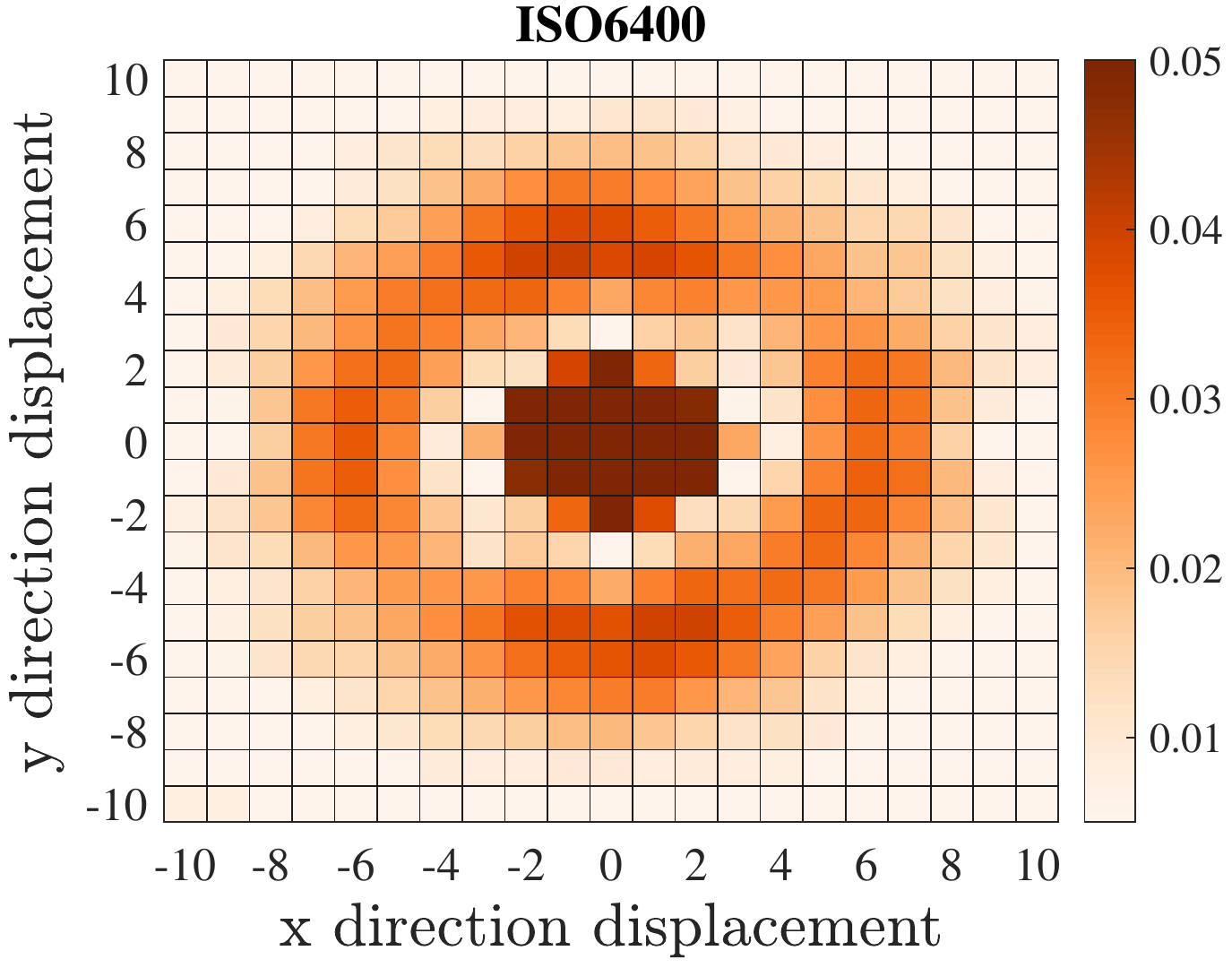} \\
\multicolumn{3}{c}{Sigma SdQuattro}\\
\includegraphics[width = .28\textwidth]{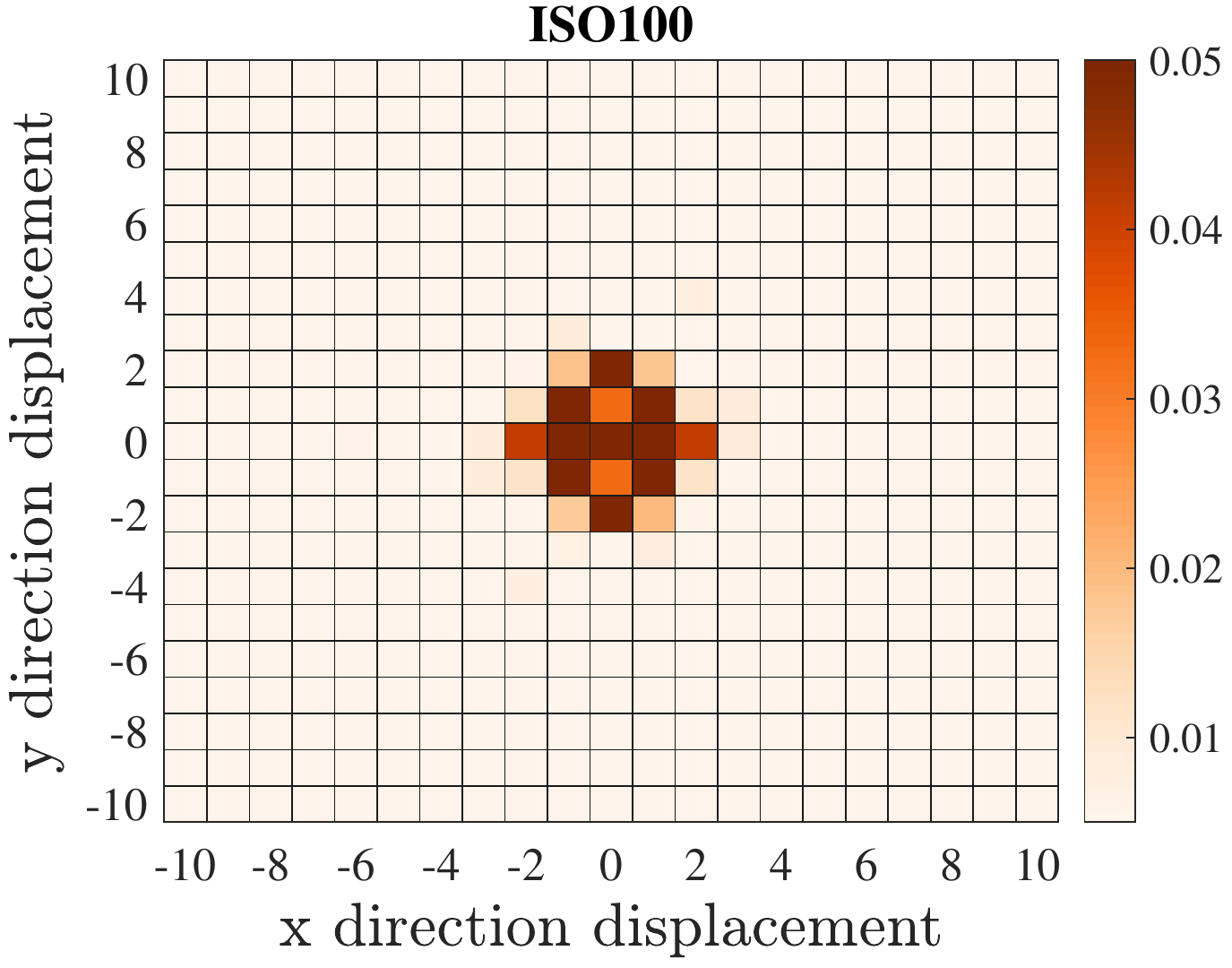} & \includegraphics[width = .28\textwidth]{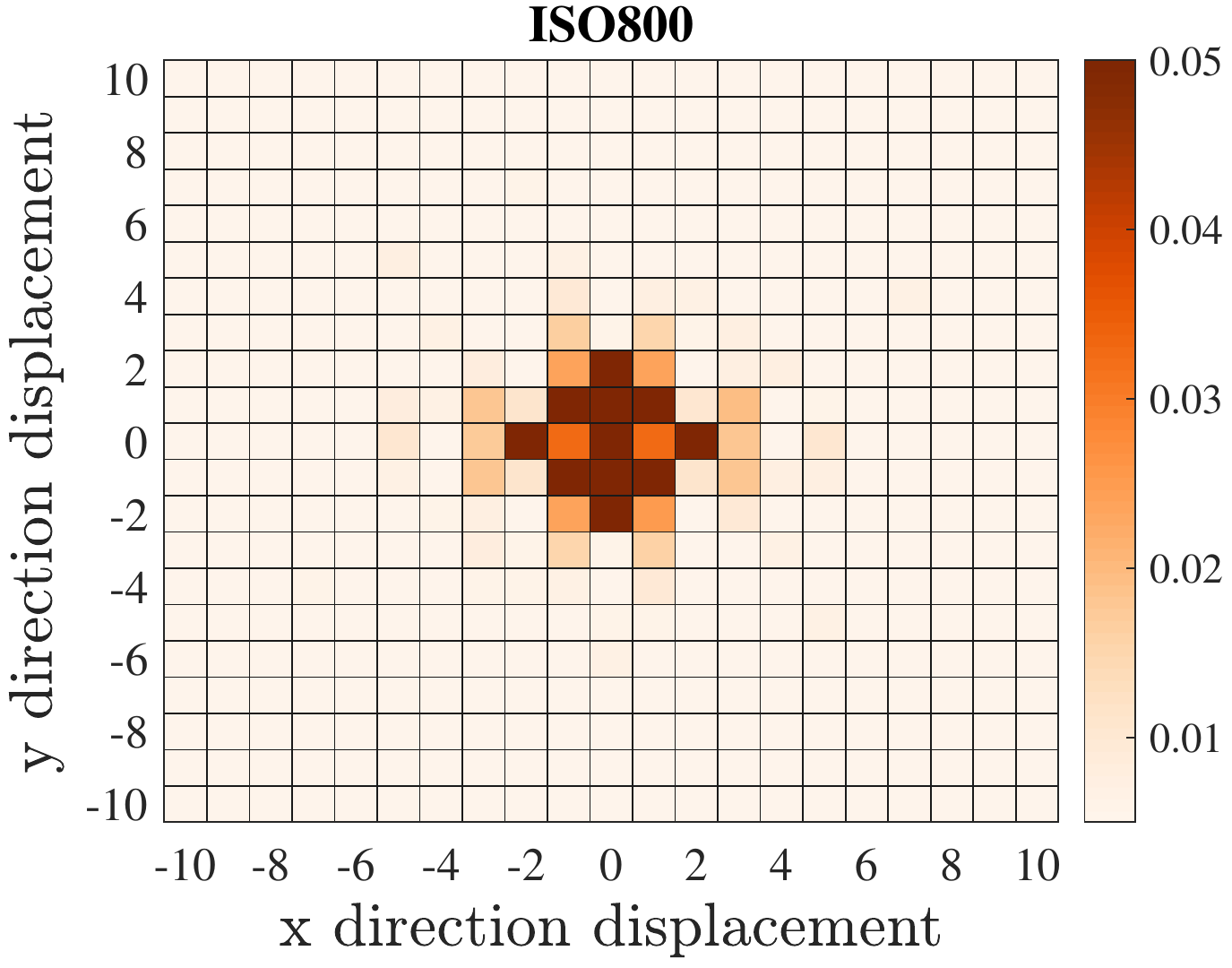} & \includegraphics[width = .28\textwidth]{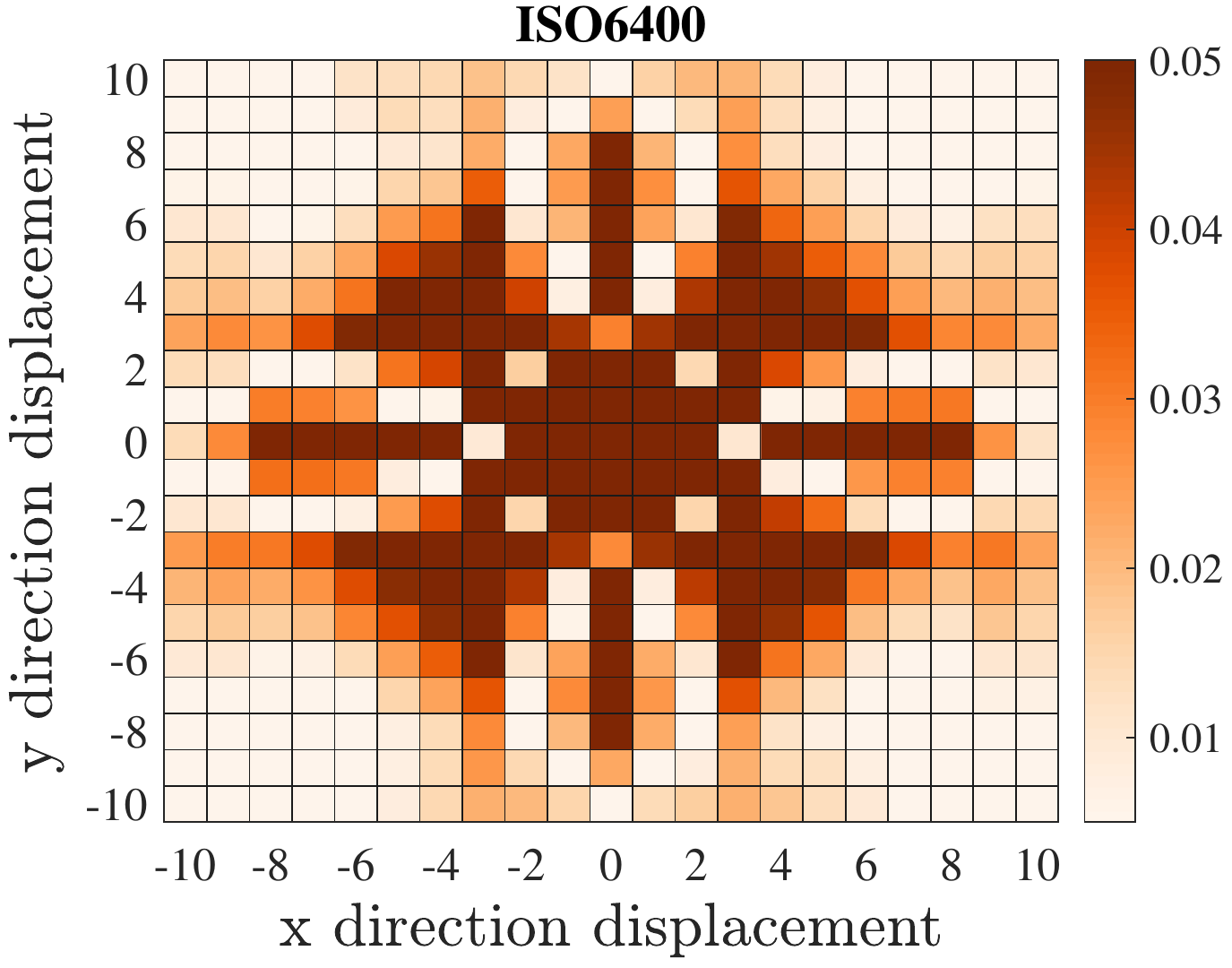}
\end{tabular}
\caption{The auto-correlation of noise residuals from images of different ISO speeds from 3 cameras. Rather than a single peak at $(0,0)$, auto-correlations have values spread over multiple pixel ranges. As the figure focuses on how far the spreading of auto-correlation reaches, the color bar focus on the range of $[0,0.05]$. Values bigger than the upper limit $0.05$ are also colored in dark brown. }
\label{fig:auto_corrs}
\vspace{-.2cm}
\end{center}
\end{figure*}
As different ISO speeds can introduce different levels of noise to the images, such behavior would impact the reference PRNU extraction process as well. A typical method to extract a device's reference PRNU is averaging the noise residuals from flat-field images (images of flattened content, e.g. pure color boards, etc.). The use of flat-field images can mostly avoid the distortion due to image content (e.g. texture, edges, etc.). For a flat-field RAW image, we can approximate its noise model according to Equation (1), which means its noise residual consists of both the PRNU and PRNU-irrelevant parts. By averaging the noise residuals of multiple flat-field images from the same device with similar quality of the PRNU, their PRNU-irrelevant part can get attenuated and thus a better approximation of the PRNU can be obtained.

In real-life forensics, the images available for the reference extraction may not be RAW images but in some compressed format, e.g. JPEG images, similar behavior is expected. Also, due to the influence of ISO speed, it is reasonable for us to expect that, with the same number of images, the reference PRNU extracted from lower ISO speed images would be of better quality than the one extracted from images with higher ISO speeds. To verify this, we test the PRNU extracted from varying numbers (from 1 to 50) of flat-field images with different ISO speeds (100, 800 and 6400) from three cameras, namely a Canon 6D MKII, a Nikon D7200 and a Sigma SdQuattro. The images used in this test are JPEG images of a flat color panel and are straight out of the three cameras. To ensure a fair comparison between different ISO speeds, we set the JPEG compression quality to the best available setting on each camera for every ISO speed. The quality of the extracted PRNUs is examined by computing the correlation between them and another reference PRNU of the same camera, which in our case is computed from 100 images with ISO speed of 100. We call the PRNUs generated from the one hundred ISO 100 images as the \textit{sample PRNUs}.

In theory, the three sample PRNUs may still differ from the ground truth slightly, the correlation between them and the one extracted from the test images are still representative to tell the difference between the quality of PRNU generated from images of different ISO speeds, as we can see from Fig. \ref{fig:converge}. From the figures, we can confirm that the lower ISO speed generates PRNU of better quality. For each ISO speed, the correlation increases as the number of images used to extract the reference PRNU increases.

Furthermore, for different ISO speeds from the same camera, the correlation curves shown in Fig.\ref{fig:converge} tend to converge to different values. It means that no matter how many images are used to extract the reference PRNU, the ones from images of higher ISO speeds can be of worse quality than the ones from a sufficient number of images of lower ISO speeds. Such a phenomenon suggests the incompatibility of PRNU's extracted from higher ISO speed images with the sample PRNU.

 We found that this is mainly due to the reason that the PRNU signal remaining in higher ISO images is more prone to the low-pass filtering like JPEG compression despite the images are saved under the same JPEG compression quality factor. As the higher ISO flat-field images are noisier, they have more high frequency signals in the image. Thus, when a low-pass filter is applied to them to reduce the amount of high frequency signal remaining in the images, the PRNU signal, which is also a high frequency signal, is more likely to be vitiated. 
 
 In Fig. \ref{fig:auto_corrs}, we use the auto-correlations of the flat-field images' noise residual to demonstrate such an effect. For a random noise, as the value of each pixel is independent, its auto-correlation should have a single peak at $(0, 0)$ and is zero elsewhere. However, due to post-processing, especially the JPEG compression, the auto-correlation will spread over multiple pixels and the extend of this spreading can be an indicator of how severe the post-processing may distort the extracted noise residual. From Fig. \ref{fig:auto_corrs}, for each of the three cameras, we clearly see the trend that as the ISO speed increases, the spreading reaches further. Furthermore, the symmetric spreading shapes as we observed from the plots for the ISO 6400 images are more likely to be from JPEG compression, which compresses signals of a certain frequency in the images. Color interpolation (also known as demosaicking) at each pixel involves the colors of the pixels within a neighborhood, which means the color at each pixel does ``spread" across a certain neighborhood. Interestingly, unlike the Bayer filter used on the sensors in Canon 6D MKII and Nikon D7200, the Foveon X3 sensor in the Sigma SdQuattro has a stacked color filtering array, which does not require color interpolation. The spreading of the auto-correlation can still be observed with the Sigma SdQuattro. This evidence further justifies that the further spreading of the auto-correlation is more likely to be caused by JPEG compression instead of color interpolation.